\documentclass[useAMS,usenatbib,usegraphicx]{mn2e}

\usepackage{color}
\usepackage{amssymb}
\usepackage{multirow}

\newcommand{\lexp}{\mathop{\langle}}    % left bracket for expectation value
\newcommand{\rexp}{\mathop{\rangle}}    % right bracket for expectation value
\newcommand\Eqn[1]     {Eq.\,(\ref{#1})}
\newcommand\Eqns[2]    {Eqs\,(\ref{#1}) and~(\ref{#2})}

\newcommand{\be}{\begin{equation}}
\newcommand{\ee}{\end{equation}}
\newcommand{\ba}{\begin{eqnarray}}
\newcommand{\ea}{\end{eqnarray}}

\newcommand\Fig[1]     {Fig.\,\ref{#1}}

\def\bk{{\bf k}}
\def\nn{\nonumber}

\def\m{\rm m}
\def\h{\rm h}

\def\bx{{\bf x}}
\def\bk{{\bf k}}

\def\bq{{\bf q}}

\def\dx{{\rm d}^3\!{\bf x}}
\def\dk{{\rm d}^3{\bf k}}
\def\dq{{\rm d}^3{\bf q}}

\def\de{\delta}
\def\dek{\delta}

\def\dD{\delta^{\rm D}}

\def\NL{\rm NL}

\def\cyc{\rm cyc}

\def\Vu{V_{\mu}}

\def\Msol{h^{-1}M_{\odot}}

\def\Mpc{\, h^{-1}{\rm Mpc}}

\def\Gpccube{\, h^{-3} \, {\rm Gpc}^3}
\def\kMpc{\, h \, {\rm Mpc}^{-1}}

\newcommand{\nbar}{\overline{n}}

\newcommand{\rhobar}{\overline{\rho}}

\def\vn{{\bf n}}

\def\vx{{\bf x}}

\def\vq{{\bf q}}

\def\vk{{\bf k}}

\def\hh{\rm hh}
\def\hm{\rm hm}
\def\mmm{\rm mmm}
\def\hmm{\rm hmm}
\def\hhm{\rm hhm}
\def\hhh{\rm hhh}

\def\bmm{\rm bmm}

\def\brr{\begin{array}}
\def\err{\end{array}}

\def\be{\begin{equation}}
\def\ee{\end{equation}}
\def\bea{\begin{eqnarray}}
\def\eea{\end{eqnarray}}
\def\ba{\begin{eqnarray}}
\def\ea{\end{eqnarray}}

\title[Modelling halo bias using the bispectrum]
{Modelling large-scale halo bias using the bispectrum}

\author[J. E. Pollack, R. E. Smith \& C. Porciani]
{Jennifer E. Pollack$^{1}$\thanks{E-mail: jpollack@astro.uni-bonn.de}, 
Robert E. Smith$^{2,1}$\thanks{res@physik.unizh.ch}, \&
Cristiano Porciani$^{1}$\thanks{porciani@astro.uni-bonn.de}\\
$^{1}$Argelander Institut f\"ur Astronomie der Universit\"at Bonn, Auf dem
H\"ugel 71, D-53121 Bonn, Germany\\
$^{2}$Institute for Theoretical Physics, University of Zurich, Zurich, CH 8037}

\begin{document}

\maketitle

%%%%%%%%%%%%%%%%%%%%%%%%%%%%%%%%%%%%%%%%%%%%%%%%%%%%%%%

\begin{abstract} 
  We study the relation between the density distribution of tracers
  for large-scale structure and the underlying matter distribution --
  commonly termed bias -- in the $\Lambda$CDM framework.  In
  particular, we examine the validity of the local model of biasing at
  quadratic order in the matter density.  This model is characterized
  by parameters $b_1$ and $b_2$.  Using an ensemble of $N$-body
  simulations, we apply several statistical methods to estimate the
  parameters. We measure halo and matter fluctuations smoothed on
  various scales. We find that, whilst the fits are reasonably good,
  the parameters vary with smoothing scale. We argue that, for
  real-space measurements, owing to the mixing of wavemodes, no
  smoothing scale can be found for which the parameters are
  independent of smoothing.  However, this is not the case in Fourier
  space. We measure halo and halo-mass power spectra and from these
  construct estimates of the effective large-scale bias as a guide for
  $b_1$. We measure the configuration dependence of the halo bispectra
  $B_{\rm hhh}$ and reduced bispectra $Q_{\rm hhh}$ for very
  large-scale $k$-space triangles.  From this data we constrain $b_1$
  and $b_2$, taking into account the full bispectrum covariance
  matrix. Using the lowest-order perturbation theory, we find that for
  $B_{\rm hhh}$ the best-fit parameters are in reasonable agreement
  with one another as the triangle scale is varied; although, the fits
  become poor as smaller scales are included. The same is true for
  $Q_{\rm hhh}$. The best-fit values were found to depend on the discreteness
  correction. This led us to consider halo-mass cross-bispectra. The
  results from these statistics supported our earlier findings. We
  then developed a test to explore whether the inconsistency in the
  recovered bias parameters could be attributed to missing
  higher-order corrections in the models.  We prove that low-order
  expansions are not sufficiently accurate to model the data, even on
  scales $k_1\sim0.04\kMpc$. If robust inferences concerning bias are
  to be drawn from future galaxy surveys, then accurate models for the
  full nonlinear bispectrum and trispectrum will be essential.
\end{abstract}

%%%%%%%%%%%%%%%%%%%%%%%%%%%%%%%%%%%%%%%%%%%%%%%%%%%%%%%


%%%%%%%%%%%%%%%%%%%%%%%%%%%%%%%%%%%%%%%%%%%%%%%%%%%%%%%
%%%%%%%%%%%%%%%%%%%%%%%%%%%%%%%%%%%%%%%%%%%%%%%%%%%%%%%

\section{Introduction}

The accurate estimation and modelling of higher-order clustering
statistics in current and future galaxy redshift surveys has the
potential to act as a powerful probe for cosmological physics.  The
higher-order connected moments, beginning at lowest order with the
three-point correlation function and its Fourier analogue, the
bispectrum, when interpreted within the gravitational instability
paradigm, encode important information regarding the growth of
large-scale structure \citep{Matarreseetal1997,Scoccimarroetal1998}.
Their measurements also provide insight into the statistical nature of
the primordial fluctuations
\citep{FryScherrer1994,SefusattiKomatsu2007,Nishimichietal2010,Baldaufetal2011}
and the cosmological parameters \citep{Sefusattietal2006}.  Another
attribute of three-point statistics, and the focus of our study, is
their capability to probe the manner in which an observable tracer
population of objects, such as galaxies, is related to the
unobservable matter distribution -- termed the `bias'
\citep{Kaiser1984,DekelRees1987,FryGaztanaga1993,DekelLahav1999,Catelanetal2000}.

If the primordial fluctuations were Gaussian, as appears to be the
case \citep{Komatsuetal2010short}, then the statistical properties of
the initial fields are fully characterized by the power spectrum, with
all higher-order connected correlators vanishing. However,
gravitational instability leads to the coupling of Fourier modes and
this generates a hierarchy of non-vanishing connected correlators,
each of which has a precise characteristic mathematical structure. The
matter bispectrum is thus an inherently nonlinear quantity, whose
signal depends on closed triangles in Fourier space. In theory, this
should vanish at early times and on scales large enough where linear
theory is valid. If galaxy bias is local and linear, then the
bispectrum of the observable tracers is proportional to the matter
bispectrum. If on the other hand, bias is local and nonlinear then the
triangle configuration dependence of the signal is modified, and this
happens in a very precise and calculable way. Thus the bispectrum can
be used to constrain the bias
\citep{FryGaztanaga1993,Matarreseetal1997,Scoccimarroetal1999a}.

There is a long and rich history of measurements of three-point
statistics from galaxy surveys, going all the way back to
\citet{PeeblesGroth1975}.  However, attempts to constrain the
nonlinearity of galaxy bias from galaxy redshift surveys have only
been performed over the last decade. \citet{Feldmanetal2001} and
\citet{Scoccimarroetal2001b} both analyzed the IRAS survey using the
bispectrum, and found a negative quadratic bias; although, due to
small sample size the constraints were rather
weak. \citet{Verdeetal2002short} analyzed the 2dFGRS survey, also
using the bispectrum approach, and claimed that the flux-limited
sample was an unbiased tracer of the dark matter. A subsequent
analysis of the final 2dFGRS data set by \citet{Gaztanagaetal2005},
using the 3-point correlation function, contradicted this: using
information from weakly non-linear scales ($R\sim 6-27$ $\Mpc$) the
unbiased case ($b_1=1$ and $b_2=0$) was excluded at the order of
9$\sigma$. More recently a number of authors have analyzed various
data releases of the SDSS (\citealt{Nishimichietal2007} -- DR3,
\citealt{McBrideetal2011} -- DR6 and \citealt{Marin2010} --
DR7). These all claim a non-zero quadratic bias term for most of the
samples within the dataset. Obviously, these variations of the results
with survey and statistical method require an explanation.

Whilst the local bias model can be used to test whether the bias is
linear or nonlinear, a significant detection of non-zero nonlinear
bias does not imply that we have understood the bias. In order to
believe that these measurements are meaningful, we need to be sure
that the local model is indeed the correct model for interpreting
data. This is currently an open question. Attempting to shed light on
this subject is one of the aims of this paper. Over the past few
years, the local model of galaxy bias has been scrutinized by a number
of authors
\citep{Heavensetal1998,GaztanagaScoccimarro2005,Smithetal2007,GuoJing2009,Maneraetal2010,
  RothPorciani2011,ManeraGaztanaga2011}. However, no firm conclusions
have yet been reached.

In this paper, we use a large ensemble of 40 mid-resolution, large
volume, pure dark matter $N$-body simulations, to test the validity of
the local bias model.  In this study we compare a selection of
different methods for determining the bias. We first present a
point-wise comparison of the halo and matter density fields smoothed
on certain scales. We also utilize the power spectrum to estimate a
large-scale effective bias parameter. Then we expend most of our
efforts on using the bispectrum and reduced bispectrum approach for
constraining the bias. Besides the auto-bispectra, we also present,
for the first time, measurements of the halo-matter cross-bispectra:
$B_{\h\h\m}$ and $B_{\h\m\m}$, and $Q_{\h\h\m}$ and $Q_{\h\m\m}$. The
value of these new statistics becomes apparent when correcting for
shot-noise effects. Finally, we perform a numerical test that allows
us to sharply illuminate the importance of terms in the theory that
are beyond the tree-level expansions typically used.

The paper is divided up as follows: in \S\ref{sec:SPT} we present the
theory for the matter bispectrum and the local bias model.  In
\S\ref{sec:sims}, we provide details of the numerical simulations used
in this work. In \S\ref{sec:HaloBiasMethods}, we present the results
for the bias parameters from the various commonly used simple
estimators. In \S\ref{sec:BispHaloEst} we present the estimation of
the bias from the bispectrum. Then, in \S\ref{sec:CrossSpec} we
present measurements of bias from the cross-bispectra. In
\S\ref{sec:biasbyhand} we present a test of the importance of terms in
the theory that are beyond tree level. Finally in
\S\ref{sec:discussion} and \S\ref{sec:conclusions} we discuss our
findings and present our conclusions.

%%%%%%%%%%%%%%%%%%%%%%%%%%%%%%%%%%%%%%%%%%%%%%%%%%%%%%%
%%%%%%%%%%%%%%%%%%%%%%%%%%%%%%%%%%%%%%%%%%%%%%%%%%%%%%%

\section{Theoretical Overview}\label{sec:SPT}

%%%%%%%%%%%%%%%%%%%%%%%%%%%%%%%%%%%%%%%%%%%%%%%%%%%%%%%

\subsection{Standard perturbation theory dynamics}

In the fluid approximation, the gravitational collapse of
collisionless cold matter structures in the expanding Universe, can be
fully characterized by specifying the evolution of the density
$\delta\rho(\bx)$ and the peculiar velocity $\delta{\bf v}(\bx)$
perturbations \citep{Bernardeauetal2002}. Focusing primarily on the density field, we work with models of the matter density contrast:
\be 
\de(\vx,t)\equiv \frac{\rho(\vx,t)-\rhobar(t)}{\rhobar(t)}\ ,
\ee
where $\bar{\rho}(t)$ is the mean matter density of the Universe.  In Fourier space, we define its corresponding Fourier representation,
$\dek(\vk)$, accordingly, as
\be
\dek(\bx) = \int \frac{\dk}{(2\pi)^3} \de(\bk)e^{-i\vk\cdot\vx} 
\ \ 
\Leftrightarrow 
\ \ 
 \dek(\vk) = \! \int \dx\, \de(\vx)e^{i\vk\cdot\vx} \ . \label{eq:FT}
\ee

It can be shown that the nonlinear equations of motion for
$\delta(\vk)$ can be solved exactly by perturbative expansions of the
type \citep{Juszkiewicz1981,Vishniac1983,Goroffetal1986}:
\begin{equation}
 \dek(\vk)   = \sum_{n = 1}^{\infty} a^n(t)\dek_{n}(\vk)\ ,
\end{equation}
and, where $\dek_n(\vk)$ is given by,
\ba
\dek_n(\vk) \!\!& = &\!\! \int \frac{{\rm d}^3\vq_1}{(2\pi)^3} \ldots \int \frac{{\rm d}^3\vq_n}{(2\pi)^3}\, (2\pi)^3 \dD(\vk - \vq_{1} - \cdots -\vq_n) \nn \\ 
&& \times F_n(\vq_1,\ldots ,\vq_n) \dek_1(\vq_1) \cdots \dek_1(\vq_n) \  .
\ea
The density kernel $F_n$ is the dimensionless, homogeneous, mode
coupling function that couples together the amplitudes and phases of
$n$ initial Fourier wavemodes $\{\delta(\vq_1), \ldots
,\delta(\vq_n)\}$. As was shown by
\citep{Goroffetal1986,Makinoetal1992,JainBertschinger1994}, the $n$th
order kernel may be constructed recursively from the lower-order
solutions.  Linear theory is thus represented by $F_1(\vq_1)=1$, and
the first nonlinear correction by $F_2$, where 
\ba
F_2(\vk_1,\vk_2) \!\! & = & \!\! \frac{5}{7}\ + \frac{\vk_1 \cdot \vk_2}{k_1 k_2} 
\left(\frac{k_1}{k_2} + \frac{k_2}{k_1}\right) + 
\frac{2}{7} \frac{(\vk_1 \cdot \vk_2)^2}{k_1^2 k_2^2} \label{eq:F2}\ .
\ea
The above approach defines the standard perturbation theory (hereafter
SPT). Before moving on, we note that the above statements are only
exactly true for the Einstein-de Sitter model. However, it has been
shown that the $F_2$ kernel is almost independent of
cosmology \citep{Fry1994,Bouchetetal1995,Hivonetal1995}. We therefore
adopt \Eqn{eq:F2} when dealing with the density at second order.

%%%%%%%%%%%%%%%%%%%%%%%%%%%%%%%%%%%%%%%%%%%%%%%%%%%%%%%

\subsection{From dynamics to statistics}

Owing to the stochastic nature of the density field, we are not
interested in reproducing a specific density field {\it per se}, but
instead in characterizing its statistical properties. In this work we
focus on 2- and 3-point correlation functions in Fourier space. These
we may write as:
\ba
\lexp \dek(\vk_1) \dek(\vk_2) \rexp  \!\! & \equiv & \!\!\!  
(2\pi)^3\,\dD(\vk_{12})P_{\rm mm}(\bk_1) \ ; \\
\lexp \dek(\vk_1) \dek(\vk_2) \dek(\vk_3) \rexp \! \! & \equiv & \! \! \! (2\pi)^3\,\dD(\vk_{123}) 
B_{\rm mmm}(\vk_1,\vk_2,\vk_3), 
\ea
where $P_{\rm mm}(\bk)$ and $B_{\rm mmm}(\bk_1,\bk_2,\bk_3)$
constitute definitions of the power and bispectrum. For the Dirac
delta functions we used the short-hand notation $\dD(\vk_{1\dots
  n})\equiv \delta^{D}(\bk_1+\dots+\bk_n)$ and these guarantee that
$P$ and $B$ are translationally invariant. This is an important
property for estimation, since it means that we should consider only
closed pairs and triangles in Fourier space: $\sum\vk_i=0$.

The perturbative expansion of the density field described in the
previous section implies that $P_{\rm mm}$ and $B_{\rm mmm}$ may also
be described in a perturbative fashion. Hence,
\ba
\left<\delta(\bk_1)\delta(\bk_2)\right> & = &
\left<
\left[\delta_1(\bk_1)+\delta_2(\bk_1)+\dots\right]\right.\nn\\
& & \times \left. \left[\delta_1(\bk_2)+\delta_2(\bk_2)+\dots\right]
\right>\ ; \\
\left<\delta(\bk_1)\dots\delta(\bk_3)\right> & = &
\left<
\left[\delta_1(\bk_1)+\delta_2(\bk_1)+\dots\right]\dots\right.\nn\\
& & \times \left. \left[\delta_1(\bk_3)+\delta_2(\bk_3)+\dots\right]
\right> \ .
\ea
Since we are assuming that the initial Fourier modes are Gaussianly
distributed, i.e. the phase of each initial mode is uniformly random
$\phi\in[0,2\pi]$, modes must cancel in pairs. Hence, Wick's theorem
applies, and so odd products of initial Fourier modes must vanish:
$ \left<\delta_1(\bk_1)\delta_2(\bk_2)\right>=
\left<\delta_1(\bk_1)\delta_1(\bk_2)\delta_1(\bk_3)\right>=0\ .  $
This leads us to write the perturbative expansions for $P_{\rm mm}$ and $B_{\rm mmm}$ as:
\ba
P_{\rm mm}(k) & = & P_{\rm mm}^{(0)}(k)+P_{\rm mm}^{(1)}(k)+\dots \ ; \\
B_{\rm mmm}(\bk_1,\bk_2) & = & B_{\rm mmm}^{(0)}(\bk_1,\bk_2) +B_{\rm mmm}^{(1)}(\bk_1,\bk_2)+\dots \nn \\
\ea
We shall refer to the lowest order terms in the expansions as
`tree-level' terms. For $P$, $P^{(0)}$ is simply the linear spectrum,
while for $B$ the tree-level term can be written:
\be B^{(0)}_{\rm mmm}(\bk_1, \bk_2) = 2\,P_{\rm mm}^{(0)}(k_1)\,P_{\rm mm}^{(0)}(k_2)\,F_2(\vk_1, \vk_2)
+ 2 \ {\rm cyc} \ .
\label{eq:bisptree}
\ee
In this work we shall mainly be dealing with
tree-level quantities; we now set: $P_{\rm mm}^{(0)}=P_{\rm mm}$ and
$B_{\rm mmm}^{(0)}=B_{\rm mmm}$, unless otherwise stated.

Another statistical quantity commonly used to explore galaxy
clustering is the reduced bispectrum
\citep{PeeblesGroth1975,Scoccimarroetal1998}, which can be defined:
\be Q_{\rm mmm}(\vk_1,\vk_2,\vk_3) \equiv \frac{B_{\rm mmm}(\vk_1,\vk_2,\vk_3)}{P_{\rm mm}(k_1)P_{\rm mm}(k_2) +
  2\,\cyc\,} . \label{eq:Qm} \ee
As will be made clear below, the importance of this statistic becomes
apparent when one considers non-Gaussian terms that are generated by
simple quadratic products of Gaussian fields. In this case $Q_{\rm
  mmm}$ simply scales as a constant.

%%%%%%%%%%%%%%%%%%%%%%%%%%%%%%%%%%%%%%%%%%%%%%%%%%%%%%%

\subsection{Halo Bias: Local Form}\label{ssec:localmodel}

In this study we investigate the relation between the
clustering of dark matter haloes and total matter. If galaxies are only
formed in dark matter haloes, as is the usual assumption for all
models of galaxy formation \citep{WhiteRees1978}, then understanding
the clustering of haloes is an essential component of any theory of
galaxy biasing \citep{Smithetal2007}. In the local model of halo
biasing, the number density of dark matter haloes of mass scale $M$,
smoothed over a scale $R$, can be expressed as a function of the local
matter density, smoothed on the scale $R$. This function may then be
Taylor expanded to give
\citep{FryGaztanaga1993,Coles1993,Moetal1997,Smithetal2007}:
\be
 \de_{\rm h}(\bx|M,R) = \sum_{j=0}^{\infty}  \frac{b_{j}(M)}{j!}[\delta(\bx|R)]^j \ ,
\label{eq:Eulbias}
\ee
where we defined the smoothed halo over-density to be,
$\de_{\rm h}(\vx|M,R) \equiv [\nbar_{\rm h}(\vx|M,R) - \nbar_{\rm
    h}(M)]/\nbar_{\rm h}(M)$. Owing to the fact that $\lexp \de_{\rm
  h} \rexp = 0$, the constant coefficient $b_0(M) = -
\sum_{j=2}^{\infty} b_j(M) \lexp \delta^j \rexp/j!$
\citep{FryGaztanaga1993}. Note that on Fourier transforming $\de_{\rm
  h}(\vx|M,R)$ the constant $b_0$ only contributes to the
unmeasurable $k=0$ mode. The terms $b_1(M)$ and $b_2(M)$ represent
the linear and first nonlinear bias parameters, respectively.  

In Fourier space, \Eqn{eq:Eulbias} can be written as:
\ba
\delta_{\rm h}(\bk|M,R) & = & b_1(M)\delta(\bk|R) \nn \\
& & \hspace{-1cm}+\frac{b_2(M)}{2}\int \frac{\dq_1 }{(2\pi)^3}
\delta(\bq_1|R)\delta(\bk-\bq_1|R) +\dots \ , \label{eq:biasLoc}\ea
where $\delta_i(\bq_j|R)\equiv W(|\bq_j|R)\delta_i(\bq_j)$.  If one
inserts the SPT expansions for the density into the local model, then,
up to second order in the density and bias, one finds:
\ba \delta^{\h}(\bk|M,R) & = & b_1(M)\left[\delta_1(\bk|R)+
  \delta_2(\bk|R)\right]\nonumber\\ 
& & \hspace{-0.5cm}+\frac{b_2(M)}{2}\int \frac{\dq_1}{(2\pi)^3}
\delta_1(\bq_1|R)\delta_1(\bk-\bq_1|R)\ .
\label{eq:localPT}  
\ea
Using this approach one may then find a perturbative expansion for the
halo power and bispectra:
\ba 
P_{\h\h}(M) \!& \!= \!& P^{(0)}_{\h\h}(M)+P^{(1)}_{\h\h}(M)+\dots \\
B_{\h\h\h}(M) \! & \!= \!& B^{(0)}_{\h\h\h}(M)+B^{(1)}_{\h\h\h}(M)+\dots \ .
\ea
Again, we refer to the lowest order terms in these expansions as
tree-level terms, and for these we have:
\ba
\widetilde{{\mathcal P}}^{(0)}_{\h\h}(k|M) & = & b_1^2(M) 
\widetilde{\mathcal P}_{\rm mm}(k)  \label{eq:PhhS}\ ;\\
\widetilde{\mathcal B}^{(0)}_{\hhh}(\vk_1, \vk_2|M) & = & 
b_{1}^3(M)\widetilde{\mathcal B}_{\rm mmm}(\vk_1, \vk_2)+ b_1^2(M)b_2(M)  \, \nn \\
& & \times  \left[\widetilde{\mathcal P}_{\rm mm}(k_1)  
\widetilde{\mathcal P}_{\rm mm}(k_2) +2\,\cyc\ \right] , \label{eq:BihhhS}
\ea
where in the above expressions we have derived the spectra of the
smoothed fields: $\widetilde{\mathcal P}\equiv W^2(kR)P(k)$, and
$\widetilde{\mathcal B}\equiv W(k_1R)W(k_2R)W(k_3R)B$. However, when
we estimate the bispectrum from data we do not smooth the fields apart from the CIC assignment scheme
used to obtain the density contrast field. As
pointed out by \citet{Smithetal2007,Smithetal2008b} and
\citet{Sefusatti2009}, one way to overcome this is to adopt the
ansatz:
\ba
P^{(0)}_{\h\h}(k|M) & = & \frac{\widetilde{\mathcal P}^{(0)}_{\h\h}(k|M,R)}{W^2(kR)} 
\ ; \\
B^{(0)}_{\hhh}(\vk_1, \vk_2, \vk_3|M) & = & 
\frac{\widetilde{\mathcal B}^{(0)}_{\hhh}(\vk_1, \vk_2, \vk_3|M,R)}
{W(k_1R)W(k_2R)W(k_3R)} \ .\label{eq:treeBihhh}
\ea
On applying this `de-smoothing' operation to
\Eqns{eq:PhhS}{eq:BihhhS}, one finds:
\ba
P^{(0)}_{\h\h}(k|M) & = & b_1^2(M) P_{\rm mm}(k) \ ; \label{eq:LPhh}\\
B^{(0)}_{\hhh}(\vk_1, \vk_2|M) & = & b_{1}^3(M)B_{\rm mmm}(\vk_1, \vk_2) + b_1^2(M)\,b_2(M)\nn \\
& & \hspace{-0cm} \times \widetilde{W}_{\bk_1,\bk_2} P_{\rm mm}(k_1) P_{\rm mm}(k_2) + 2\,\cyc\ \ . \nn \\
\label{eq:Bihhh}
\ea
where we have defined the function \citep{Sefusatti2009}:
\be 
\widetilde{W}_{\bk_1,\bk_2}\equiv \frac{W(|\bk_1|R)W(|\bk_2|R)}{W(|\bk_1+\bk_2|R)}\ .
\ee
Note that in the limit of very large scales or arbitrarily small
smoothing scales, $k_iR\rightarrow0$ for $i\in\{1,2,3\}$,
\Eqn{eq:Bihhh} approximates to:
\ba B^{(0)}_{\hhh}(\vk_1, \vk_2|M) & \approx & b_{1}^3(M)B_{\rm
  mmm}(\vk_1, \vk_2) + b_1^2(M)\,b_2(M)\nn \\ & & \hspace{-0cm} \times
\left[P_{\rm mm}(k_1) P_{\rm mm}(k_2) + 2\,\cyc\ \right]
\ .\label{eq:Bihhh2} \ea
Again, since in this paper we are only considering tree-level
expressions we shall take $P_{\h\h}^{(0)}\rightarrow P_{\h\h}$ and
$B_{\hhh}^{(0)}\rightarrow B_{\hhh}$.

Considering now the reduced halo bispectrum, it may be
defined in a similar fashion to \Eqn{eq:Qm}:
\be
 Q_{\h\h\h}(\vk_1, \vk_{2}, \vk_{3}|M)\, \equiv 
\frac{B_{\hhh}(\vk_1, \vk_2|M)}{P_{\h\h}(k_1|M)P_{\h\h}(k_2|M)+2\, \cyc }\ .
\ee
On inserting our tree-level expressions from \Eqns{eq:LPhh}{eq:Bihhh},
we find that
\be
 Q_{\h\h\h}(M) =  
\frac{Q_{\rm mmm}}{b_1(M)} + \frac{b_{2}(M)}{b_{1}^{2}(M)} \alpha(\bk_1,\bk_2,\bk_3)
\label{eq:QQspEul}\ ,
\ee
where we have defined,
\be 
\alpha(\bk_1,\bk_2,\bk_3)\equiv
\frac{\widetilde{W}_{\bk_1,\bk_2}P(k_1)P(k_2)+2\,\cyc\,}{P(k_1)P(k_2)+2\,\cyc\,}\ \label{eq:alpha} .
\ee
Again, in the limit of very large scales or arbitrarily small
smoothing scales and \mbox{$\alpha(\bk_1,\bk_2,\bk_3)\rightarrow 1$}, the above expression approximates to:
\be
 Q_{\h\h\h}(\vk_1, \vk_{2}, \vk_{3}|M) \approx \frac{Q_{\rm mmm}(\vk_1,\vk_2,\vk_3)}{b_1(M)} 
+ \frac{b_{2}(M)}{b_{1}^{2}(M)} \ .
\label{eq:QQspEulApprox}
\ee
We now see the utility of the reduced bispectrum: if one constructs
halo/galaxy density fields from a local transformations of the matter
density, then the lowest order nonlinear corrections will lead to a
function that is a scaled version of the matter $Q_{\rm mmm}$, plus a
constant offset. Moreover, if the density field were simply Gaussian,
then estimates of $Q_{\h\h\h}$ on large scales would directly measure
$b_2/b_1^2$.

%%%%%%%%%%%%%%%%%%%%%%%%%%%%%%%%%%%%%%%%%%%%%%%%%%%%%%
%%%%%%%%%%%%%%%%%%%%%%%%%%%%%%%%%%%%%%%%%%%%%%%%%%%%%%

\section{$N$-body simulations}\label{sec:sims}

For our investigations of the bias, we use an ensemble of 40 large
$N$-body simulations, executed on the \mbox{zBOX-2} and \mbox{zBOX-3}
supercomputers at the University of Z\"{u}rich.  We use only the $z=0$
outputs from the simulations. Each simulation was performed using the
publicly available {\tt Gadget-2} code \citep{Springel2005}, and
followed the nonlinear evolution under gravity of $N=750^3$ equal-mass
particles in a comoving cube of length $L_{\rm sim}=1500\Mpc$. 

The cosmological model that we simulated was analogous to the basic
vanilla $\Lambda$CDM model determined by the WMAP experiment
\citep{Komatsuetal2009}: matter density $\Omega_m=0.25$, vacuum
density $\Omega_{\Lambda}=0.75$, power spectrum normalization
$\sigma_8=0.8$, power spectral index $n=1$, and dimensionless Hubble
parameter $h=0.7$. The transfer function for the simulations was
generated using the publicly available {\tt cmbfast} code
\citep{SeljakZaldarriaga1996,Seljaketal2003b}, with high sampling of
the spatial frequencies on large scales. Initial conditions were set
at redshift $z=49$ using the serial version of the publicly available
{\tt 2LPT} code \citep{Scoccimarro1998,Crocceetal2006}.

Dark matter halo catalogues were generated for each simulation using
the Friends-of-Friends (FoF) algorithm \citep{Davisetal1985}, with the
linking-length parameter $b=0.2$, where $b$ is the fraction of the
inter-particle spacing. For this we employed the fast parallel {\tt
  B-FoF} code, provided to us by V.~Springel. The minimum number of
particles for which an object is considered to be a bound halo was set
at 20 particles. This gave a minimum host halo mass of $M_{\rm min}=1.11\times10^{13}
\Msol$.  For our analysis of the bias, we use the full sample of
haloes and this corresponded to roughly $N^{\rm h}\approx1.26\times
10^6$ haloes per simulation.  Further details of the simulations may
be found in \citet{Smith2009}.

%%%%%%%%%%%%%%%%%%%%%%%%%%%%%%%%%%%%%%%%%%%%%%%%%%%%%%%

\begin{figure*}
 \centering{
 \includegraphics[width = 17cm]{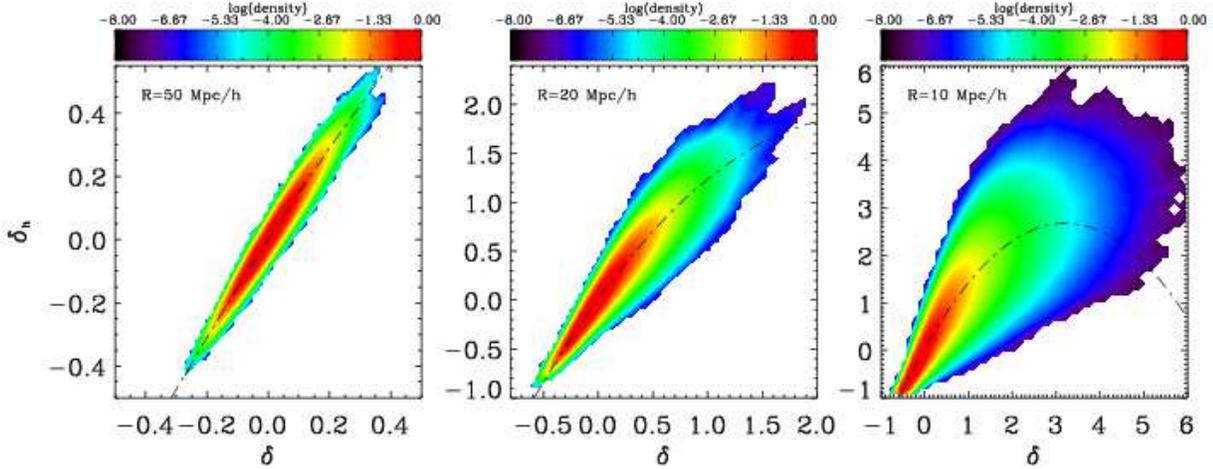}}
 \caption{\small{Scatter plots of $\delta_{\rm h}(\vx)$ versus
     $\delta(\vx)$ smoothed with a Gaussian filter of various scales
     averaged over the realizations. From left to right, the panels
     correspond to the smoothing scales \mbox{$R =
       \{50,\,20,\,10\}\Mpc$}. The color coding denotes the log of the
     population density, i.e. the red region corresponds to the
     largest concentration of points and the white background to null
     values.  The dot-dashed line in each panel denotes the local halo
     bias model up to second-order with the best-fitting bias
     parameters averaged over 28 realizations.}\label{fig:deltabias}}
\end{figure*}

%%%%%%%%%%%%%%%%%%%%%%%%%%%%%%%%%%%%%%%%%%%%%%%%%%%%%%%

\section{Simple estimates of bias}\label{sec:HaloBiasMethods}

Before we examine halo bias in the context of the bispectrum, we
explore two alternative methods for studying the bias. We first
evaluate the second-order local biasing model directly, by comparing
in a point-wise fashion the halo and matter density fields, smoothed
over a range of scales. Then, we use power spectra to determine an
effective large-scale bias.

%%%%%%%%%%%%%%%%%%%%%%%%%%%%%%%%%%%%%%%%%%%%%%%%%%%%%%%

\subsection{Analyzing Density Fields}\label{ssec:DensityFields}

One obvious way to examine the local model of biasing is to simply
construct a scatter plot of the local density of dark matter haloes
against the local density of dark matter in the simulations \citep[see
  for example][]{ShethLemson1999,DekelLahav1999}. As was discussed in
\S\ref{ssec:localmodel}, this model only makes sense in the context of
smooth fields. We shall therefore also inspect how the model
parameters depend on the adopted smoothing scale $R$.

We generate the smoothed density fields as follows: we assign
particles/haloes to a Fourier grid using the CIC algorithm
(c.f.~\S\ref{ssec:BispEst}); then we Fourier transform the grid using
the FFT algorithm; each Fourier mode is then multiplied by a Gaussian
filter of the form:
\be
W(kR) \equiv \exp\left[-(kR)^2/2\right]\ .
\ee
Finally, on taking the inverse Fourier transform, we obtain the
smoothed $\delta(\bx|R)$ and $\delta^{\h}(\bx|R)$. We perform the
above procedure for 28 of the ensemble of simulations and consider the
filter scales: $R=\{50\,,20\,,\,10\}\Mpc$. 

In Figure~\ref{fig:deltabias} we present the bin averaged scatter
plots of $\delta^{\h}(\vx|R)$ vs. $\delta(\vx|R)$, averaged over the
realizations. The colour contours are shaded by the normalized
population density of that pixel, e.g. the central red region
indicates that most of the points in the simulation are regions of
density close to average. We also see that as the smoothing scale is
decreased (panels going from left-to-right), that the scatter
increases and that there are more points that have higher and lower
density. Conversely, as the filter scale is increased the relation
becomes tighter and more linear. One obvious conclusion that may be
drawn from this behaviour is that the bias relation is certainly not
deterministic.

%%%%%%%%%%%%%%%%%%%%%%%%%%%%%%%%%%%%%%%%%%%%%%%%%%%%%%%

\begin{table}
\centering
\begin{tabular}{|c|c|c|c|} \hline
$R$ & $b_{0}$ $\pm$ $\sigma_{b_0}$ & 
$b_{1}$ $\pm$ $\sigma_{b_1}$ & $b_{2}$ $\pm$ $\sigma_{b_2}$  \\ 
$[\Mpc]$ & $\times 10^{-3}$ \\
\hline 
50   & 1.3  $\pm$ 0.1 & 1.497 $\pm$ 0.002 & -0.577 $\pm$ 0.031 \\
20   & 12.0 $\pm$ 0.1 & 1.542 $\pm$ 0.006 & -0.635 $\pm$ 0.004 \\   
10   & 37.2 $\pm$ 0.1 & 1.644 $\pm$ 0.005 & -0.512 $\pm$ 0.001 \\
\hline
\end{tabular}
\caption{\small{Average of the mean bias parameters and the root-mean
    square errors for the local halo bias model up to second-order
    averaged over 28 realizations determined from fitting the scatter
    plots of the halo and matter density fields smoothed on scales
    $k_s = \{0.02, 0.05, 0.1\}\kMpc$.}}
\label{tab:deltabias}
\end{table}

%%%%%%%%%%%%%%%%%%%%%%%%%%%%%%%%%%%%%%%%%%%%%%%%%%%%%%%

In order to obtain a more quantitative understanding, we next consider
fitting for the parameters of the local bias model at
second-order. From \Eqn{eq:Eulbias} we have:
\be
\de_{\rm h}(\bx|M,R) = b_0(M) + b_1(M)\,\de(\bx|R) + \frac{b_2(M)}{2}\,[\de(\bx|R)]^2\,.
\label{eq:eulbias2}
\ee
We perform a least-squares analysis on each realization, and then
average over the resulting set of bias parameters to obtain the mean
parameters: $b_0(M)$, $b_1(M)$, and $b_2(M)$. The $1\sigma$-errors are
then estimated in the usual way, as quadratic deviations from the
sample mean.  In \Fig{fig:deltabias} we plot the resultant best-fit
local model as the dot-dashed line in each of the three panels.

The information on the parameters is summarized in Table
\ref{tab:deltabias} as a function of the filter scale $R$. This
clearly shows that the estimates of $b_1$ increase as the smoothing
scale is decreased, whereas those for $b_2$ appear to be parabolical.
Naively, one might expect that the nonlinear bias terms should
approach zero as the amount of smoothing is increased and
nonlinearities are washed out, however, at $R=50\Mpc$ even with
$\sigma(x|R)<1$, the fluctuations are still significant enough to
yield a non-zero $b_2$. Note also that in all cases $b_0\ne0$.

The local model, as written in \Eqn{eq:Eulbias}, asserts
that the parameters $b_i$ are independent of the
smoothing scale $R$, and we, therefore, consider the implications
as follows. Suppose that
nonlinear bias is exactly as described by \Eqn{eq:eulbias2}, but that
the coefficients are not independent of the smoothing scale. Let us
now consider the results that would be obtained from measurements for
two smoothing scales $R_a$ and $R_b$. From \Eqn{eq:eulbias2} we would
have:
\ba
\delta_{\rm h}(\bx|R_a) & = & b^a_0 + b^a_1\,\de(\bx|R_a) 
+ \frac{b_2^a}{2}\,[\de(\bx|R_a)]^2 \ ;\\
\delta_{\rm h}(\bx|R_b) & = & b_0^b + b_1^b\,\de(\bx|R_b) 
+ \frac{b_2^b}{2}\,[\de(\bx|R_b)]^2 \ .
\ea
Supposing now that we desmoothed each of the fields, by Fourier
transforming and dividing out the appropriate window function. We
would then have:
\ba
\delta_{\rm h}(\bk) & = & 
b^a_1\,\de(\bk)
+ \frac{b_2^a}{2}\int \frac{\dq}{(2\pi)^3} \de(\bq)\de(|\bk-\bq|)
\widetilde{W}_{\bq,\bk-\bq}(R_a) \nn \\
\delta_{\rm h}(\bk) & = & b_1^b\,\de(\bk)
+ \frac{b_2^b}{2}\int \frac{\dq}{(2\pi)^3} \de(\bq)\de(|\bk-\bq|)
\widetilde{W}_{\bq,\bk-\bq}(R_b)\nn \ .
\ea
In order for the above equations to be equivalent, then we must have
\ba 
b_1^a & = & b_1^b \\
b_2^a & = & b_2^{b}
\left[ 
\frac{W(qR_b)}{W(qR_a)}
\frac{W(|\bk-\bq|R_a)}{W(|\bk-\bq|R_b)}
\frac{W(kR_a)}{W(kR_b)}
\right] \ .
\label{eq:biasequal}
\ea
The last of the two equations may only be satisfied if and only if
$R_a=R_b$ or $\{kR, qR, |\bk-\bq|R\}\ll 1$. Since the
$\delta^{\h}(\vx|R)$ vs. $\delta(\vx|R)$ method is inherently a real
space measure it involves contributions from all Fourier modes. It is
therefore difficult to ensure that $b_2^a=b_2^b$. 

We conclude that the above method will not be a safe way to recover
bias parameters independent of the smoothing scale. We now turn to
Fourier space methods.

%%%%%%%%%%%%%%%%%%%%%%%%%%%%%%%%%%%%%%%%%%%%%%%%%%%%%%%
%%%%%%%%%%%%%%%%%%%%%%%%%%%%%%%%%%%%%%%%%%%%%%%%%%%%%%%

\begin{figure*}
  \centering{
    \includegraphics[width = 2.3in,keepaspectratio=true]{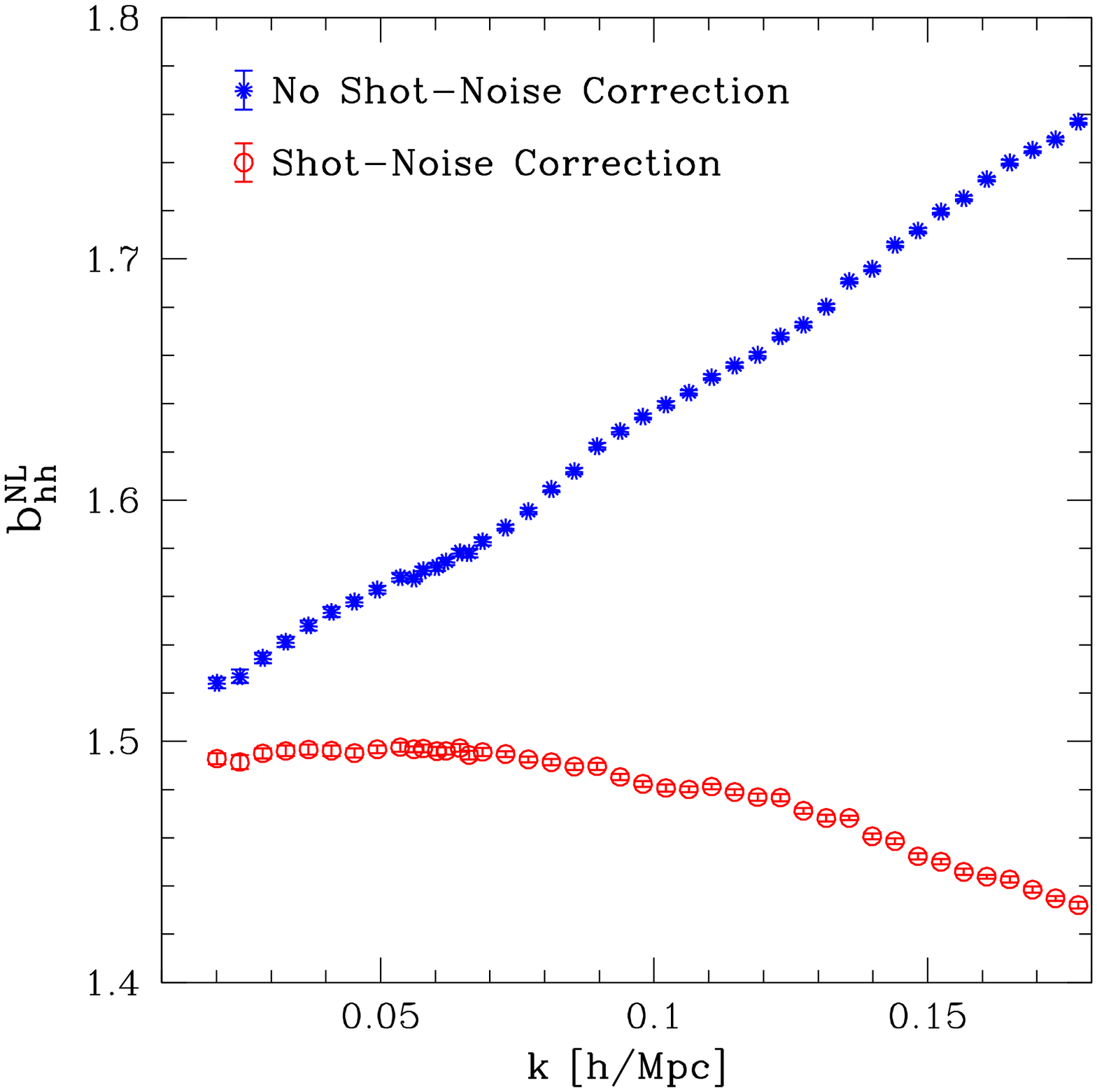}}
  \hspace{-10pt}
  \centering{
    \includegraphics[width = 2.3in,keepaspectratio=true]{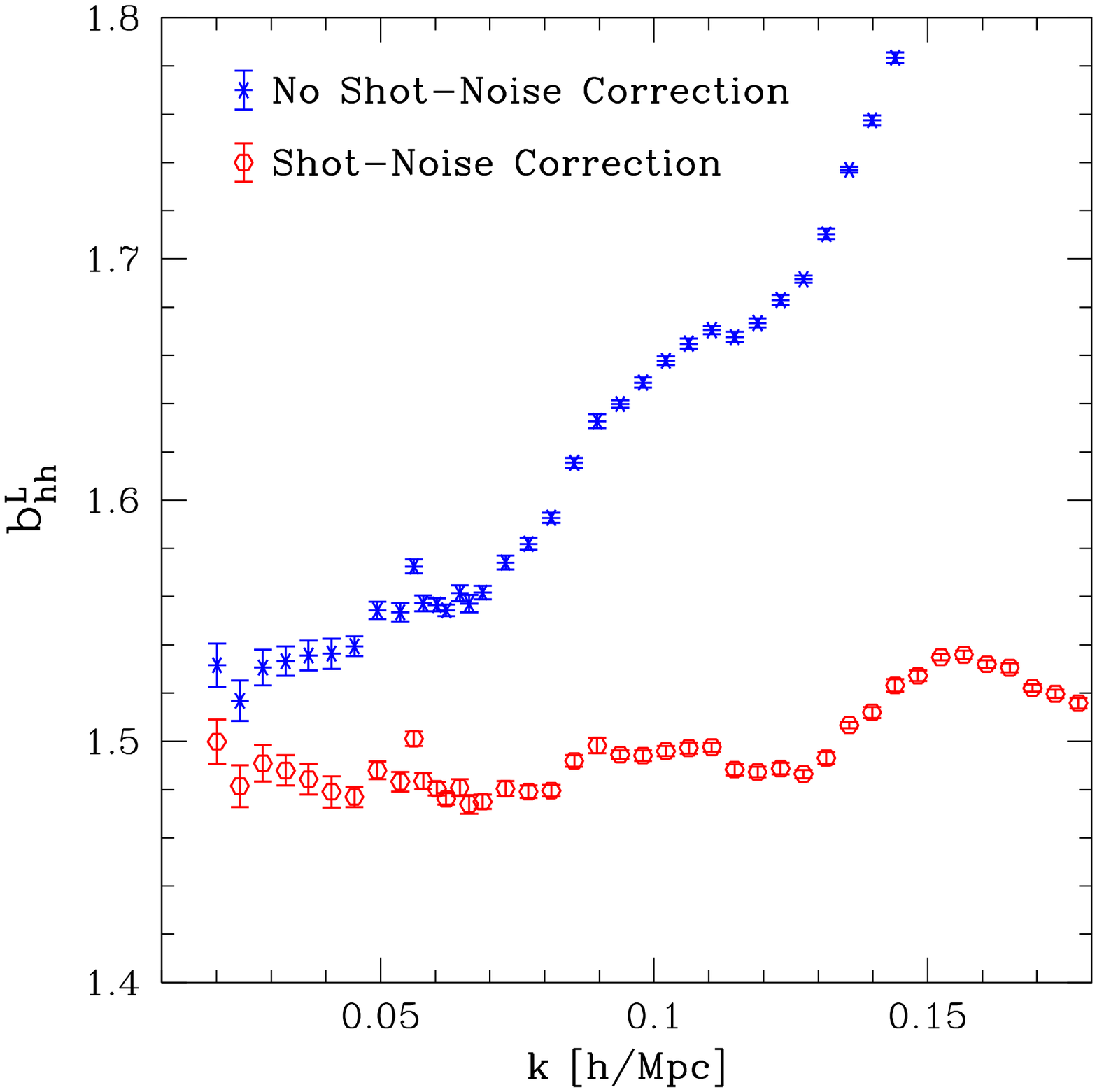}}	
  \hspace{-10pt}
  \centering{
    \includegraphics[width = 2.3in,keepaspectratio=true]{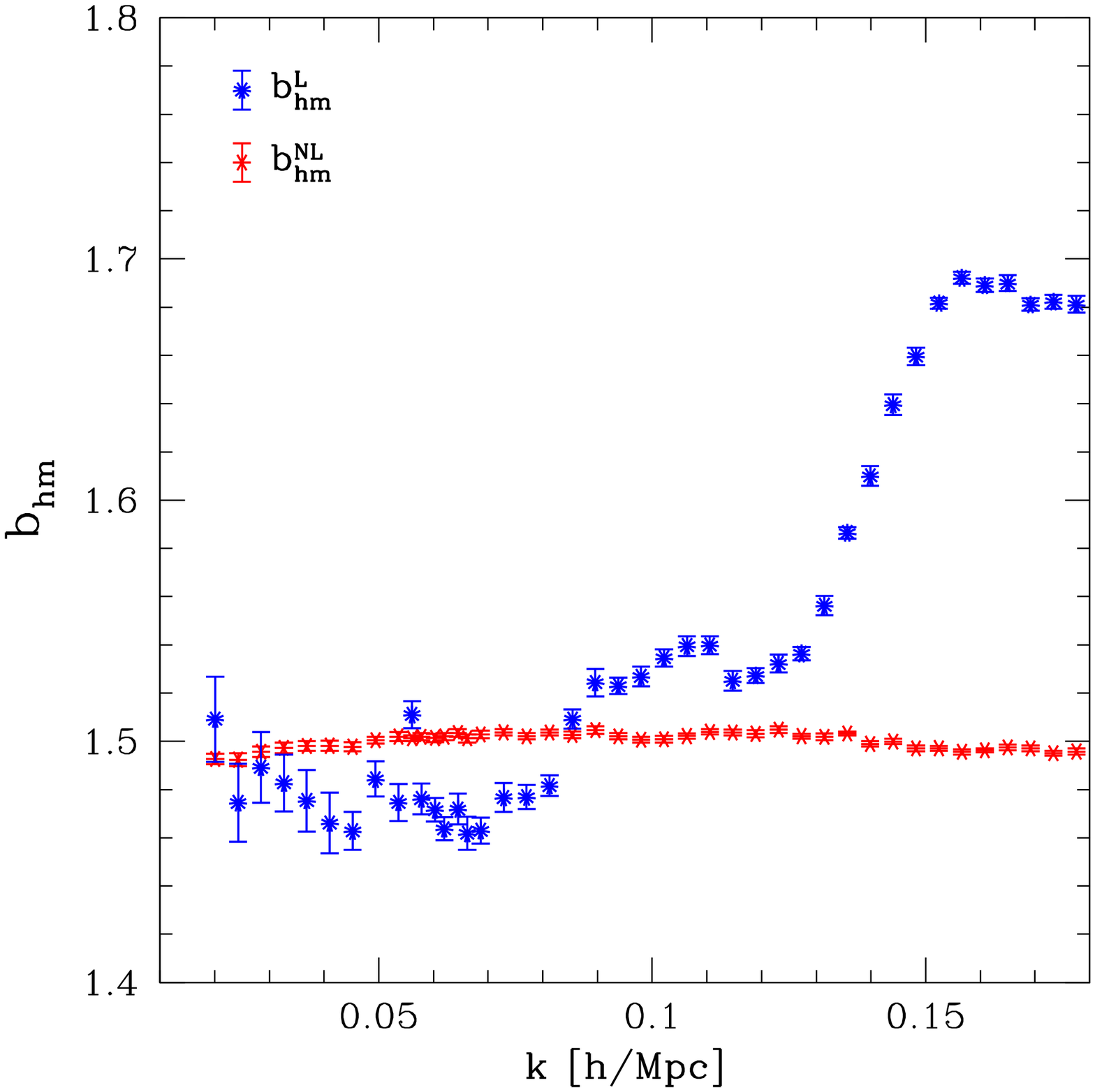}}	
  \caption{\small{Scale dependence of effective bias parameters
      $b^{\rm NL}_{\rm hh}$, $b^{\rm L}_{\rm hh}$, $b^{\rm NL}_{\rm
        hm}$ and $b^{\rm L}_{\rm hm}$
      (c.f. \Eqns{eq:effbias}{eq:bhm}), estimated from the auto- and
      cross-power spectrum as a function of wavemode.  For the left
      and central panels: solid blue and open red symbols denote the
      bias when $P_{\hh}$ is not and is shot noise corrected,
      respectively. The first panel shows $b_{\rm hh}$ when the
      nonlinear matter power spectrum is used; the second panel shows
      the same but when the linear matter power spectrum is used; the
      third panel shows $b_{\rm hm}$, where the red stars and blue
      points denote the case where the nonlinear and linear matter
      power spectra are used, respectively. }}
\label{fig:effbias}
\end{figure*}

\begin{table*}
\centering
\begin{tabular}{|c|c|c|c|c|} \hline
 $k\,[\kMpc]$ & $b^{\rm NL}_{\hh}$   & $b^{\rm NL, SC}_{\hh}$   &   $b^{\rm L}_{\hh}$    &   $b^{\rm L, SC}_{\hh}$   \\ \hline
 0.03-0.09   &  1.589 $\pm$ 0.002 & 1.493 $\pm$ 0.002 & 1.589 $\pm$ 0.004 & 1.487 $\pm$ 0.004  \\ 
 0.04-0.12   &  1.624 $\pm$ 0.002 & 1.486 $\pm$ 0.002 & 1.638 $\pm$ 0.003 & 1.489 $\pm$ 0.003  \\  	
 0.05-0.15   &  1.663 $\pm$ 0.002 & 1.474 $\pm$ 0.002 & 1.709 $\pm$ 0.003 & 1.503 $\pm$ 0.003  \\ 
 0.06-0.18   &  1.695 $\pm$ 0.001 & 1.460 $\pm$ 0.002 & 1.775 $\pm$ 0.002 & 1.511 $\pm$ 0.003  \\ \hline
\end{tabular}
\caption{\small{Weighted average estimates of the effective bias,
    $b_{\hh}$ (see \Eqn{eq:effbias} for a defintions).}}
\label{tab:effbias}
\end{table*}
\subsection{Effective large-scale bias from power spectra}\label{ssec:Pwspec}

We now use various halo power spectra to derive estimates for an {\it
  effective} large-scale halo bias.

In order to do this, we first measure the Fourier transform of the
matter and halo density fields as described in Appendix~\ref{app:bispec}. The
halo-halo, halo-mass and mass-mass power spectra, $\{P_{\rm hh},P_{\rm
  hm},P_{\rm mm}\}$, are then estimated from the data by performing
the following sums:
\be\widehat{P}_{\mu\nu}(k_l)=\frac{\Vu}{N(k)}\sum_{m=1}^{N(k)}
\delta_{\mu}(\bk_l)\delta_{\nu}^*(\bk_l) \ ,\ee 
where $\{\mu,\nu\}\in\{\rm m,\rm h\}$, $\Vu$ is the sample volume (which in our case is the simulation volume), 
and where $N(k)$ are the number
of Fourier modes in a shell of thickness $\Delta k$.

Following \citet{Smithetal2007}, we next construct the estimators:
\ba
& &  \hspace{-0.6cm} \widehat{b}^{\rm NL}_{\hh}  =  \frac{1}{N_{\rm s}}\sum_{i=1}^{N_{\rm s}}
\sqrt{\frac{\widehat{P}_{\rm hh}(k_i)}{\widehat{P}_{\rm mm}(k_i)}} \ ; \ 
\widehat{b}^{\rm L}_{\hh} =  \frac{1}{N_{\rm s}}\sum_{i=1}^{N_{\rm s}}
\sqrt{\frac{\widehat{P}_{\rm hh}(k_i)}{P^{\rm L}_{\rm mm}(k_i)}}
\label{eq:effbias} \ ;\\
& &  \hspace{-0.6cm} \widehat{b}^{\rm NL}_{\rm hm}  =  \frac{1}{N_{\rm s}}\sum_{i=1}^{N_{\rm s}}
\frac{\widehat{P}_{\rm hm}(k_i)}{\widehat{P}_{\rm mm}(k_i)} \ ; \ \ \ \ 
\widehat{b}^{\rm L}_{\rm hm}  =  \frac{1}{N_{\rm s}}\sum_{i=1}^{N_{\rm s}}
\frac{\widehat{P}_{\rm hm}(k_i)}{P^{\rm L}_{\rm mm}(k_i)},
\label{eq:bhm}
\ea
where $N_{\rm s}$ is the number of simulations and $P^{\rm L}_{\rm
  mm}$ is the linear matter power spectrum. Note that in the case of
$b_{\h\h}$ we also consider shot-noise corrected versions of these two
estimators, i.e. we correct $P_{\h\h}$ using \Eqn{eq:shotcorr}. We
denote these bias estimates by $b_{\h\h}^{\rm NL,SC}$ and
$b_{\h\h}^{\rm L,SC}$, respectively. Finally, we determine the
$1\sigma$ errors by evaluating the variance of each realization
against the mean.

The first panel of \Fig{fig:effbias} shows $b^{\NL}_{\hh}$ (solid blue
data points) and $b^{\NL,SC}_{\hh}$ (open red points). The bias for
the shot-noise corrected terms remains roughly constant at $\sim1.49$
down to scales $k\sim0.08\kMpc$ and with very small errors, indicating
that the result is highly constrained by the data. On scales smaller
than this the bias is a decreasing function of $k$. Without shot-noise
correction, we find that the bias is strongly scale dependent, and the
bias rapidly increases with increasing $k$.

The second panel of \Fig{fig:effbias} shows the results obtained from
using $b^{\rm L}_{\hh}$ (solid blue points) and $b^{\rm L, SC}_{\hh}$
(open hexagonal points).  The results are similar to those for $b^{\rm
  NL,SC}_{\hh}$, but with increased cosmic variance on large scales.
An oscillation structure is also present, this can be understood as
explained in \citet{GuzikBernsteinSmith2007}. Nevertheless, comparing
the two provides a clear indication of the validity of the tree-level
power spectrum up to $k=0.08$ $\kMpc$.

The third panel of \Fig{fig:effbias} shows the bias results
$b^{\NL}_{\rm hm}$ (solid red points) and $b^{\rm L}_{\hm}$ (solid
blue points). The value of $b^{\NL}_{\hm}$ stays roughly constant for
the whole scale range considered in the estimate, while $b^{\rm
  L}_{\hm}$ is not smooth and clearly shows the imprint of the oscillation
structure. Nevertheless, we find $b^{\rm L}_{\hm}\sim1.48$, to within
the errors for $k<0.08 \kMpc$.  On comparing the results for
$b^{\NL,SC}_{\h\h}$ and $b^{\NL}_{\hm}$, we see that for scales
\mbox{$k\lesssim0.08\kMpc$}, these estimates are compatible and that
the effective large-scale bias is roughly $b\sim1.49$. Interestingly,
these findings are consistent with real space measures of the
effective large scale bias from cell variances
\citep{SmithMarian2011}.

In Table~\ref{tab:effbias}, we report the weighted average and
corresponding 1$\sigma$ error on the effective bias, $b_{\hh}$,
computed over the same $k$-modes corresponding to the magnitude of the
third wavevector $k_3$ for each triangle configuration. The tabulated
results for the analysis of the uncorrected data confirms the results
shown in Figure \ref{fig:effbias}, that bias is indeed
scale-dependent.  Applying the shot-noise correction yields a more
constant effective bias, even for the range of $k$-modes entering into
our bispectrum estimation.  Interestingly, the value
$b_1=1.49\pm0.002$ found for $k\in[0.03,0.09]$, is in good agreement
with the result for $b_1$ obtained from fitting the density fields
smoothed on scales $R=50\Mpc$.  Therefore, if we opt to infer that the
effective bias is equivalent to $b_1$ over these scales, then the
bispectrum (reduced bispectrum) should also yield this value for $b_1$
when fitted over the same scale ranges (c.f.~Table
\ref{tab:effbias}). That is, if the local bias model is correct and
the tree-level bispectrum (reduced bispectrum) is a sufficient
description of the nonlinearities on these scales.

%%%%%%%%%%%%%%%%%%%%%%%%%%%%%%%%%%%%%%%%%%%%%%%%%%%%%%%

Before moving on, we point out that one can also use the halo power
spectra to define an effective $b_2$ \citep{Smithetal2009}, however we
shall not explore this here.

%%%%%%%%%%%%%%%%%%%%%%%%%%%%%%%%%%%%%%%%%%%%%%%%%%%%%%%

\section{Halo bias from Bispectra}\label{sec:BispHaloEst}

In this section, we present our main results from the analysis of the
halo bispectra.

%%%%%%%%%%%%%%%%%%%%%%%%%%%%%%%%%%%%%%%%%%%%%%%%%%%%%%%

\subsection{Bispectrum Estimation}\label{ssec:BispEst}

The computational code used to estimate the matter and halo power and
bispectra is a modified version of the code developed in
\citet{Smithetal2008b}, which itself is based on the algorithm of
\citet{Scoccimarroetal1998}. The major modification to that code,
which we have implemented, is that no random subsampling of the
Fourier modes is performed to estimate the bispectrum. Instead, all modes
that contribute to a particular triangle configuration are used.  In
this work we use a FFT grid of size $N_{\rm g}=512^3$ to estimate the
power and bispectra. We only evaluate triangles that have $k_2=2k_1$,
but consider the variation of $B$ with the angular separation of the two
vectors. The largest scale at which we estimate the bispectrum is
$k_1=0.03\kMpc$, and this is $\approx 7.5k_{\rm f}$, where $k_{\rm
  f}=2\pi/L\approx 0.004 \kMpc$. Further details of the bispectrum
estimation procedure may be found in Appendix~\ref{app:bispec}.

Figure~\ref{fig:Bisp} shows the ensemble-averaged shot-noise corrected
results for the halo bispectra $B_{\rm hhh}$ (open red squares) and
matter bispectra $B_{\rm mmm}$ (solid blue diamonds), measured from
the ensemble of $N$-body simulations.  The four panels show the
results obtained for the scales:
$k_1=\{0.03,0.04,0.05,0.06\}\kMpc$. The error bars are the 1-$\sigma$
errors on the mean, derived from the ensemble to ensemble variation.
The solid red line represents the tree-level prediction for $B_{\rm
  mmm}$ as given by \Eqn{eq:bisptree}. We see that this appears to be
a good description of the $B_{\rm mmm}$ estimates for the scales that
we have considered. We notice that, for the case $k_1=0.06\kMpc$, the
theory systematically under-predicts the measurements for
$\theta/\pi\sim0.5$ (but see \S\ref{ssec:treelevel} for a more
quantitative discussion of the goodness of fit).

Figure~\ref{fig:QQsp} shows the same as in \Fig{fig:Bisp}, however
this time for $Q_{\rm hhh}$ and $Q_{\rm mmm}$.

%%%%%%%%%%%%%%%%%%%%%%%%%%%%%%%%%%%%%%%%%%%%%%%%%%%%%%%

\begin{figure*}
\centering{
  \includegraphics[width=8cm,height=7cm]{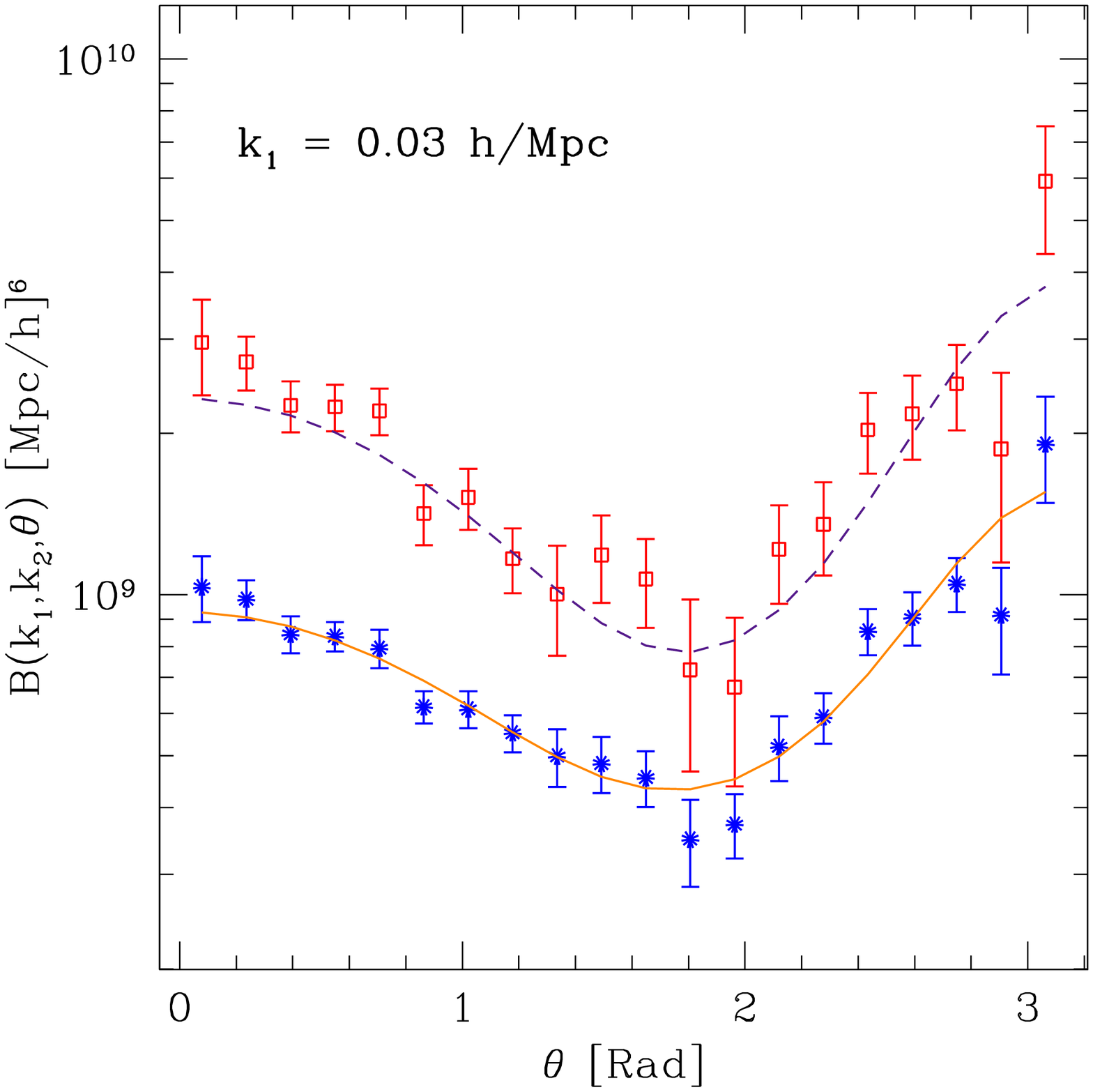}\hspace{0.5cm}
  \includegraphics[width=8cm,height=7cm]{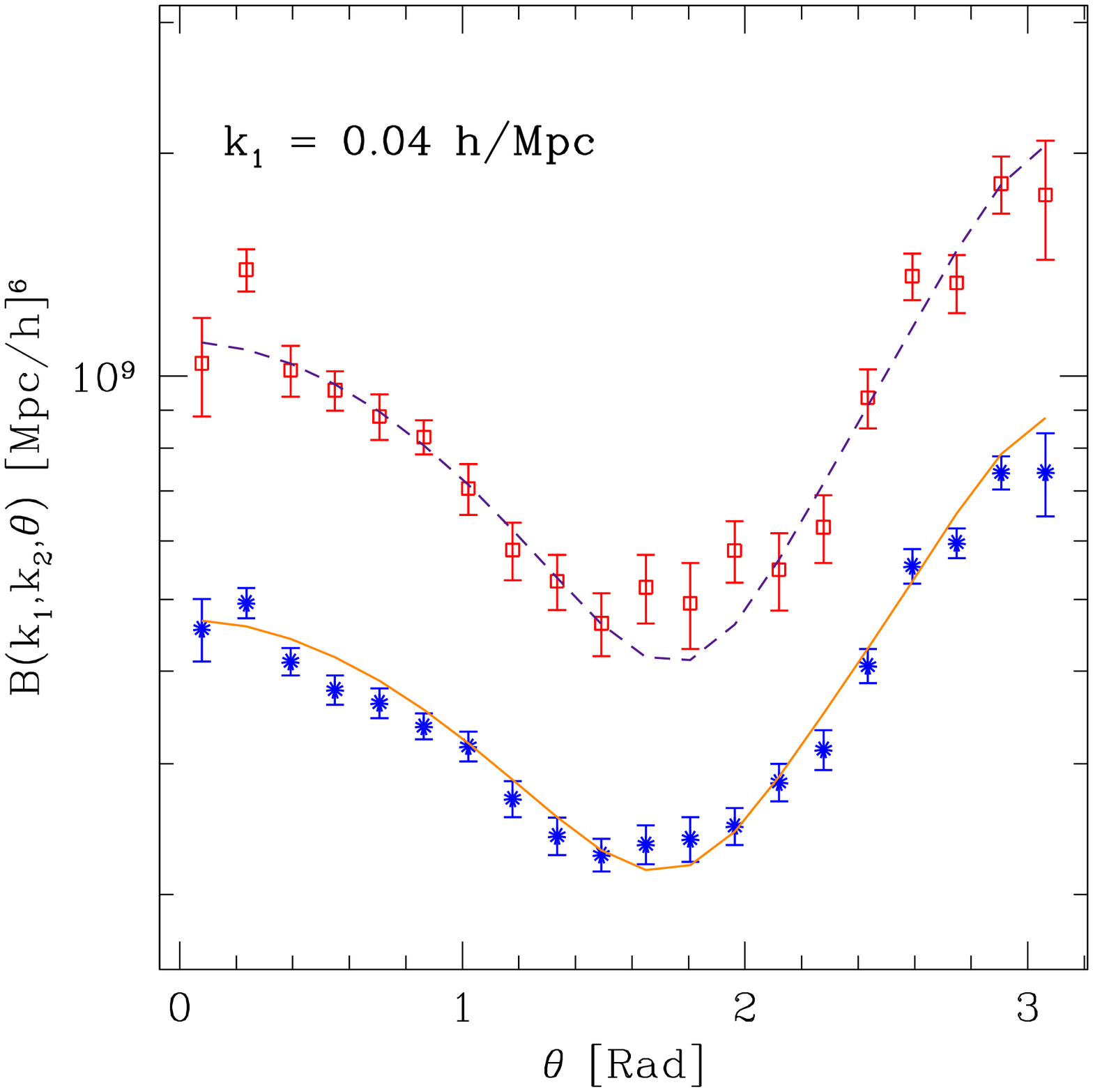}}
\centering{
  \includegraphics[width=8cm,height=7cm]{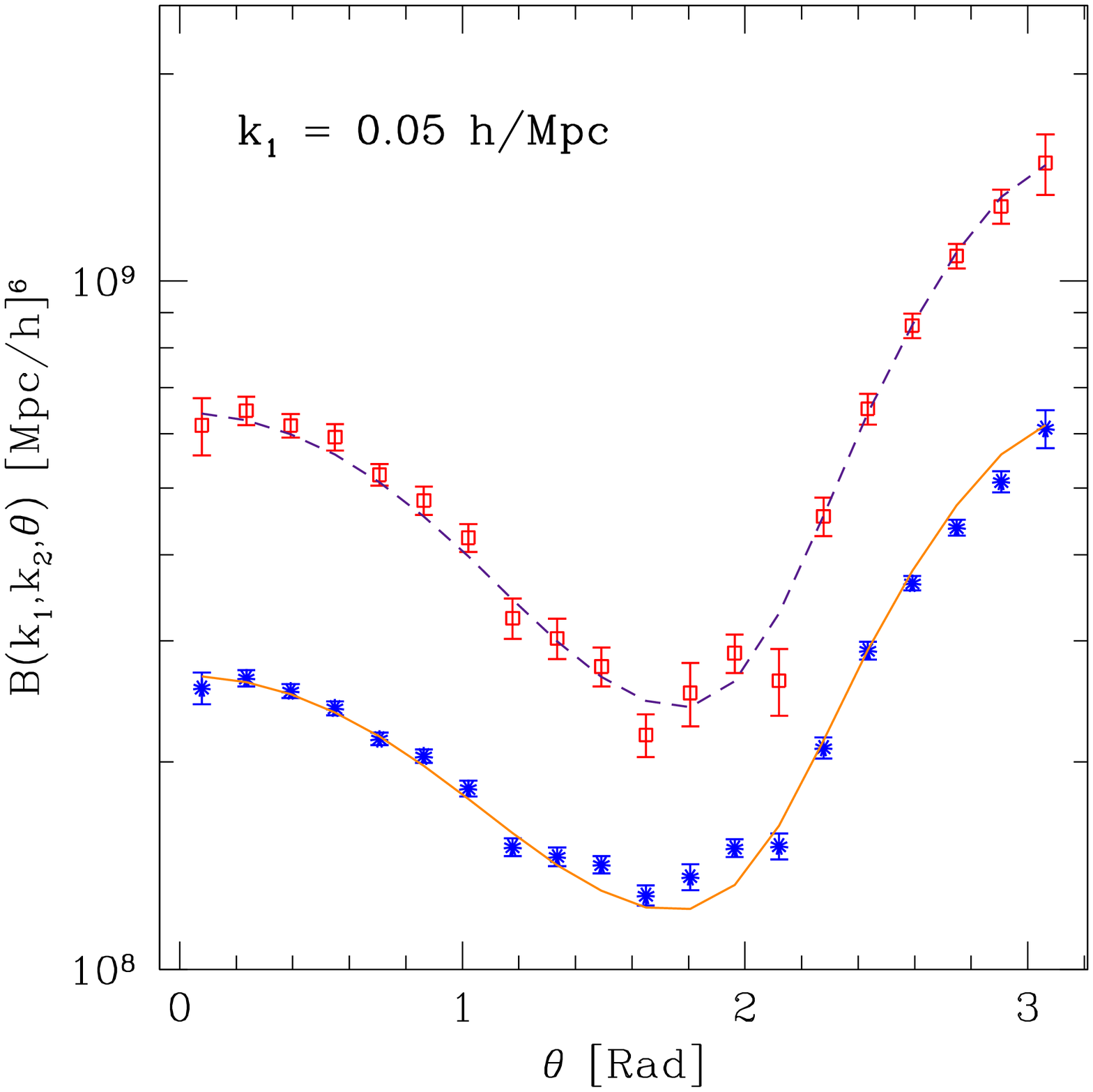}\hspace{0.5cm}
  \includegraphics[width=8cm,height=7cm]{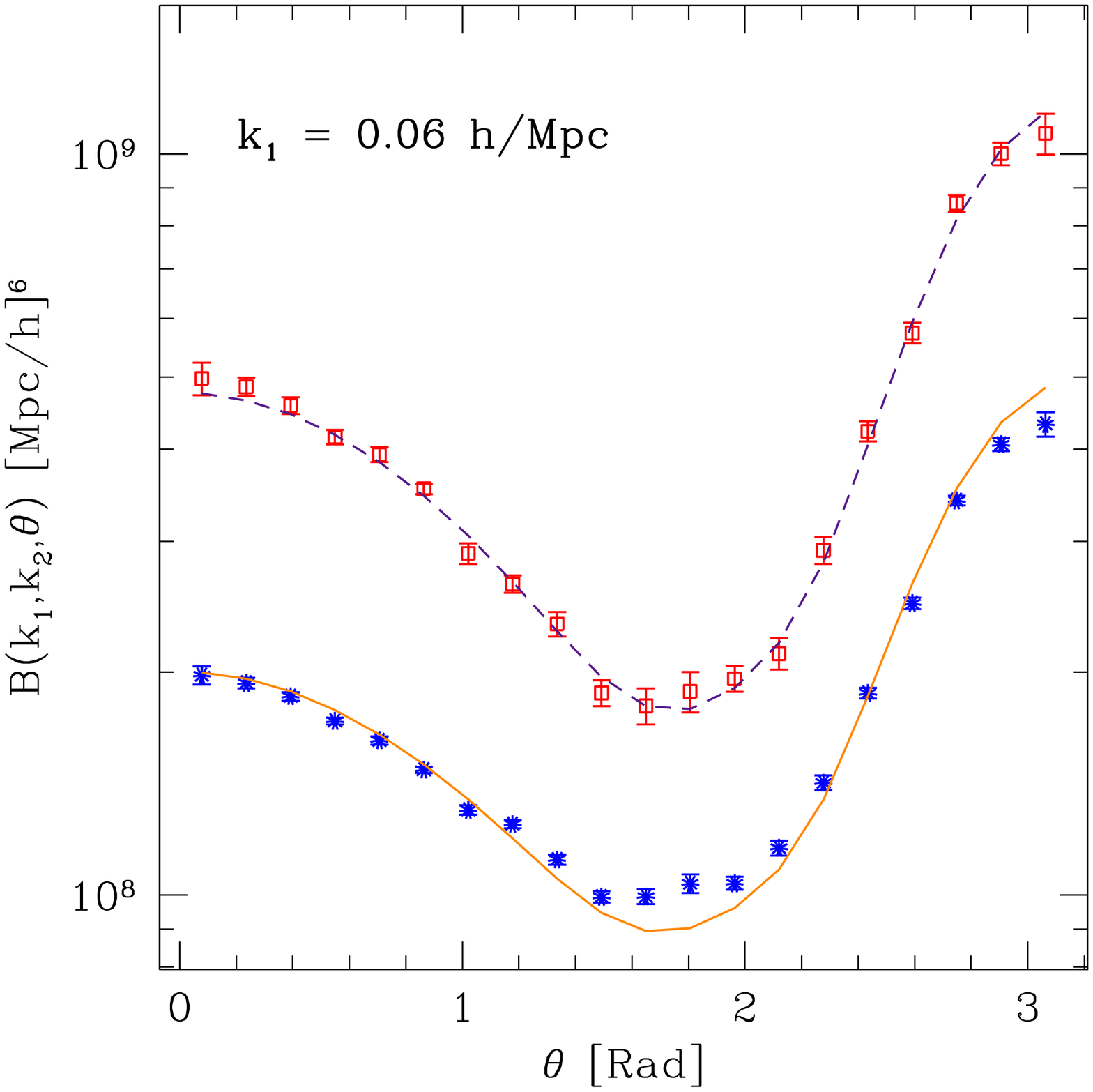}}
\caption{\small{Ensemble-averaged matter and halo bispectrum
    measurements for 40 LCDM $N$-body simulations in real-space in
    comparison with the PT models at tree-level. Each panel shows the
    shot-noise corrected bispectrum measurements as a function of
    angle for a variety of triangle configurations at different scale
    ranges set by $k_1 = \{0.03\,\,0.04\,\,0.05\,\,0.06\} \kMpc$ and
    $k_2 = 2 k_1$.  The blue solid symbols represent the matter
    bispectrum, whereas the open squares denote the halo bispectrum.
    The tree-level bispectrum is represented by the solid orange line
    and the local halo bias model with the best-fitting parameters
    listed in Table \ref{tab:auto} is the dashed violet line.}}
\label{fig:Bisp}
\end{figure*}

%%%%%%%%%%%%%%%%%%%%%%%%%%%%%%%%%%%%%%%%%%%%%%%%%%%%%%%

\begin{figure*}
\centering{
  \includegraphics[width=8cm,height=7cm]{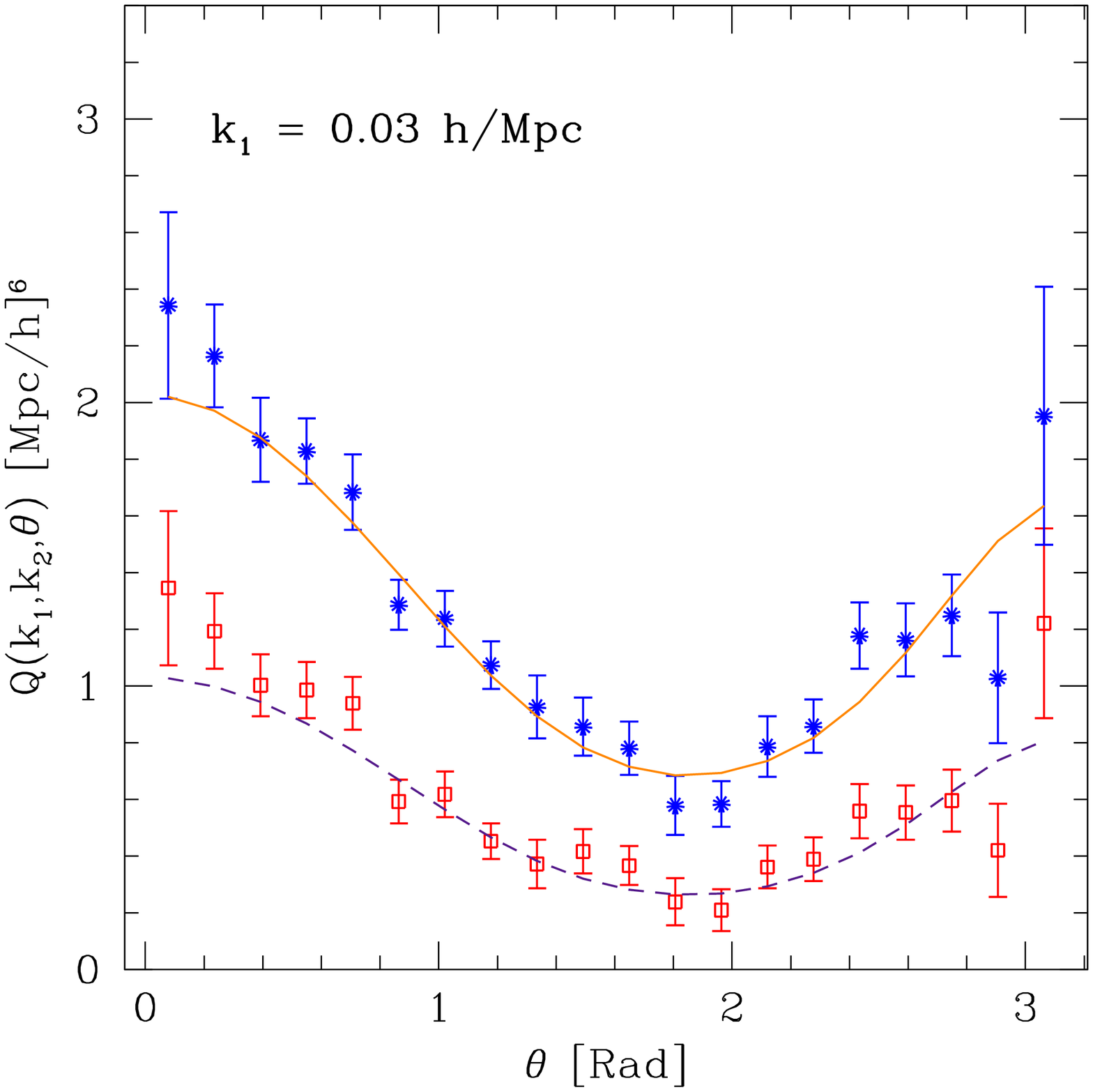}\hspace{0.5cm}
  \includegraphics[width=8cm,height=7cm]{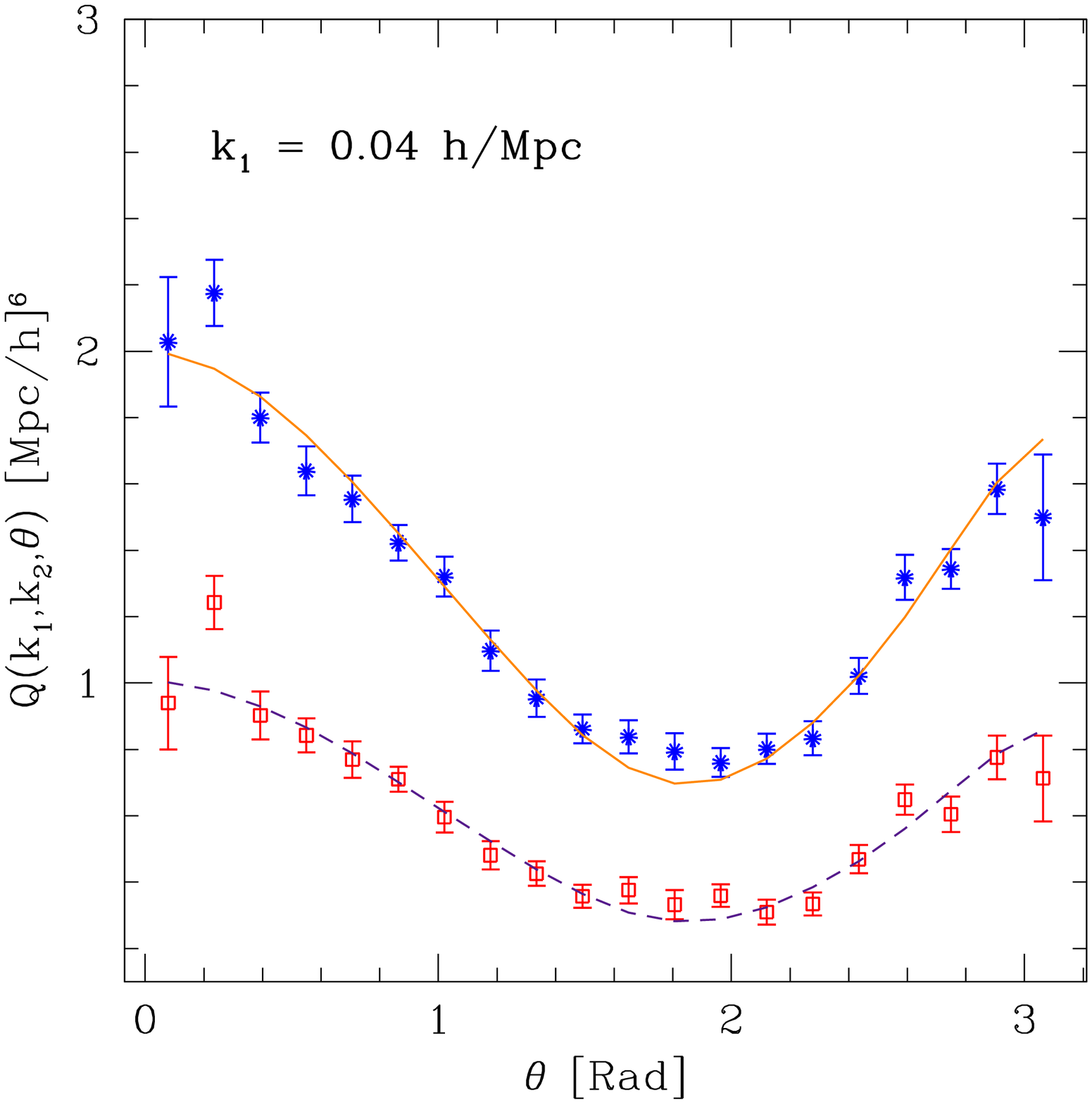}}
\centering{
  \includegraphics[width=8cm,height=7cm]{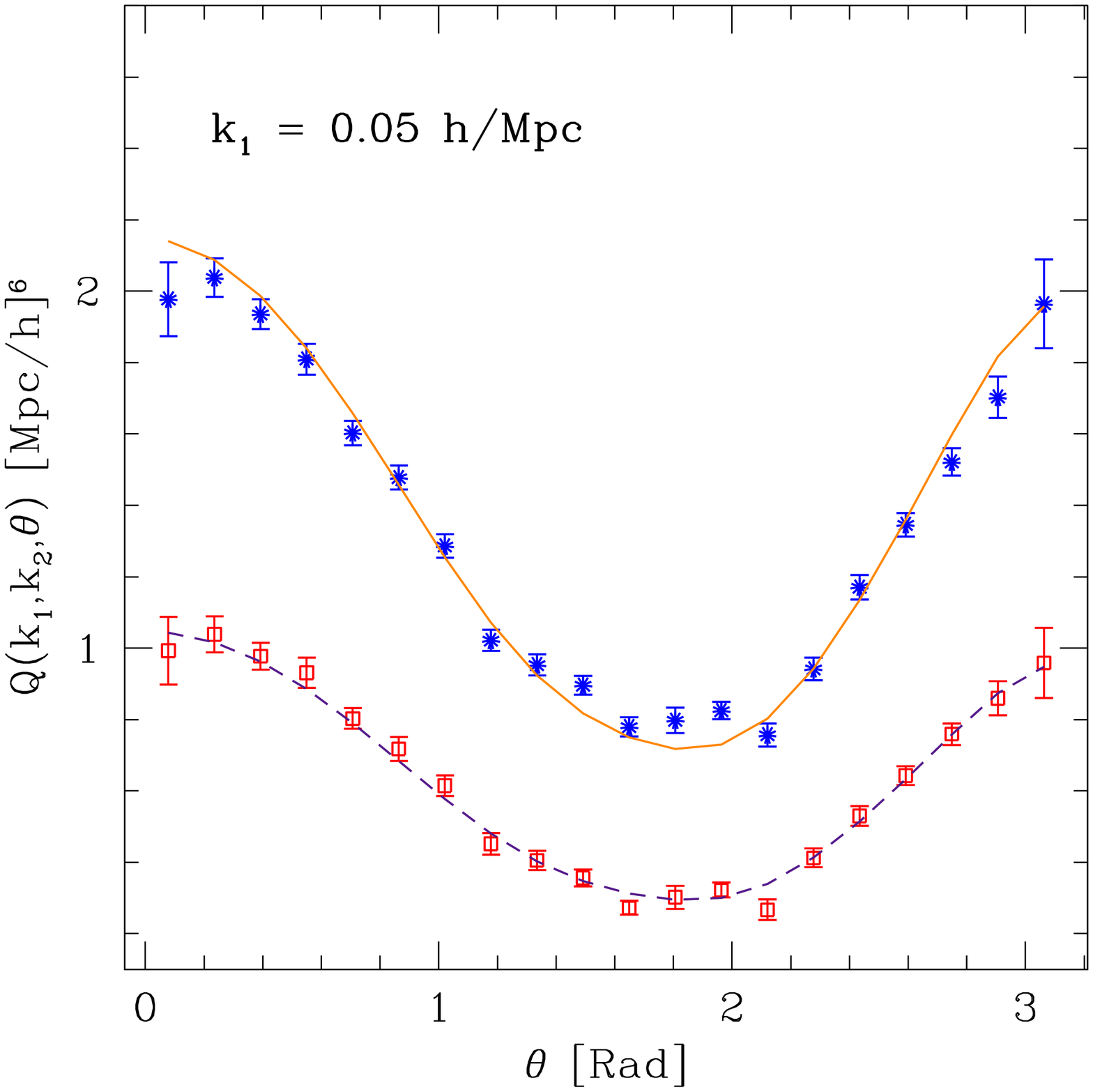}\hspace{0.5cm}
  \includegraphics[width=8cm,height=7cm]{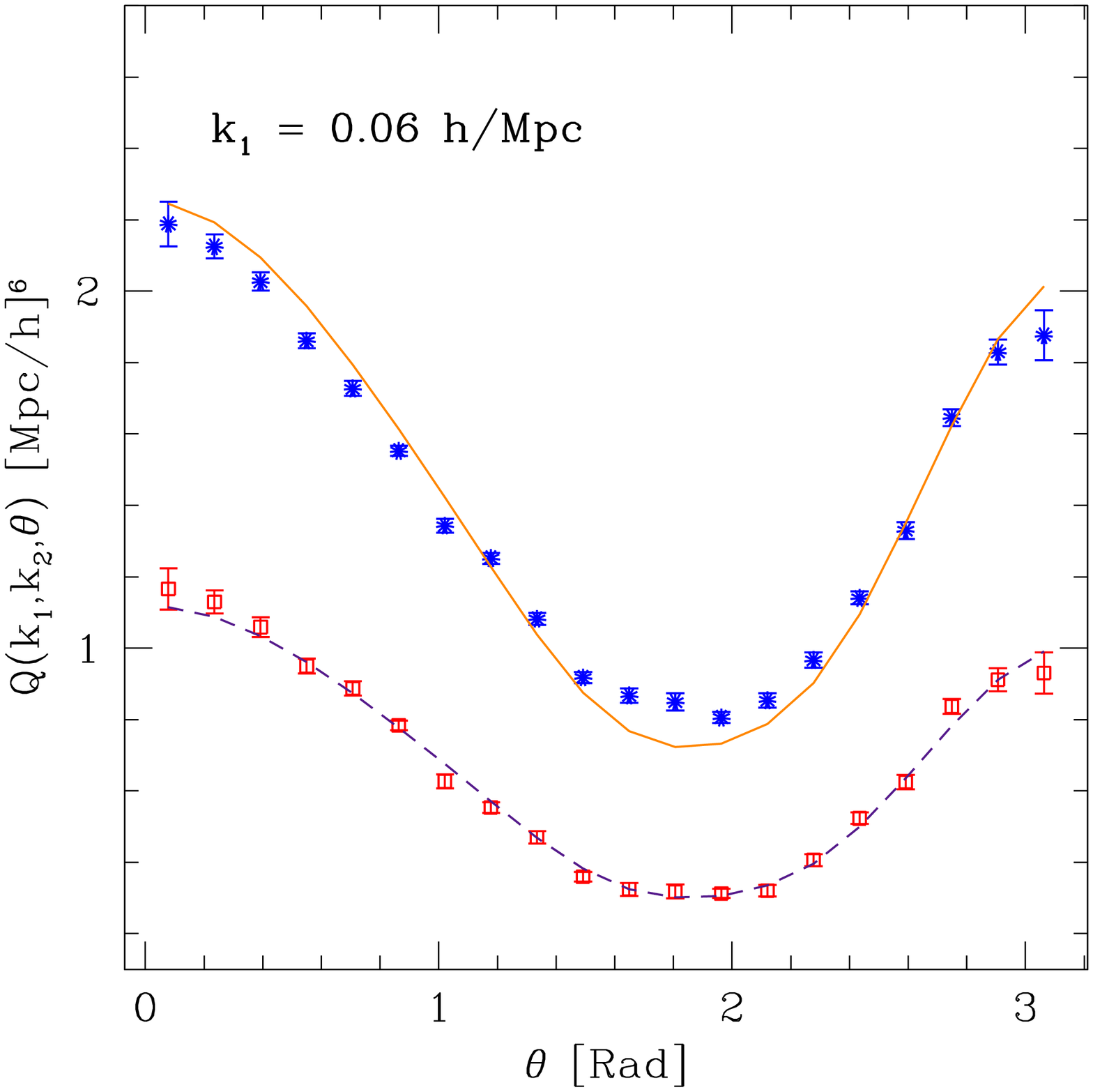}}
\caption{\small{Ensemble-averaged matter and halo reduced bispectrum
    measurements of the 40 LCDM $N$-body simulations in real-space in
    comparison with the PT models at tree-level. Point and line styles
    as in Figure~\ref{fig:Bisp}.}}
\label{fig:QQsp} 
\end{figure*}

%%%%%%%%%%%%%%%%%%%%%%%%%%%%%%%%%%%%%%%%%%%%%%%%%%%%%%%
%%%%%%%%%%%%%%%%%%%%%%%%%%%%%%%%%%%%%%%%%%%%%%%%%%%%%%%

\subsection{Bias estimation}\label{ssec:biasest}

Our method for estimating the bias parameters follows an approach
similar to that presented by \citet{Scoccimarro2000b} and \citet{PorcianiGiavalisco2002}.  
To start, we take a $\chi^2$
function that is a quadratic form of the type:
\be \chi^2(b_1,b_2) = 
\sum_{i=1}^{N_{\theta}} 
\sum_{j=1}^{N_{\theta}} \Delta_i(b_1,b_2)
\widehat{r^{-1}}_{ij} 
\Delta_j(b_1,b_2) \ , \label{eq:chisq}
\ee
where $N_{\theta}$ is the number of angular bins considered and
\be
\Delta_i\equiv
\frac{\widehat{\overline{B}}_{\h\h\h}(k_1,k_2,\theta_i)-B^{\rm mod}_{\h\h\h}(k_1,k_2,\theta_i|b_1,b_2)}
{\sigma_{\h\h\h}(k_1,k_2,\theta_{i})} \label{eq:norm}
\ .\ee
Note that in dividing the difference between the estimate of the
ensemble average bispectrum ($\widehat{\overline{B}}$) and the model
prediction ($B^{\rm mod}$) by the standard deviation
($\sigma_{\h\h\h}$), $\widehat{r^{-1}}_{ij}$ is in fact the inverse
correlation matrix. Recall that the correlation and covariance
matrices are related by: $r_{ij}=C_{ij}/\sqrt{C_{ii}C_{jj}}$.

In order to minimize this $\chi^2$ function and so recover the
best-fit bias parameters, we need an estimate of
$\widehat{r^{-1}}_{ij}$, and we do this using the standard unbiased
estimator:
\be
\widehat{r}_{ij} = \frac{1}{(N_{\rm sim}-1)}
\sum_{k=1}^{N_{\rm sim}}
\widetilde{\Delta}_i^k
\widetilde{\Delta}_j^k
\label{eq:corr} \ ,
\ee
where $N_{\rm sim}$ is the number of simulations and where
\be \widetilde{\Delta}_i\equiv
\frac{\widehat{B}_{\h\h\h}^{(k)}(k_1,k_2,\theta_i)-\widehat{\overline{B}}_{\h\h\h}(k_1,k_2,\theta_i)}
     {\sigma_{\h\h\h}(k_1,k_2,\theta_{i})} \label{eq:normjk} \ ,\ee
where in the above $\widehat{B}_{\h\h\h}^{(k)}$ is the $k^{\rm th}$
estimate of the bispectrum and where $\widehat{\overline{B}}_{\h\h\h}$
is the mean of the ensemble.

Next, we use singular-value-decomposition (SVD) to invert the
$\widehat{r}_{ij}$. For the estimate of the inverse correlation
matrix, we utilize principal component analysis (PCA) to remove some
of the noisy eigenvectors. We select the fraction of principal
components that account for 95\% of the variance.  According to this
selection criteria we typically retain 15 out of 20 of the most
`dominant' eigenmodes. Note that as pointed out in
\citet{Hartlapetal2007}, $\widehat{C^{-1}} \ne \widehat{C}^{-1}$.
However, since we are using PCA, this should be a subdominant
correction.  Thus, we may approximate \Eqn{eq:chisq} as:
\ba 
\chi^2 & = &
\sum_{i=1}^{N_{\theta}} 
\sum_{j=1}^{N_{\theta}} 
{\Delta}_i(b_1,b_2) [R^{T} \Lambda R]^{-1}_{ij}
{\Delta}_j(b_1,b_2) \ , \nn \\
& = &
\sum_{i=1}^{N_{\theta}} 
\sum_{j=1}^{N_{\theta}} 
{\Delta}_i(b_1,b_2) \sum_{l=1}^{N_{\theta}} R^{T}_{il} \Lambda^{-1}_{ll} R_{lj}
{\Delta}_j(b_1,b_2) \ , \nn \\
& \approx &
\sum_{l=1}^{N_{\theta}}
 \Lambda^{-1}_{ll} Y_l^2 \Theta_{ll}\ ,
\ea
where the correlation matrix $r$ was diagonalized by rotation into
its eigenbasis, i.e. $r=R^{T} \Lambda R$, with $\Lambda$ representing
a diagonal matrix of eigenvalues. We also defined $Y_l\equiv
\sum_{i=1}^{N_{\theta}} R_{li} {\Delta}_i(b_1,b_2)$. Note that in the
final approximate expression we include a matrix $\Theta_{ll}$, this a
diagonal matrix with entries either 1 or 0, depending on whether the
eigenvector is to be retained or cut from the PCA reconstruction.

Finally, the $\chi^2(b_1,b_2)$ function was minimized using the
Levenberg-Marquardt routine for non-linear least squares fitting. 

%%%%%%%%%%%%%%%%%%%%%%%%%%%%%%%%%%%%%%%%%%%%%%%%%%%%%%%

\subsection{Errors in parameter estimates}\label{ssec:biaserr}

To the best-fit parameters $(b_1,b_2)$, we assign both systematic and
statistical errors.

In our context, the systematic errors correspond to the errors induced
in the best-fit parameters from fitting the data with a noisy inverse
covariance matrix (or correlation matrix). Owing to the relatively low
number of simulations $(N_{\rm sim}=40)$, we expect that \Eqn{eq:corr}
provides a noisy estimate of $r_{ij}^{-1}$. In order to estimate the
errors this has on the best-fit parameters we employ the
jackknife subsampling method \citep[see for
  example][]{Norbergetal2009}. This involves slicing the total data set into $N_{\rm sub}$
subsamples. Then a resampling of the data is obtained by excluding one
of the subsamples from the set. From this resampling we then estimate
the mean statistic of interest and the inverse correlation matrix as
described in the previous section. The resampled data set is then used
to determine a new estimate of the best-fit bias parameters. This
procedure is then repeated for all of the possible $N_{\rm sub}$ 
resamplings of the data. In our particular case we treat the
measurements from each simulation as the regions to be included or
excluded, and this gives us 40 jackknife estimates of the bias
parameters $(b_1,b_2)$.  The parameter covariance matrix for the
systematic errors can be computed as \citep{Norbergetal2009}:
\be \widehat{C}^{\rm JK}[b_i,b_j] = \frac{N_{\rm sub}-1}{N_{\rm
    sub}}\sum_{k=1}^{N_{\rm sub}} (b_{i,k}-\hat{\bar{b}}_{j})
(b_{j,k}-\hat{\bar{b}}_j) \ , \label{eq:jkparam}\ee
where $b_{i,k}$ is the estimate of $b_i$ from the $k^{\rm th}$
resampling of the data and $\hat{\bar{b}}_{i}$ is the estimate of the
mean $b_i$ obtained from all of the resamplings.

The statistical error is obtained directly from the nonlinear
least-squares analysis. The routine {\tt mrqmin} provides an
approximation to the errors on the parameters that corresponds to a
$\Delta \chi^2\approx1$ for a one-parameter model.  However, the
confidence regions we present in the forthcoming plots correspond to
either $\Delta \chi^2=(2.30,6.17)$, which roughly denote the
($\sim1\sigma$, $\sim2\sigma$) errors for a two-parameter model. 

Given that we consider the two forms of error: systematic and
statistical), and that one is never
consistently larger than the other, in all forthcoming tables, we
choose to report only the total error. This is obtained simply from
the two errors added in quadrature.

%%%%%%%%%%%%%%%%%%%%%%%%%%%%%%%%%%%%%%%%%%%%%%%%%%%%%%%

\subsection{Testing the validity of the tree-level matter $B$}\label{ssec:treelevel}

%%%%%%%%%%%%%%%%%%%%%%%%%%%%%%%%%%%%%%%%%%%%%%%%%%%%%%%

Before we report the estimates of the halo bias parameters, we first
present a test of the validity of the tree-level model for the matter
bispectrum.  We do this by applying the $\chi^2$ test described above,
to the $B_{\rm mmm}$ and $Q_{\rm mmm}$ data, and so fit for $b_1$ and
$b_2$. Note that since the total number of principal components retained
equals 15, then for a two-parameter model the number of degrees-of-freedom 
equals 13. If the tree-level expressions in the large-scale limit as given
by \Eqns{eq:Bihhh2}{eq:QQspEulApprox} are correct, then we should
expect to find $b_1=1$ and $b_2=0$.

Table \ref{tab:treelevel} presents the best-fit nonlinear bias
parameters for the four different bispectrum scale ranges discussed
earlier. In the analysis we fit the shot-noise corrected
bispectra. The $\chi^2$ values (end column of the table), confirms
that the tree-level expressions $B^0_{\rm mmm}$ and $Q^0_{\rm mmm}$,
provide good fits for the triangle configurations with
$k_1=\{0.03,0.04\}\kMpc$. However, for $k_1=\{0.05,0.06\}\kMpc$ the
fits are poor given the $\chi^2$ estimates, and we see that, for both
$B$ and $Q$, they yield non-zero values for $b_2$ at 1$\sigma$.  The
results also imply that the failure of the tree-level model on these
scales is more severe for $Q$ than for $B$. This can be understood by
noting that $b_1$ from $Q_{\rm mmm}$ shows a prominent departure from
unity, whereas $B_{\rm mmm}$ does not (although the deviation still
exceeds 2$\sigma$). We thus conclude that it is likely that the
tree-level expressions for the halo bispectra will only be valid for
$k_1\le0.04\kMpc$, for our chosen bispectra configurations.

%%%%%%%%%%%%%%%%%%%%%%%%%%%%%%%%%%%%%%%%%%%%%%%%%%%%%%%

\begin{table}
\centering
\begin{tabular}{|c|c|c|c|c|} \hline
$\!k_1$ $[\kMpc]$ &  & $b_{1}$ $\pm$ $\sigma_{b_1}$ & $b_{2}$ $\pm$ $\sigma_{b_2}$ & $\chi^2$ \\ \hline 
\multirow{2}{*}{0.03} & $B_{\rm mmm}$	& 1.01	$\pm$	0.07  & -0.04 $\pm$ 0.25 & 19.08 \\
		      & $Q_{\rm mmm}$	& 0.93	$\pm$	0.19  & -0.05 $\pm$ 0.30 & 19.03 \\
\hline
\multirow{2}{*}{0.04} & $B_{\rm mmm}$	& 0.98	$\pm$	0.03  & 0.04 $\pm$ 0.10 & 14.31  \\
		      & $Q_{\rm mmm}$	& 1.05	$\pm$	0.09 & 0.07 $\pm$ 0.16 & 14.95 \\
\hline
\multirow{2}{*}{0.05} & $B_{\rm mmm}$	& 0.97	$\pm$	0.02  & 0.13 $\pm$ 0.08 & 38.83	\\
		      & $Q_{\rm mmm}$	& 1.14	$\pm$   0.07  &	0.19 $\pm$ 0.13 & 19.97 \\
\hline
\multirow{2}{*}{0.06} & $B_{\rm mmm}$	& 0.98	$\pm$	0.02  & 0.10 $\pm$ 0.10 & 34.09 \\
		      & $Q_{\rm mmm}$	& 1.15	$\pm$	0.04  & 0.22 $\pm$ 0.08 & 29.58 \\
\hline
\end{tabular}
\caption{\small{Assessment of the validity of the tree-level modelling
    by fitting the matter bispectra and reduced bispectra.  Column 1:
    bispectrum triangle scale; column 2: statistic, where $B_{\mmm}$
    and $Q_{\mmm}$ are shot-noise corrected; columns 3 and 4: best-fit
    $b_1$ and $b_2$ along with $1\sigma$ errors; column 5:
    $\chi^2$.}}
\label{tab:treelevel}
\end{table}

\subsection{Constraints on $b_1$ and $b_2$ from halo bispectra}\label{ssec:b1b2}

Table~\ref{tab:auto} presents the best-fit nonlinear bias parameters
and their respective 1$\sigma$ errors in quadrature, obtained from the $\chi^2$
analysis of $B_{\hhh}$ and $Q_{\hhh}$. Note that we present the
results for both the uncorrected and shot-noise corrected
measurements, indicated in the table by superscript `SC'.
Table~\ref{tab:auto} also shows the $\chi^2$ value of these best-fit
parameters as an indication of the goodness-of-fit.

%%%%%%%%%%%%%%%%%%%%%%%%%%%%%%%%%%%%%%%%%%%%%%%%%%%%%%%
In Figures~\ref{fig:Bisp} and \ref{fig:QQsp} we also show the
tree-level theoretical models for $B_{\hhh}$ and $Q_{\hhh}$ (dashed
lines), where the best-fit bias parameters from Table~\ref{tab:auto}
have been used. These figures demonstrate that, at least by-eye, the
tree-level models provide a reasonable description of the
data. However a more detailed inspection of Table~\ref{tab:auto}
reveals some important discrepancies.

For the case of fitting $B$, the shot-noise correction is less
important, as we see that the estimates of $b_1$ for all bispectra
configurations with and without shot-noise corrections are consistent
to within the errors, and have $b_1\sim1.4$. However, $b_2$ shows
systematic differences, being more negative if the correction is made,
and for this we find that $b_2\sim-0.25$.  On the other hand, for the
case of $Q$, the results clearly show that the shot-noise subtraction
has an important effect on the recovered values for the bias
parameters. If the shot-noise is not corrected, then we see that the
estimates for $b_1$ increase systematically as we go from triangle
configurations with $k_1=0.03\kMpc$ to $k_1=0.06\kMpc$. Whereas if it
is corrected, then we find $b_1\sim1.8$ and $b_2\sim-0.3$ to within
the errors.  On comparing the results from $B$ and $Q$ we see that,
whilst the values for $b_1$ disagree significantly, surprisingly those
for $b_2$ remain consistent at the 1$\sigma$ level.

%%%%%%%%%%%%%%%%%%%%%%%%%%%%%%%%%%%%%%%%%%%%%%%%%%%%%%%

\begin{table}
\centering
\begin{tabular}{|c|c|c|c|c|} \hline
$k_1$ $[\kMpc]$ &  & $b_{1}$ $\pm$ $\sigma_{b_1}$ & $b_{2}$ $\pm$ $\sigma_{b_2}$ & $\chi^2$  \\ \hline
\multirow{4}{*}{0.03} & $B_{\rm hhh}$		& 1.43 $\pm$ 0.11 & -0.18 $\pm$ 0.40 & 17.20 \\%&	0.002 & -0.009\\
		      & $B^{\rm SC}_{\rm hhh}$	& 1.42 $\pm$ 0.11 & -0.36 $\pm$ 0.38 & 17.08 \\%&	0.001 & -0.006 \\
			& $Q_{\rm hhh}$		& 2.09 $\pm$ 0.55 & -0.12 $\pm$ 0.76 & 16.08 \\%&	0.009 &  0.009\\
		       & $Q^{\rm SC}_{\rm hhh}$	& 1.75 $\pm$ 0.47 & -0.39 $\pm$ 0.56 & 16.66 \\%&  0.006 &	 0.006 \\
\hline
\multirow{4}{*}{0.04} & $B_{\rm hhh}$		& 1.41 $\pm$ 0.08 & -0.05 $\pm$ 0.26 & 26.92 \\%& -0.037 & 0.024	\\
		       & $B^{\rm SC}_{\rm hhh}$	& 1.38 $\pm$ 0.08 & -0.27 $\pm$ 0.25 & 26.16 \\%& -0.030 & 0.018	\\
		       & $Q_{\rm hhh}$		& 2.32 $\pm$ 0.39 & 0.14 $\pm$ 0.58 & 26.40  \\%&	0.010 & 0.018 \\
		       & $Q^{\rm SC}_{\rm hhh}$	& 1.80 $\pm$ 0.29 & -0.34 $\pm$ 0.36 & 26.96 \\%& 	0.019 & 0.025\\
		   \hline
\multirow{4}{*}{0.05} & $B_{\rm hhh}$		& 1.40 $\pm$ 0.06 & 0.15 $\pm$ 0.21 & 31.92  \\%& -0.430 & -0.021	\\
		      &	$B^{\rm SC}_{\rm hhh}$	& 1.38 $\pm$ 0.05 & -0.25 $\pm$ 0.15 & 12.63 \\%& -0.011 & 0.032 \\
		      & $Q_{\rm hhh}$		& 2.66 $\pm$ 0.26 & 0.57 $\pm$ 0.42 & 11.53  \\%&	0.048 & 0.069 \\
		      &	$Q^{\rm SC}_{\rm hhh}$	& 1.90 $\pm$ 0.19 & -0.30 $\pm$ 0.22 & 11.60 \\%&	0.037 & 0.027 \\
\hline
\multirow{4}{*}{0.06} & $B_{\rm hhh}$		& 1.41 $\pm$ 0.05 & 0.19 $\pm$ 0.24 & 63.64  \\%& -0.432 &	-0.100 \\
		      &	$B^{\rm SC}_{\rm hhh}$	& 1.37 $\pm$ 0.03 & -0.23 $\pm$ 0.13 & 19.30 \\%&  0.088 &	-0.030 \\
		      &	$Q_{\rm hhh}$		& 2.84 $\pm$ 0.20 & 0.88 $\pm$ 0.40 & 20.70  \\%& -0.042 & -0.031 \\
		      &	$Q^{\rm SC}_{\rm hhh}$	& 1.87 $\pm$ 0.14 & -0.30 $\pm$ 0.19 & 19.47 \\%& -0.031 & -0.029\\

\hline
\end{tabular}
\caption{\small{Best-fit bias parameters from fitting the
    halo-halo-halo bispectra and reduced bispectra. Column 1:
    bispectrum triangle scale; column 2: statistic, where $B_{\hhh}$
    and $Q_{\hhh}$ are raw, and where $B_{\hhh}^{\rm SC}$ and
    $Q_{\hhh}^{\rm SC}$ are shot-noise corrected; columns 3 and 4:
    best-fit $b_1$ and $b_2$ along with $1\sigma$ errors; column 5:
    $\chi^2$.}}
\label{tab:auto}
\end{table}

%%%%%%%%%%%%%%%%%%%%%%%%%%%%%%%%%%%%%%%%%%%%%%%%%%%%%%%

\begin{figure*}
 \centering{
   \includegraphics[width=8cm,height=7cm]{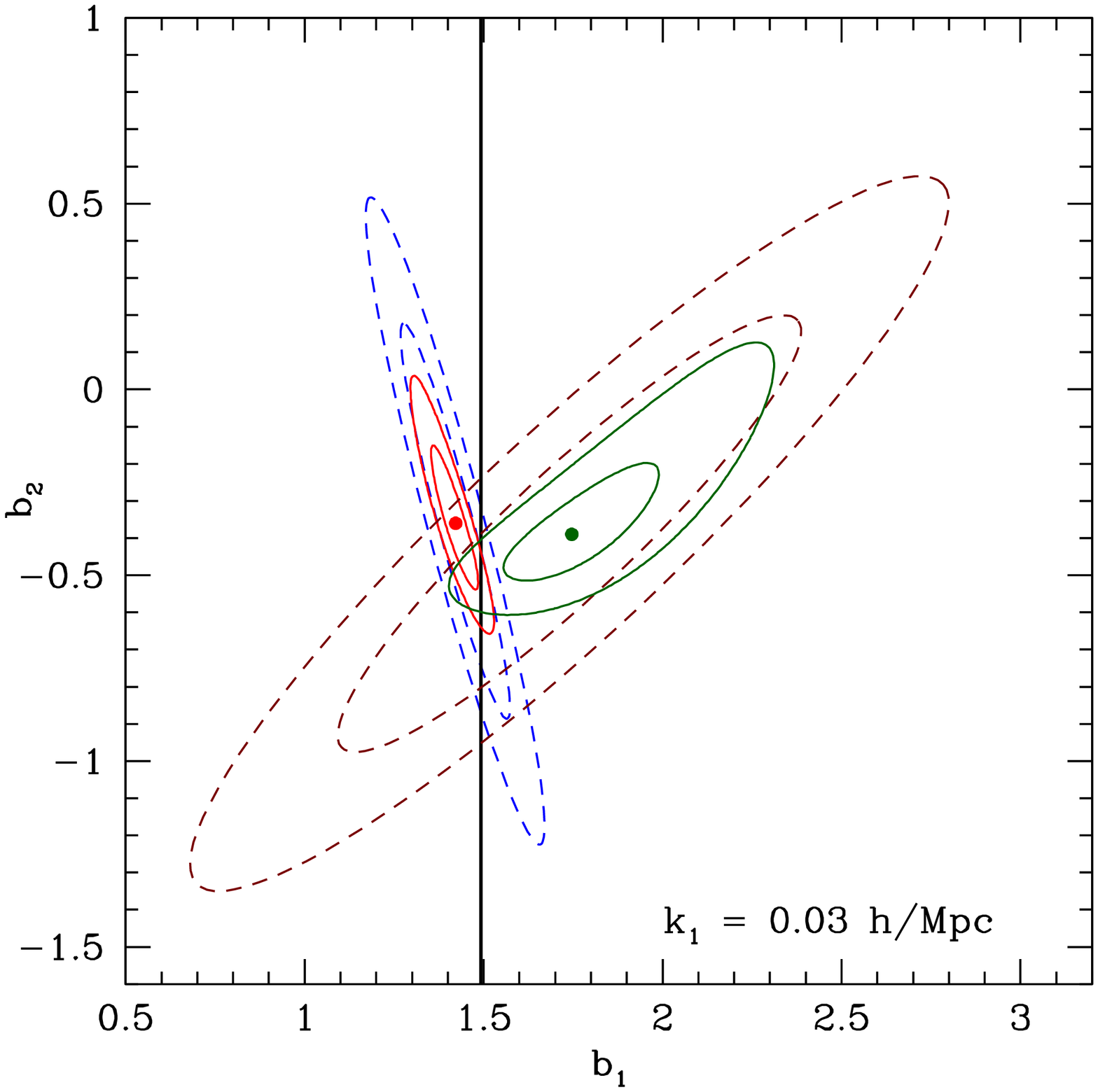}\hspace{-5pt}
   \includegraphics[width=8cm,height=7cm]{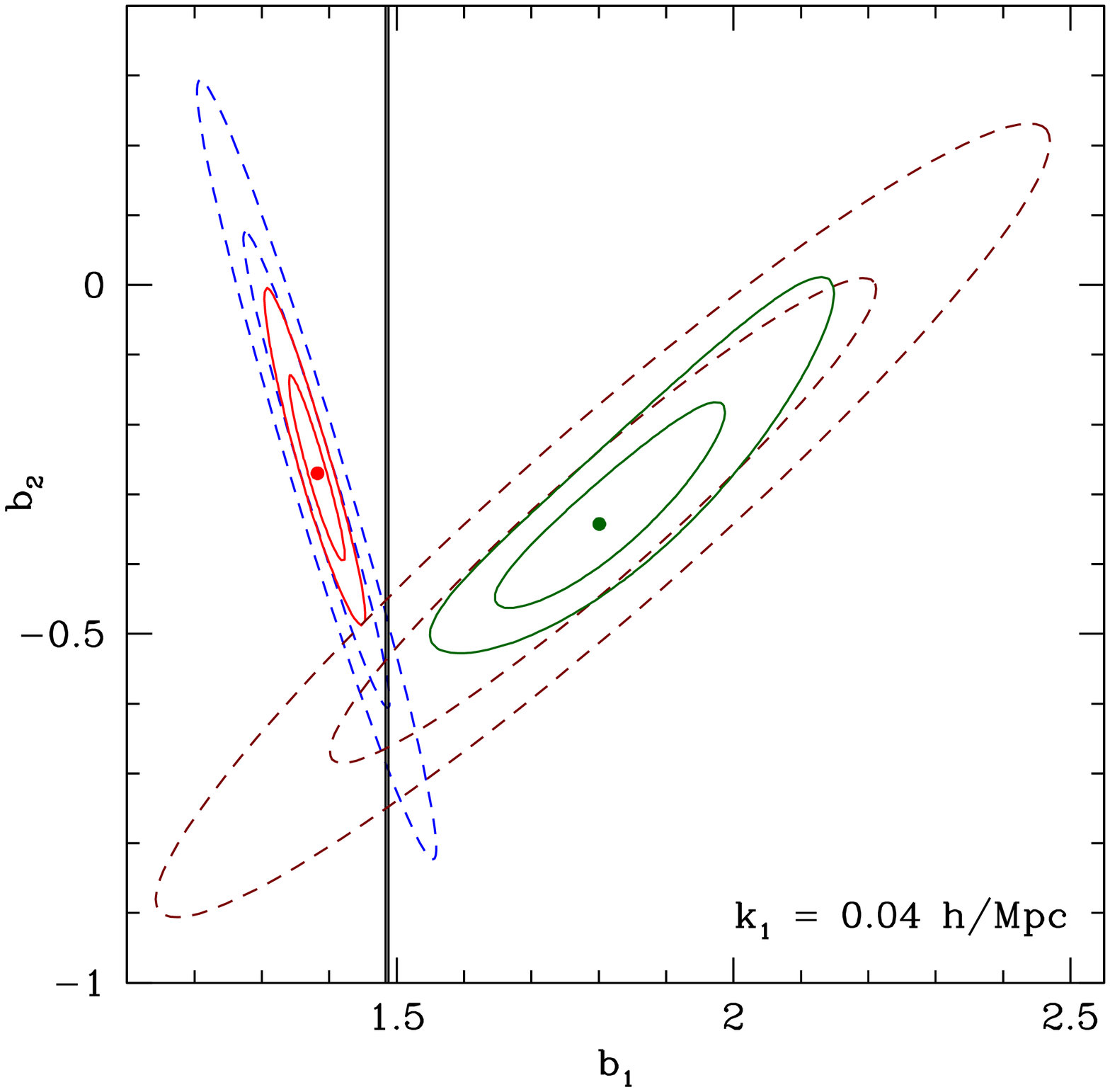}}\vspace{-10pt}
 \centering{ 
   \includegraphics[width=8cm,height=7cm]{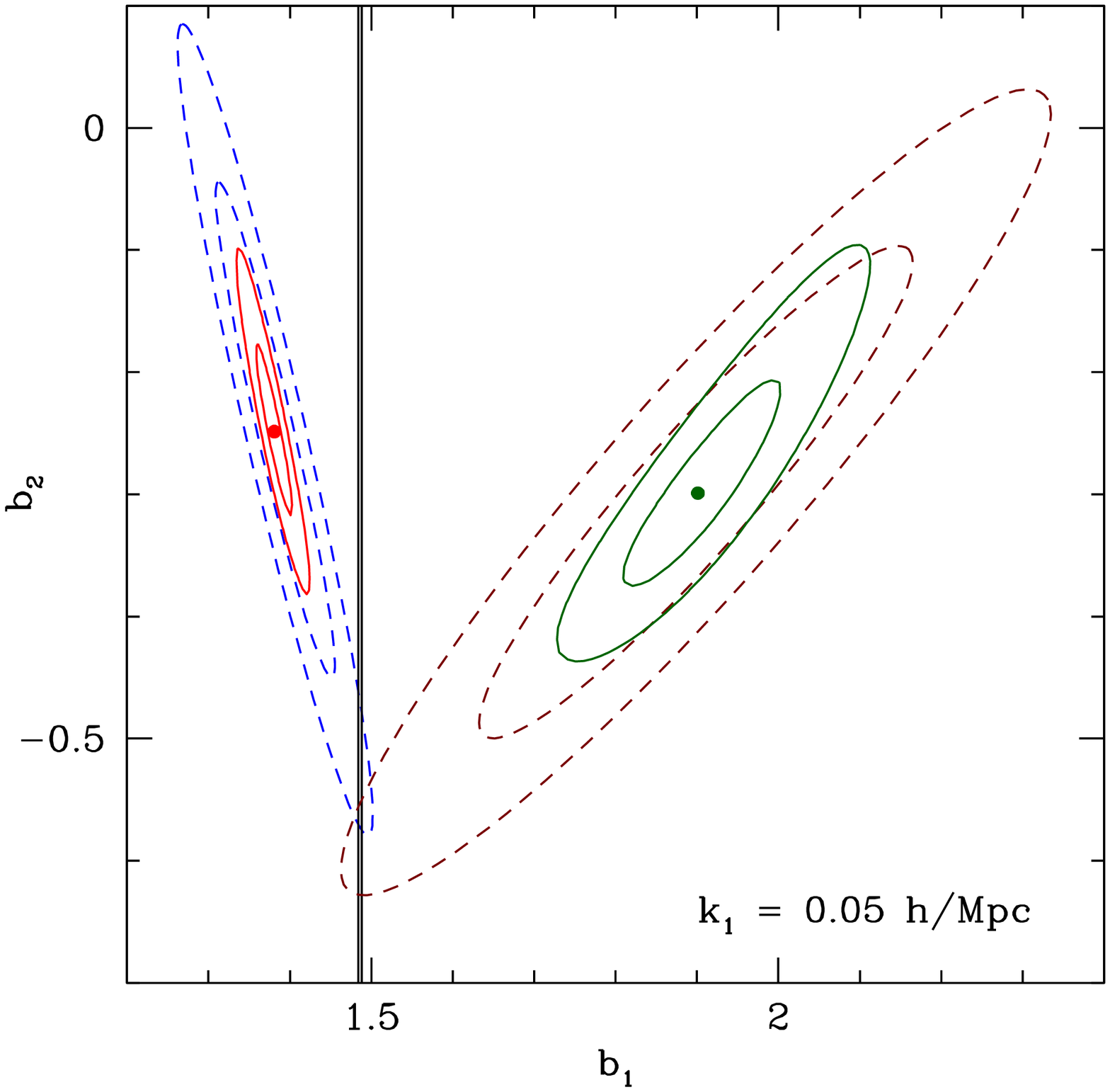}\hspace{-5pt}
   \includegraphics[width=8cm,height=7cm]{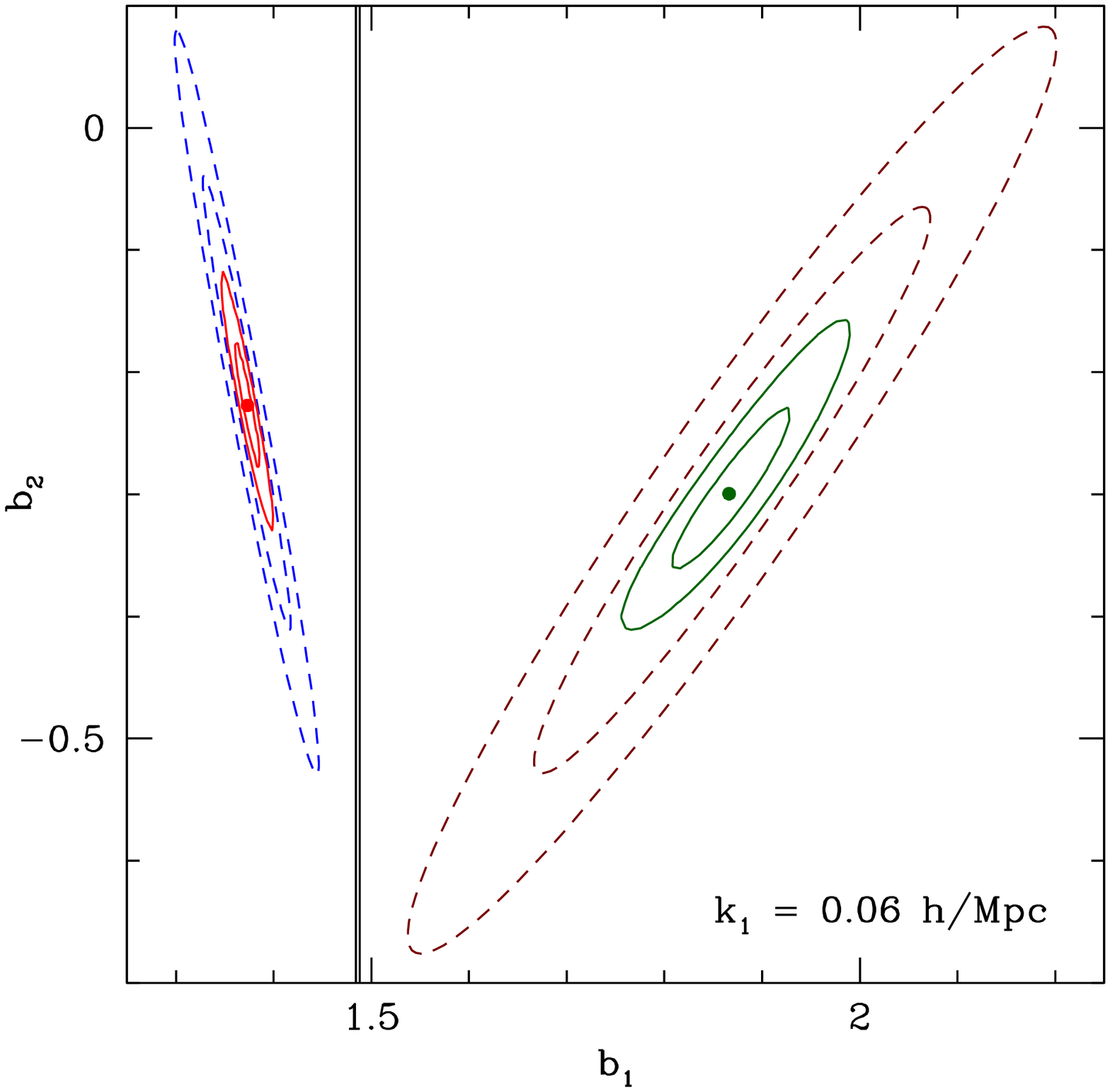}}
\caption{\small{Evolution of the likelihood contours for the bias
    parameters $b_1$ and $b_2$, estimated from $B_{\hhh}$ and
    $Q_{\hhh}$, with scale. The solid lines denote the 68$\%$ and
    95$\%$ confidence intervals, obtained from a full exploration of
    the likelihood surface around the best fit values; the dashed
    lines denote the same, but where the jackknife parameter
    covariance matrix from \Eqn{eq:jkparam} has been used to determine
    the error contours. The top left, top right, bottom left and
    bottom right panels show the results for triangle configurations
    with $k_1=\{0.03,\,0.04,\,0.05,\,0.06\}\kMpc$, respectively. The
    vertical black lines denote the effective bias parameter $b^{\rm
      NL,SC}_{\hh}$, using the same wavemodes that enter into the
    bispectrum estimates. }}
\label{fig:contours3scales}
\end{figure*}

%%%%%%%%%%%%%%%%%%%%%%%%%%%%%%%%%%%%%%%%%%%%%%%%%%%%%%%

The $\chi^2$ function of \Eqn{eq:norm} may be interpreted as a
Gaussian likelihood if we make the transformation, ${\mathcal
  L}(\{B_{\rm hhh}\}|b_1,b_2)\propto \exp[-\chi^2/2]$. Once suitably
normalized and on assuming a set of prior probabilities we may then
explore the shape of the confidence regions in the posterior probability
$p(b_1,b_2|\{B_{\rm hhh}\})$. \\

Figure~\ref{fig:contours3scales} shows the 1$\sigma$ likelihood
confidence contours in the posterior probability for the nonlinear
bias parameters for the four scales considered according to our method
of analysis described above.  The solid lines denote the size of the
confidence regions at the 68$\%$ and 95$\%$ level (i.e. $\Delta \chi^2
\approx2.3,6.17$) when we construct a correlation matrix from the 40
realizations without regard for the systematic uncertainty.  The
dashed-lines demonstrate the magnitude at which the 68$\%$ and 95$\%$
confidence regions expand following our generation of a set of
jackknife subsamples to monitor the effect due to the implicit error
associated with the estimated correlation matrix. Hence, we clearly
see the relevance of accounting for the uncertainty of the correlation
matrix when obtaining the bias parameter constraints.  The discrepancy
between the resulting jackknife error ellipses for $B$ and $Q$ is less
severe than the likelihood contours obtained from the complete sample
where the level of agreement improves progressing to large scales, yet
this might be due to the fact that the statistical error is more
prominent at larger scales.  Interestingly, the overlap of the two
likelihood regions at 2$\sigma$ for $k_1$=0.03, 0.04 and 0.05 $\kMpc$
occurs with the rectangular region or strip denoting the effective
bias measure, $b^{\NL, SC}_{\hh}$, at 1$\sigma$. These set of panels in
Figure~\ref{fig:contours3scales} convey pictorially the information
obtained from the results in Table \ref{tab:auto} that the likelihood
contours from analysis of $Q$ show an evolution with decreasing scale
toward larger and larger $b_1$, whereas the constraints on $b_2$
remain consistent. Lastly, the constraints obtained from analyzing the
$Q$-amplitudes are much weaker than those coming from the bispectrum.

%%%%%%%%%%%%%%%%%%%%%%%%%%%%%%%%%%%%%%%%%%%%%%%%%%%%%%%

\section{Halo bias from cross-bispectra}\label{sec:CrossSpec}

In \S\ref{ssec:b1b2} we saw that shot-noise corrections influenced the
recovery of the bias parameters particularly for $Q$. In this
section, we attempt to develop the use of cross-bispectra, as measures
of the bias that are less susceptible to discreteness effects. The use of
cross-correlations in large-scale structure work has long been known
as a way of reducing discreteness corrections
\citep{Peebles1980}. However, it is only relatively recent that it
has been applied to study bias
\citep{Smithetal2007,Dalaletal2008,Smith2009,Padmanabhanetal2009,
Desjacquesetal2009,Pillepichetal2010}.

%%%%%%%%%%%%%%%%%%%%%%%%%%%%%%%%%%%%%%%%%%%%%%%%%%%%%%%

\subsection{Definitions and theory}

We may define the halo cross-bispectra as follows:
\ba
\lexp \dek_{\rm h}(\vk_1)\dek_{\rm h} (\vk_2)\dek(\vk_3) \rexp \!\!\!\!
& \!\! = & \!\!\!\!\! (2\pi)^3\,\dD(\vk_{123})B_{\rm hhm}(\vk_1,\vk_2,\vk_3) \ ; \\
\lexp \dek_{\rm h}(\vk_1)\dek(\vk_2)\dek(\vk_3) \rexp \!\!\!\!
& \!\! = & \!\!\!\!\! (2\pi)^3\,\dD(\vk_{123})B_{\rm hmm}(\vk_1,\vk_2,\vk_3) \ .
\ea
We then symmeterize these quantities by the operations:
\ba 
B_{\hhm}^{(\rm sym)} & = &
\left[B_{\hhm}+B_{\rm hmh}+B_{\rm mhh}\right]/3 \ ;\\
B_{\hmm}^{(\rm sym)} & = &
\left[B_{\hmm}+B_{\rm mhm}+B_{\rm mmh}\right]/3  \ .
\ea
For ease of notation we shall now simply take $B_{\hhm}^{(\rm
  sym)}\equiv B_{\hhm}$ and $B^{(\rm sym)}_{\hmm}\equiv B_{\hmm}$,
unless otherwise indicated.  We may now also define the cross-reduced
bispectra as:
\ba 
Q_{\rm hhm} & \equiv & B_{\rm hhm}/PP_{\hhm} \ ;\\
Q_{\rm hmm} & \equiv & B_{\rm hmm}/PP_{\hmm} \ ,
\ea
where we have for the denominators (again symmetrized):
\ba
PP_{\hhm} & = & 
\frac{2}{3}
\left[P_{\rm hh}(k_1)P_{\rm hm}(k_2)+2\ \cyc\, \right] \nn \\
& & +\frac{1}{3}
\left[P_{\rm hm}(k_1)P_{\rm hm}(k_2)+2\ \cyc\, \right] \ ; \label{eq:PPhhm}\\
PP_{\hmm} & = & 
\frac{2}{3}
\left[P_{\rm hm}(k_1)P_{\rm mm}(k_2)+2\ \cyc\, \right] \nn \\
& & +\frac{1}{3}
\left[P_{\rm hm}(k_1)P_{\rm hm}(k_2)+2\ \cyc\, \right] \ . \label{eq:PPhmm}
\ea
The relations for $PP_{\hhm}$ and $PP_{\hmm}$ can easily be
constructed using a graphical approach. Let us consider three nodes
two of which are the same and the third is different (we shall think
of the nodes as the density fields). Label these nodes 1, 2, and
3. Now consider all possible ways to connect the three nodes together
by two edges. When two nodes, which are the same, connect together this
gives us an auto-power spectrum with a delta function, and when two
nodes that are different connect together this gives us a cross-power
spectrum and delta function. One may then symmetrize the results by
considering all possible relabellings of the nodes and dividing by
three.

In Appendix~\ref{app:halobispectra} we calculate the tree-level
cross-bispectra, $B_{\hmm}$ and $B_{\hhm}$, in the local model of halo
biasing. The main results are:
\ba
B^{(0)}_{\rm hmm}(\bk_1,\bk_2,\bk_3) & \approx & b_1(M)B_{\rm mmm}^{(0)}(\bk_1,\bk_2,\bk_3) \nn \\ 
& & \hspace{-2.3cm}+
\frac{b_2(M)}{3}\left[\widetilde{W}_{\bk_1,\bk_2} P^{(0)}_{\rm mm}(k_1)P^{(0)}_{\rm mm}(k_2)+2\,\cyc \right] \ ;\\
B^{(0)}_{\rm hhm}(\bk_1,\bk_2,\bk_3) & \approx &
b_1^2(M) B_{\rm mmm}^{(0)}(\bk_1,\bk_2,\bk_3) +\frac{1}{3}b_1(M) \nn \\ 
& & \hspace{-2.3cm} \times 
b_2(M)\big[ \widetilde{W}_{\bk_1,\bk_2} P_{\rm mm}^{(0)}(k_1)P_{\rm mm}^{(0)}(k_2)+2\,\cyc\,\big]\ .
\ea
In the limit of large scales, and or small smoothing scales, the
filter functions $\widetilde{W}_{\bk_1,\bk_2}\rightarrow 1$ and we
have 
\ba
&  & \!\!\!\!\!\!\!B^{(0)}_{\rm hmm} \approx b_1B_{\rm mmm}^{(0)} +
\frac{b_2}{3}\left[P^{(0)}_{\rm mm}(k_1)P^{(0)}_{\rm mm}(k_2)+2\,\cyc \right] \ ;\\
& & \!\!\!\!\!\!\!B^{(0)}_{\rm hhm} \approx 
b_1^2 B_{\rm mmm}^{(0)} +\frac{1}{3}b_1b_2 \left[ P_{\rm mm}^{(0)}(k_1)P_{\rm mm}^{(0)}(k_2) 
+2\,\cyc\,\right] \ . \nn \\
\ea

At second order in the nonlinear bias and PT, the cross-reduced
bispectra are:
\ba
Q^{(0)}_{\hmm} \!\! & \approx  &\!\!  
\frac{3 Q_{\rm mmm}}{2+b_1(M)} + \frac{b_2(M)}{2 b_1(M)+b^2_1(M)}
\alpha(\bk_1,\bk_2,\bk_3) \ ;  \\
Q^{(0)}_{\hhm} \!\! & \approx & \!\! 
\frac{3\,Q_{\rm mmm}}{2 b_1(M) + 1} + \frac{2 b_2(M)}{2 b_1^2(M) + b_1(M)}
\alpha(\bk_1,\bk_2,\bk_3)\ . \nn \\
\ea
In the large-scale limit, $\alpha\rightarrow1$, these expressions
become:
\ba
Q^{(0)}_{\hmm} & \approx  &  \frac{3 Q_{\rm mmm}}{2+b_1(M)} + \frac{b_2(M)}{2 b_1(M)+b^2_1(M)}\ ;\\
Q^{(0)}_{\hhm} & \approx &  \frac{3\,Q_{\rm mmm}}{2 b_1(M) + 1} + \frac{2 b_2(M)}{2 b_1^2(M) + b_1(M)}\ .
\ea

%%%%%%%%%%%%%%%%%%%%%%%%%%%%%%%%%%%%%%%%%%%%%%%%%%%%%%%

\begin{table}
\centering
%\begin{tabular}{|c|c|c|c|c|c|c|} \hline
\begin{tabular}{|c|c|c|c|c|} \hline
$k_1\,[\kMpc]$ & & $b_{1}$ $\pm$ $\sigma_{b_1}$ & $b_{2}$ $\pm$ $\sigma_{b_2}$ & $\chi^2$ \\% & $\Delta b_1$ & $\Delta b_2$ \\ 
\hline 
\multirow{4}{*}{0.03} & $B_{\hmm}$	    &  1.37 $\pm$ 1.18 & -0.36 $\pm$ 0.32 & 16.86  \\% & -0.030 & 0.056  \\
		       & $B^{\rm SC}_{\hmm}$&  1.37 $\pm$ 1.18 & -0.36 $\pm$ 0.32 & 16.86    \\% & -0.030 & 0.056	 \\
		       & $Q_{\hmm}$	    &  1.98 $\pm$ 0.75 & -0.75 $\pm$ 0.99 & 19.09  \\% & -0.031 & -0.031	 \\
		       & $Q^{\rm SC}_{\hmm}$&  1.98 $\pm$ 0.75 & -0.75 $\pm$ 0.99 & 19.09    \\% &	-0.031 & -0.040 \\
\hline
\multirow{4}{*}{0.04} & $B_{\hmm}$	    &  1.33 $\pm$ 0.16 & -0.33 $\pm$ 0.51 & 16.83 \\%  &	 0.022	& -0.032 \\
		       & $B^{\rm SC}_{\hmm}$&  1.33 $\pm$ 0.16 & -0.33 $\pm$ 0.51 & 16.83 \\%  &	 0.022	& -0.032 \\
		       & $Q_{\hmm}$	    &  2.50 $\pm$ 0.56 & -0.31 $\pm$ 0.89 & 17.96 \\% &	-0.094	& -0.032 \\
		       & $Q^{\rm SC}_{\hmm}$&  2.51 $\pm$ 0.56 & -0.32 $\pm$ 0.89 & 17.97 \\% &	-0.094	& -0.032 \\
\hline
\multirow{4}{*}{0.05} & $B_{\hmm}$	    &  1.30 $\pm$ 0.07 & -0.01 $\pm$ 0.23 & 26.98 \\% &	-0.002  &  0.010 \\
		       & $B^{\rm SC}_{\hmm}$&  1.30 $\pm$ 0.07 & -0.02 $\pm$ 0.23 & 26.82 \\% &	-0.002	&  0.009 \\
		       & $Q_{\hmm}$	    &  2.94 $\pm$ 0.25 &  0.24 $\pm$ 0.57 & 16.90 \\% &	-0.002	& -0.062 \\
		       & $Q^{\rm SC}_{\hmm}$&  2.94 $\pm$ 0.25 &  0.23 $\pm$ 0.57 & 16.95 \\% &	-0.002	& -0.062 \\
\hline
\multirow{4}{*}{0.06} & $B_{\hmm}$	    &  1.29 $\pm$ 0.07 & -0.001 $\pm$ 0.28 & 13.99 \\% & -0.006	& 0.036	 \\
		       & $B^{\rm SC}_{\hmm}$&  1.29 $\pm$ 0.07 & -0.004 $\pm$ 0.28 & 13.93 \\% & -0.006	& 0.036	 \\
		       & $Q_{\hmm}$	    &  3.45 $\pm$ 0.30 &  1.43  $\pm$ 0.94 & 46.60 \\% & 0.013	& 0.048	  \\
                       & $Q^{\rm SC}_{\hmm}$&  3.45 $\pm$ 0.30 &  1.42  $\pm$ 0.94 & 46.87 \\% & 0.013	& 0.048	\\
\hline
\end{tabular}
\caption{\small{Best-fit bias parameters from halo-mass-mass bispectra
    and reduced bispectra. Column 1: bispectrum triangle scale; column
    2: statistic, where $B_{\hmm}$ and $Q_{\hmm}$ are raw and where
    $B_{\hmm}^{\rm SC}$ and $Q_{\hmm}^{\rm SC}$ are shot-noise
    corrected; columns 3 and 4: best-fit $b_1$ and $b_2$ along with
    $1\sigma$ errors; column 5: $\chi^2$.}}
\label{tab:crosshmm}
\end{table}

\begin{table}
\centering
%\begin{tabular}{|c|c|c|c|c|c|c|c|} 
\begin{tabular}{|c|c|c|c|c|c|} 
\hline
$k_1\,[\kMpc]$ & & $b_{1}$ $\pm$ $\sigma_{b_1}$ & $b_{2}$ $\pm$ $\sigma_{b_2}$ & $\chi^2$ \\ % & $\Delta b_1$ & $\Delta b_2$ \\ 
\hline 
\multirow{4}{*}{0.03}  & $B_{\hhm}$          & 1.42 $\pm$ 0.25 & -0.32 $\pm$ 0.60 & 17.13 \\ %& -0.016 & 0.030  \\
		       & $B^{\rm SC}_{\hhm}$ & 1.42 $\pm$ 0.24 & -0.45 $\pm$ 0.47 & 17.49 \\ %& -0.020 & 0.039 \\
		       & $Q_{\hhm}$	     & 2.08 $\pm$ 0.40 & -0.49 $\pm$ 0.46 & 17.45 \\ %& -0.014 & -0.013  \\
		       & $Q^{\rm SC}_{\hhm}$ & 2.11 $\pm$ 0.40 & -0.67 $\pm$ 0.42 & 17.46 \\ %& -0.013 & -0.015 \\
\hline
\multirow{4}{*}{0.04}  & $B_{\hhm}$	     & 1.39 $\pm$ 0.14 & -0.24 $\pm$ 0.25 & 21.29 \\ %& -0.096 &  0.074  \\
		       & $B^{\rm SC}_{\hhm}$ & 1.37 $\pm$ 0.15 & -0.37 $\pm$ 0.24 & 21.42 \\ %& -0.010 &  0.079  \\
		       & $Q_{\hhm}$	     & 2.49 $\pm$ 0.46 & -0.17 $\pm$ 0.56 & 22.92 \\ %& -0.080 & -0.042  \\
		       & $Q^{\rm SC}_{\hhm}$ & 2.56 $\pm$ 0.49 & -0.47 $\pm$ 0.51 & 24.60 \\ %& -0.082 &  0.033 \\
\hline
\multirow{4}{*}{0.05} & $B_{\hhm}$	     & 1.38 $\pm$ 0.08 & -0.05 $\pm$ 0.18 & 22.46 \\ %& -0.0001 & 0.002   \\
		       & $B^{\rm SC}_{\hhm}$ & 1.36 $\pm$ 0.07 & -0.29 $\pm$ 0.14 & 13.16 \\ %& -0.0003 & 0.002  \\
		       & $Q_{\hhm}$	     & 2.87 $\pm$ 0.18 &  0.22 $\pm$ 0.29 & 12.55 \\ %&  0.003  &  -0.002  \\
		       & $Q^{\rm SC}_{\hhm}$ & 2.92 $\pm$ 0.19 & -0.37 $\pm$ 0.25 & 16.77 \\ %&  0.004  &  -0.0003 \\
\hline
\multirow{4}{*}{0.06} & $B_{\hhm}$	     & 1.39 $\pm$ 0.07 & -0.07 $\pm$ 0.25 & 20.56 \\ %& -0.010 & 0.039 \\
		       & $B^{\rm SC}_{\hhm}$ & 1.36 $\pm$ 0.07 & -0.31 $\pm$ 0.22 & 15.32 \\ %& -0.090 & 0.033 \\
		       & $Q_{\hhm}$	     & 3.12 $\pm$ 0.27 & 0.54  $\pm$ 0.52 & 37.81 \\ %&  0.060 & 0.121  \\
                       & $Q^{\rm SC}_{\hhm}$ & 3.25 $\pm$ 0.31 & -0.14 $\pm$ 0.47 & 60.54 \\ %&  0.064 & 0.102 \\
\hline
\end{tabular}
\caption{\small{Best-fit bias parameters from halo-halo-mass bispectra
    and reduced bispectra. Column 1: bispectrum triangle scale; column
    2: statistic, where $B_{\hhm}$ and $Q_{\hhm}$ are raw and where
    $B_{\hhm}^{\rm SC}$ and $Q_{\hhm}^{\rm SC}$ are shot-noise
    corrected; columns 3 and 4: best-fit $b_1$ and $b_2$ along with
    $1\sigma$ errors; column 5: $\chi^2$.}}
\label{tab:crosshhm}
\end{table}

%%%%%%%%%%%%%%%%%%%%%%%%%%%%%%%%%%%%%%%%%%%%%%%%%%%%%%%

\begin{figure*}
  \centering
    \includegraphics[height=12cm,width=16cm]{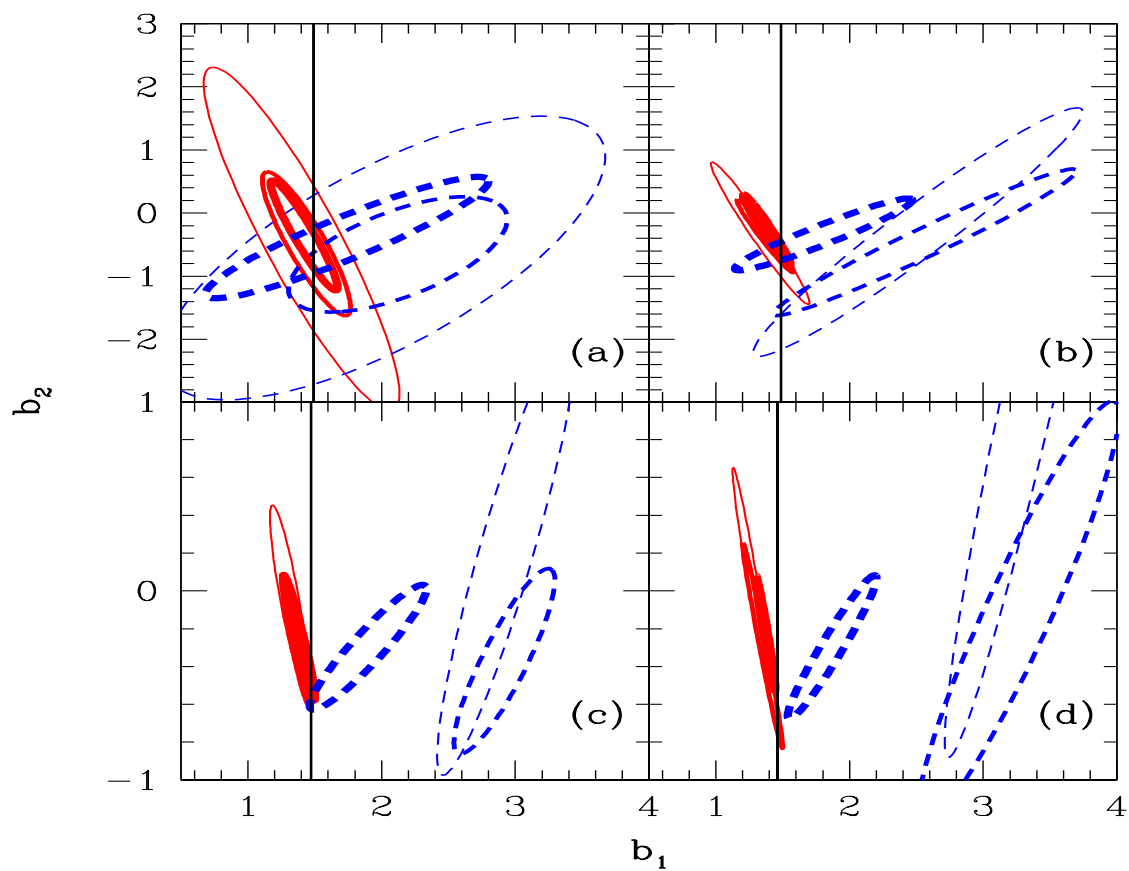}
  \caption{\small{Evolution of the 95\% likelihood contours for $b_1$
      and $b_2$ obtained from the halo auto- and cross-bispectra and
      reduced bispectra as a function of scale. In each panel, the
      solid red lines of increasing thickness denote \mbox{$\{B_{\rm
          hmm},\,B_{\rm hhm},\,B_{\rm hhh}\}$} and the dashed blue
      lines of increasing thickness denote $\{Q_{\rm hmm},\,Q_{\rm
        hhm},\,Q_{\rm hhh}\}$.  The alphabetical labels $\{\rm
      a,\,b,\,c,\,d\}$ correspond to the triangle configurations with
      $k_1=\{0.03,\,0.04,\,0.05,\,0.06\}\kMpc$, respectively.  The
      vertical black lines denote the effective bias parameter $b^{\rm
        NL,SC}_{\hh}$, using the same wavemodes that enter into the
      bispectrum estimates. }}
  \label{fig:crossSC}
\end{figure*}

%%%%%%%%%%%%%%%%%%%%%%%%%%%%%%%%%%%%%%%%%%%%%%%%%%%%%%%

\subsection{Estimation of the cross-bispectra}

The cross-bispectra $B_{\rm hhm}$ and $B_{\rm hmm}$ can be estimated
following the algorithm described in \S\ref{ssec:BispEst} with some
small modifications. Firstly, the estimates must be symmetrized, and
for the discrete form of $B_{\rm hhm}$ we have,
\ba \widehat{B}^{\rm d}_{\rm hhm}(k_1,k_2,\theta_{12}) & = & \frac{1}{3}
\frac{\Vu^2}{N_{\rm tri}} \sum_{(\vn_1, \vn_2)}^{N_{\rm tri}} \nn \\
& & \hspace{-1.5cm}\times \left\{\frac{}{}
\mathcal{R}e[\delta_{\rm h}(\vk_{\vn_1})\delta_{\rm h}(\vk_{\vn_2})\delta_{\rm m}(\vk_{\vn_3})] 
+2\,\cyc \right\}\ ,
\label{eq:biestHHM}
\ea
and with a similar relation for $\widehat{B}^{\rm d}_{\rm hmm}$. The reduced
bispectra are estimated by dividing the above bispectrum estimates by
estimates for $PP_{\rm hhm}$ and $PP_{\rm hmm}$ from
\Eqns{eq:PPhhm}{eq:PPhmm}, respectively.

One further complication is constructing the corrections for
shot-noise.  This may be performed following the counts-in-cells
approach of \citet{Peebles1980} \citep[and see also][]{Smith2009}. We
find that the symmetrized corrections for $B_{\rm hhm}$ and $B_{\rm
  hmm}$ can be written as:
\ba
\widehat{\bar{B}}_{\rm hhm,shot}  & \equiv & 
\frac{1}{3\nbar_{\rm h}}\left[P^{\rm d}_{\rm hm}(k_1)+2\,\cyc\right] \ ; \\
\widehat{\bar{B}}_{\rm hmm,shot}  & \equiv & 
\frac{1}{3\nbar_{\rm m}}\left[P^{\rm d}_{\rm hm}(k_1)+2\,\cyc\right] \ ,
\ea
where $\nbar_{\rm m}=N/\Vu$ and $\nbar_{\rm h}=N_{\rm h}/\Vu$, are the
number density of matter particles and haloes, respectively.  For the
reduced bispectra we must correct the estimates of $PP_{\rm hhm}$ and
$PP_{\rm hmm}$, which are written in the following form:
\ba 
Q^{\rm denom}_{\rm hhm,shot} & = & 
\frac{2}{3\nbar_{\rm h}}\left[\widehat{P}^{\rm d}_{\rm hm}(k_1)+2\,\cyc\right] \ ;\\
Q^{\rm denom}_{\rm hmm,shot} & = & 
\frac{2}{3\nbar_{\rm m}}\left[\widehat{P}^{\rm d}_{\rm hm}(k_1)+2\,\cyc\right] \ .
\ea
As it is the case that $\nbar_{\rm m}\gg\nbar_{\rm h}$, we expect that
the shot-noise corrections to $B_{\rm hmm}$ will be significantly
smaller than for $B_{\rm hhm}$. Hence, we shall think of this as being
an almost perfect measure independent of discreteness.

Using these estimators, we compute the ensemble average and ensemble-to-ensemble variations of the halo-mass cross-bispectra. We do this
for the same bispectra configurations that were considered in
\S\ref{ssec:BispEst}.

%%%%%%%%%%%%%%%%%%%%%%%%%%%%%%%%%%%%%%%%%%%%%%%%%%%%%%%

\subsection{Nonlinear bias from cross-bispectra}

We estimate the nonlinear bias parameters and their errors from the
cross-bispectra using the same method employed for the auto-bispectra
in \S\ref{ssec:biasest} and \S\ref{ssec:biaserr}. The results are
tabulated in Tables~\ref{tab:crosshmm} and \ref{tab:crosshhm},
respectively.
 
Figure~\ref{fig:crossSC} presents the 2-D 95\% confidence likelihood
contours for $b_1$ and $b_2$, that are obtained from fitting the
shot-noise corrected bispectra $\{B_{\hhh}^{\rm SC},\,B_{\hhm}^{\rm
  SC},\,B_{\hmm}^{\rm SC}\}$ and reduced bispectra $\{Q_{\hhh}^{\rm
  SC},\,Q_{\hhm}^{\rm SC},\,Q_{\hmm}^{\rm SC}\}$. The four panels show
the results obtained from fitting triangle configurations,
\mbox{$k_1\in\{0.03,\,0.04,\,0.05,\,0.06\}\kMpc$}, with $k_2/k_1=2$,
and these correspond to the top-left, top-right, bottom left and
bottom right panels, respectively.  For comparative purposes, the
vertical band in each panel, denotes the 1$\sigma$ constraint on
$b^{\rm NL, SC}_{\hh}$, obtained from the shot-noise corrected halo and nonlinear matter power
spectra (c.f.~\S\ref{ssec:Pwspec}).

Consider first the bispectra $\{B_{\hmm}^{\rm SC},\,B_{\hhm}^{\rm
  SC},\, B_{\hhh}^{\rm SC},\,\}$ (solid red lines of increasing
thickness), from the figure and the tables, we see that all of the
results are reasonably consistent with one another over the various
scale ranges considered. However, when smaller scales are used
(i.e. $k_1\ge0.05\kMpc$), the consistency weakens and the best-fit
parameters, obtained from $B_{\hhh}^{\rm SC}$ and $B_{\hmm}^{\rm SC}$,
differ by $\gtrsim2.5\sigma$.

Evaluating the results for the reduced bispectra $\{Q_{\hmm}^{\rm
  SC}, \,Q_{\hhm}^{\rm SC},\,Q_{\hhh}^{\rm SC}\}$ (dashed blue lines
of increasing thickness), the four panels show a strong evolution of
the error ellipses with scale. We also note that the level of
agreement between the different estimators also evolves strongly,
becoming weaker and weaker as smaller scales are considered. At the
largest scale where $k=0.03\,\kMpc$, all of the 2$\sigma$ likelihood
contour regions overlap. However, this consistency is broken for the
next scale range, \mbox{$k=0.04\,\kMpc$}, where $Q_{\hmm}$ and
$Q_{\hhm}$ are shifted downwards and further to the right favoring a
more negative $b_2$ and higher $b_1$.  The trend continues in this
same direction heading to smaller and smaller scales.

Comparing the results from both $B$ and $Q$ together, we see that only
on the largest scales is there any degree of overall consistency.  One
way to interpret the results up to now, is that, if we believe
$b_1\approx b_{\hh}^{\rm NL,SC}$, then the agent driving the
inconsistency between the parameter estimates, is the breakdown of the
local bias model at tree-level.  Furthermore, the breakdown of the
local tree-level model is more severe for the reduced bispectrum than
for the bispectrum.

%%%%%%%%%%%%%%%%%%%%%%%%%%%%%%%%%%%%%%%%%%%%%%%%%%%%%%%

\section{The need for beyond tree level bias models}\label{sec:biasbyhand}

This final set of analysis consists of a simple proof of method test,
and we determine whether, when the underlying bias model is known, the
`true' bias parameters of the model are indeed recoverable with our
approach.

%%%%%%%%%%%%%%%%%%%%%%%%%%%%%%%%%%%%%%%%%%%%%%%%%%%%%%%

\subsection{Biasing by hand}\label{ssec:biasbyhand}

For these tests, and for simplicity, we shall assume that the local
model of biasing at quadratic order is the correct underlying bias
model. Nonlinear biased density fields of this type may be obtained
through the following procedure. 

For each of the $z=0$ outputs of the 40 simulations, we assign the
nonlinear density field of matter to a cubical Fourier grid using the
CIC algorithm. This is then Fourier transformed. Each Fourier mode is
then smoothed using a Gaussian filter of scale $R$. We then inverse
Fourier transform this field and obtain the smoothed, nonlinear matter
distribution in real space. Using this we next form the sum,
\be \delta_{\rm b}(\bx|R)=b^{\rm b}_1\delta(\bx|R)+b^{\rm
  b}_2[\delta(\bx|R)]^2/2\ , \ee
where $b_1^{\rm b}$ and $b_2^{\rm b}$ are the artificial bias
parameters. Finally, this is Fourier transformed to give us
$\delta_{\rm b}(\bk|R)$. Thus, given $\delta(\bk|R)$ and $\delta_{\rm
  b}(\bk|R)$, we can now use our standard bispectrum estimators to
estimate $B_{\rm bbb}$, $B_{\rm bbm}$, $B_{\rm bmm}$ and $B_{\rm
  mmm}$. We refer to this procedure as the `biasing-by-hand' test.

The major benefits of these tests are that we are able to better gauge
the effects to which nonlinearities beyond tree-level order influence
the measured bispectra and reduced bispectra. We also note that
shot-noise plays no r\^ole here, since the biased field is created
from the matter density field which is densely sampled.

%%%%%%%%%%%%%%%%%%%%%%%%%%%%%%%%%%%%%%%%%%%%%%%%%%%%%%%

\subsection{Theoretical interpretation}

In order to interpret the results from such a construction we may use
the results presented in Appendix~\ref{app:halobispectra}, with the
small modification that we do not de-smooth the results. If we define
the smoothed bispectra as
\be {\mathcal B}(\bk_1,\bk_2,\bk_3)\equiv W(k_1R)W(k_2R)W(k_3R)B(\bk_1,\bk_2,\bk_3) \ ,\ee
then for $\{{\mathcal B}_{\rm bmm},\,{\mathcal B}_{\rm
  bbm},\,{\mathcal B}_{\rm bbb}\}$, we have:
\ba 
{\mathcal B}_{\rm bmm} \!\! & = & \!\! b_1 {\mathcal B}_{\rm mmm}   + 
\frac{b_2}{2}{\mathcal P}_{\rm 4,m} \ ;\label{eq:bmm}\\
{\mathcal B}_{\rm bbm} \!\! & = & \!\! b_1^2 {\mathcal B}_{\rm mmm} + 
\frac{b_1b_2}{3}{\mathcal P}_{\rm 4,m} +\frac{b_2^2}{12} {\mathcal P}_{\rm 5,m} \ ;\label{eq:bbm}\\
{\mathcal B}_{\rm bbb} \!\! & = & \!\! b_1^3 {\mathcal B}_{\rm mmm} + 
\frac{b_1^2b_2}{2}{\mathcal P}_{\rm 4,m} + 
\frac{b_1b_2^2}{4}{\mathcal P}_{\rm 5,m} + 
\frac{b_2^3}{8}{\mathcal P}_{\rm 6,m} \ ,\nn \\ \label{eq:bbb}
\ea
where for ease of notation we take $b_i^{\rm b}=b_i$ and in the above
we have suppressed the dependence of ${\mathcal B}$, ${\mathcal
  P}_{\rm 4,m}$, ${\mathcal P}_{\rm 5,m}$ and ${\mathcal P}_{\rm
  6,m}$, on $(\bk_1,\bk_2,-\bk_1-\bk_2)$. We have also introduced the
auxiliary functions:
\ba 
{\mathcal P}_{n,\rm m} & \equiv & W(k_1R)W(k_2R)W(k_3R) P_{n,\rm m} \ ;\\
P_{\rm 4,m} & \equiv & \int \frac{\dq_1}{(2\pi)^3} 
\widetilde{W}_{\bq_1,\bk_1-\bq_1} \nn \\ 
& & \times T(\bq_1,\bk_1-\bq_1,\bk_2,\bk_3) +2\,\cyc\, \ ;\\
P_{\rm 5,m} & \equiv & \int \frac{\dq_1}{(2\pi)^3}  \frac{\dq_2}{(2\pi)^3} 
\widetilde{W}_{\bq_1,\bk_1-\bq_1}  \widetilde{W}_{\bq_2,\bk_2-\bq_2} \nn \\
& & \times P_{\rm 5,m}(\bq_1,\bk_1-\bq_1,\bq_2,\bk_2-\bq_2,\bk_3)  \nn \\  
& & + 2\,\cyc\, \ ; \\
P_{\rm 6,m} & \equiv & \int \frac{\dq_1}{(2\pi)^3} \dots \frac{\dq_3}{(2\pi)^3}  
\widetilde{W}_{\bq_1,\bk_1-\bq_1} \dots \widetilde{W}_{\bq_3,\bk_3-\bq_3} \nn \\
& & \times P_{6}(\bq_1,\bk_1-\bq_1,\bq_2,\bk_2-\bq_2,\bq_3,\bk_3-\bq_3)  \ .
\ea
The attractive aspect of this test can now be understood: if we move
the terms in Eqns~(\ref{eq:bmm}), (\ref{eq:bbm}) and (\ref{eq:bbb}), which are
proportional to ${\mathcal B}_{\rm mmm}$ from the right to the
left-hand-side, then we may rewrite this system as the matrix
equation:
\be
\left(
\begin{array} {c}
{\mathcal Y}_{\rm bmm} \\
{\mathcal Y}_{\rm bbm} \\
{\mathcal Y}_{\rm bbb} 
\end{array}
\right)
=
\left(
\begin{array} {ccc}
b_2/2 & 0 & 0\\
b_1b_2/3 & b_2^2/12 & 0\\
b_1^2b_2/2 & b_1b_2^2/4 & b_2^3/8 
\end{array}
\right)
\left(
\begin{array} {c}
{\mathcal P}_{\rm 4,m} \\
{\mathcal P}_{\rm 5,m} \\
{\mathcal P}_{\rm 6,m} 
\end{array}
\right)\ ,
\ee
where we defined ${\mathcal Y}_{\rm bmm}\equiv {\mathcal
  B}_{\rm bmm} - b_1 {\mathcal B}_{\rm mmm}$, etc. This equation may
be inverted to give,
\be
\left(
\begin{array} {c}
{\mathcal P}_{\rm 4,m} \\
{\mathcal P}_{\rm 5,m} \\
{\mathcal P}_{\rm 6,m} 
\end{array}
\right)
=
\frac{1}{b_2^3}\left(
\begin{array} {ccc}
2b_2^2 & 0 & 0\\
-8b_1b_2 &  12b_2 & 0\\
8b_1^2 & -24b_1 & 8
\end{array}
\right)
\left(
\begin{array} {c}
{\mathcal Y}_{\rm bmm} \\
{\mathcal Y}_{\rm bbm} \\
{\mathcal Y}_{\rm bbb}  
\end{array}
\right)\ .
\ee
Hence, if we specify $b_1$, $b_2$ and measure the four bispectra
$B_{\rm mmm}$, $B_{\rm bmm}$, $B_{\rm bbm}$ and $B_{\rm bbb}$, then we
can determine exactly ${\mathcal P}_{\rm 4,m}$, ${\mathcal P}_{\rm
  5,m}$ and ${\mathcal P}_{\rm 6,m}$.  Thus we have complete knowledge
of all components of the nonlinear model at all orders in the theory.
The lowest order perturbation theory expansions of these statistics
are (c.f. Appendix~\ref{app:halobispectra}):
\ba
{\mathcal B}^{(0)}_{\rm bmm} \!\!\! & \approx &  \!\!\!
b_1 {\mathcal B}^{(0)}_{\rm mmm} +
\frac{b_2}{3}
\left[{\mathcal P}^{(0)}_{\rm mm}(k_1){\mathcal P}^{(0)}_{\rm mm}(k_2)+2\,\cyc\, \right] 
\label{eq:bmm0} \ ;\\
{\mathcal B}^{(0)}_{\rm bbm} \!\!\!  & \approx & \!\!\!
b_1^2 {\mathcal B}^{(0)}_{\rm mmm} +
\frac{b_1b_2 }{3}
\left[{\mathcal P}^{(0)}_{\rm mm}(k_1){\mathcal P}^{(0)}_{\rm mm}(k_2)+2\,\cyc\, \right] 
\label{eq:bbm0}\ ; \\
{\mathcal B}^{(0)}_{\rm bbb} \!\!\!  & \approx & \!\!\! 
b_1^3 {\mathcal B}^{(0)}_{\rm mmm} + 
b_1^2b_2 
\left[{\mathcal P}^{(0)}_{\rm mm}(k_1){\mathcal P}^{(0)}_{\rm mm}(k_2)+2\,\cyc\, \right] \ ,
\label{eq:bbb0}
\ea
where in the above, we defined ${\mathcal P}_{\rm mm}(k)\equiv
W^2(k|R)P_{\rm mm}(k)$.

%%%%%%%%%%%%%%%%%%%%%%%%%%%%%%%%%%%%%%%%%%%%%%%%%%%%%%%

\begin{table}
\centering
\begin{tabular}{|c|c|c|c|c|} \hline
$R\,[\Mpc]$ & & $b_{1}$ $\pm$ $\sigma_{b_1}$ & $b_{2}$ $\pm$ 
$\sigma_{b_2}$ & $\chi^2$ \\ \hline 
20   & $B_{\rm bbb}$   & 1.62 $\pm$ 0.07 & -0.46 $\pm$ 0.12 & 0.01 \\
20   & $B_{\rm bbm}$   & 1.62 $\pm$ 0.10 & -0.49 $\pm$ 0.19 & 0.00 \\
20   & $B_{\rm bmm}$   & 1.63 $\pm$ 0.22 & -0.53 $\pm$ 0.51 & 0.00 \\
10   & $B_{\rm bbb}$   & 1.62 $\pm$ 0.04 & -0.42 $\pm$ 0.04 & 0.15 \\
10   & $B_{\rm bbm}$   & 1.62 $\pm$ 0.06 & -0.47 $\pm$ 0.07 & 0.02 \\
10   & $B_{\rm bmm}$   & 1.63 $\pm$ 0.13 & -0.53 $\pm$ 0.20 & 0.00 \\
6.7  & $B_{\rm bbb}$   & 1.59 $\pm$ 0.04 & -0.35 $\pm$ 0.02 & 0.70 \\
6.7  & $B_{\rm bbm}$   & 1.60 $\pm$ 0.05 & -0.42 $\pm$ 0.04 & 0.14 \\
6.7  & $B_{\rm bmm}$   & 1.63 $\pm$ 0.12 & -0.53 $\pm$ 0.12 & 0.00 \\
\hline
\end{tabular}
\caption{\small{Constraints on $b_1$ and $b_2$ obtained using the {\bf
      Exact Trispectrum} model described in the text.  The actual
    input bias parameters were $b_1=1.63$ and $b_2=-0.53$. Column 1:
    the smoothing scale of the biased density field; Column 2:
    measured quantity; Column 3 and 4 best-fit values for $b_1$ and
    $b_2$ along with 1$\sigma$ errors; Column 5: the median $\chi^2$. }}
\label{tab:biasbispNG}
%%%%%%%%%%%%%%%%%%%%%%%%%%%%%%%%%%%%%%%%%%%%%%%%%%%%%%%
\vspace{0.3cm}
%%%%%%%%%%%%%%%%%%%%%%%%%%%%%%%%%%%%%%%%%%%%%%%%%%%%%%%
\centering
\begin{tabular}{|c|c|c|c|c|} \hline
$R\,[\Mpc]$ & & $b_{1}$ $\pm$ $\sigma_{b_1}$ 
& $b_{2}$ $\pm$ $\sigma_{b_2}$ & $\chi^2$ \\ \hline 
20   & $B_{\rm bbb}$   & 1.63 $\pm$ 0.11 & -0.67 $\pm$ 0.36 & 14.75 \\
20   & $B_{\rm bbm}$   & 1.58 $\pm$ 0.17 & -0.45 $\pm$ 0.65 & 15.71 \\
20   & $B_{\rm bmm}$   & 1.37 $\pm$ 0.33 &  0.56 $\pm$ 1.19 & 18.43 \\
10   & $B_{\rm bbb}$   & 1.49 $\pm$ 0.03 & -0.66 $\pm$ 0.08 & 13.32 \\
10   & $B_{\rm bbm}$   & 1.48 $\pm$ 0.04 & -0.69 $\pm$ 0.13 & 13.60  \\
10   & $B_{\rm bmm}$   & 1.46 $\pm$ 0.09 & -0.69 $\pm$ 0.29 & 14.13 \\
6.7  & $B_{\rm bbb}$   & 1.36 $\pm$ 0.02 & -0.74 $\pm$ 0.06 & 13.21 \\
6.7  & $B_{\rm bbm}$   & 1.36 $\pm$ 0.03 & -0.82 $\pm$ 0.09 & 12.87  \\
6.7  & $B_{\rm bmm}$   & 1.32 $\pm$ 0.07 & -0.86 $\pm$ 0.22 & 13.13 \\
\hline
\end{tabular}
\caption{\small{Same as Table \ref{tab:biasbispNG}, but this time the
    $\chi^2$ analysis is for the {\bf Tree-level} model described in
    the text.}}
\label{tab:biasbispPT}
\end{table}
%%%%%%%%%%%%%%%%%%%%%%%%%%%%%%%%%%%%%%%%%%%%%%%%%%%%%%%

\subsection{Results of the artificial bias test}

Following the algorithm described in \S\ref{ssec:biasbyhand}, for each
realization of our ensemble of simulations, we generate three
artificially biased density fields smoothed on scales: $R = \{20\,,
10\,, 6.67 \} \Mpc$. In all cases we apply the same nonlinear bias:
$b_1 = 1.63$ and $b_2 = -0.53$. Whilst these values are somewhat
arbitrary, they were selected to coincide with the best-fit values to
the scatter plot of $\delta_{\rm h}(\vx|R)$ vs. $\delta(\vx|R)$,
smoothed at $R \sim 10 \,\Mpc$, that we recorded in
\S\ref{ssec:DensityFields}.  

For each filtering scale, we then measure the four bispectra $B_{\rm
  mmm}$, $B_{\rm bmm}$, $B_{\rm bbm}$ and $B_{\rm bbb}$ for triangle
configurations with $k_1 = 0.04\kMpc$, $k_2/k_1 = 2$, over 20 angular
bins. From these we use the method described above, to recover the
higher-order terms: ${\mathcal P}_{\rm 4,m}$, ${\mathcal P}_{\rm 5,m}$
and ${\mathcal P}_{\rm 6,m}$.

We now define three modelling cases of interest:
\begin{itemize}
\item Case 1: {\bf All Order}: Eqns~(\ref{eq:bmm}),
  (\ref{eq:bbm}) and (\ref{eq:bbb}) are used to interpret the data.
\item Case 2: {\bf Exact Trispectrum}: Eqns~(\ref{eq:bmm}),
  (\ref{eq:bbm}) and (\ref{eq:bbb}) are exact up to ${\mathcal P}_{\rm
  4,m}$. All higher-order terms (${\mathcal P}_{\rm 5,m},\, {\mathcal
  P}_{\rm 6,m}$) are dropped from the modelling.
\item Case 3: {\bf Tree-level}: lowest order expansions given by
  Eqns~(\ref{eq:bmm0}), (\ref{eq:bbm0}) and (\ref{eq:bbb0}) are used
  to interpret the data.
\end{itemize} 
For each of the models described above, we then apply the same
$\chi^2$--fitting analysis, as described in \S\ref{ssec:biasest} to
determine the best-fit $b_1$ and $b_2$ parameters.

We begin by first examining the {\bf All Order} expansion model. We
confirm that for this case, the true bias parameters $b_1 = 1.63$ and
$b_2 = -0.53$ are recovered exactly, albeit with some uncertainty, however, with a
$\chi^2=0$, and for all the smoothing lengths considered. This null
test is important, because it gives us confidence that any departures
of the fits from the true bias values, can be attributed solely to a
break down of the theoretical modelling.

Next we focus on the {\bf Exact Trispectrum} model where ${\mathcal
  B}_{\rm mmm}$ and ${\mathcal P}_{\rm 4,m}$ are measured from the
simulations. In Table \ref{tab:biasbispNG} we report the best-fitting
bias parameters with the 1$\sigma$ errors expressed in quadrature for
the auto- and cross-bispectrum and for the four smoothing scales
examined.  For the case ${\mathcal B}_{\rm bmm}$, a quick inspection
of \Eqn{eq:bmm} tells us that the modelling should be exact, and
indeed we see that the bias parameters are correctly recovered. However, for the cases ${\mathcal B}_{\rm bbm}$ and
${\mathcal B}_{\rm bbb}$ we see that the absence of the higher-order
terms (${\mathcal P}_{\rm 5,m},\, {\mathcal P}_{\rm 6,m}$), induce
biases in the parameters. For $b_1$ the deviation from the true value
is relatively small, with the value of the parameter only slightly
decreasing in size. For $b_2$ the deviations are larger, and this
parameter becomes more positive.  We also note that the deviations
from the true values appear to increase as the smoothing scale is
decreased.

Finally, we focus on the {\bf Tree-level} model. Table
\ref{tab:biasbispPT} presents the best-fit results for $b_1$ and
$b_2$.  We see that in nearly all cases, there are systematic biases
in the recovery of the nonlinear bias parameters for all of the
measured bispectra. In particular, for the case of ${\mathcal
  B}_{\bmm}$, the results are most deviant and poorly constrained.
Whereas for ${\mathcal B}_{\rm bbb}$, only when the data has been
smoothed on scales $R=20\Mpc$ are the recovered parameters close to
the true values.

The comparison of the results from this analysis leads us to conclude
that the recovered bias parameters are very sensitive to the inclusion
of beyond leading order corrections in the modelling.  Furthermore,
accurate nonlinear modelling of, at the very least, the matter
bispectrum and trispectrum will be essential, if we are to safely
recover the nonlinear bias parameters from this approach.

%%%%%%%%%%%%%%%%%%%%%%%%%%%%%%%%%%%%%%%%%%%%%%%%%%%%%%%

\section{Discussion}\label{sec:discussion}

\begin{figure}
  \centering{
    \includegraphics[width = 8.7cm, height=9cm]{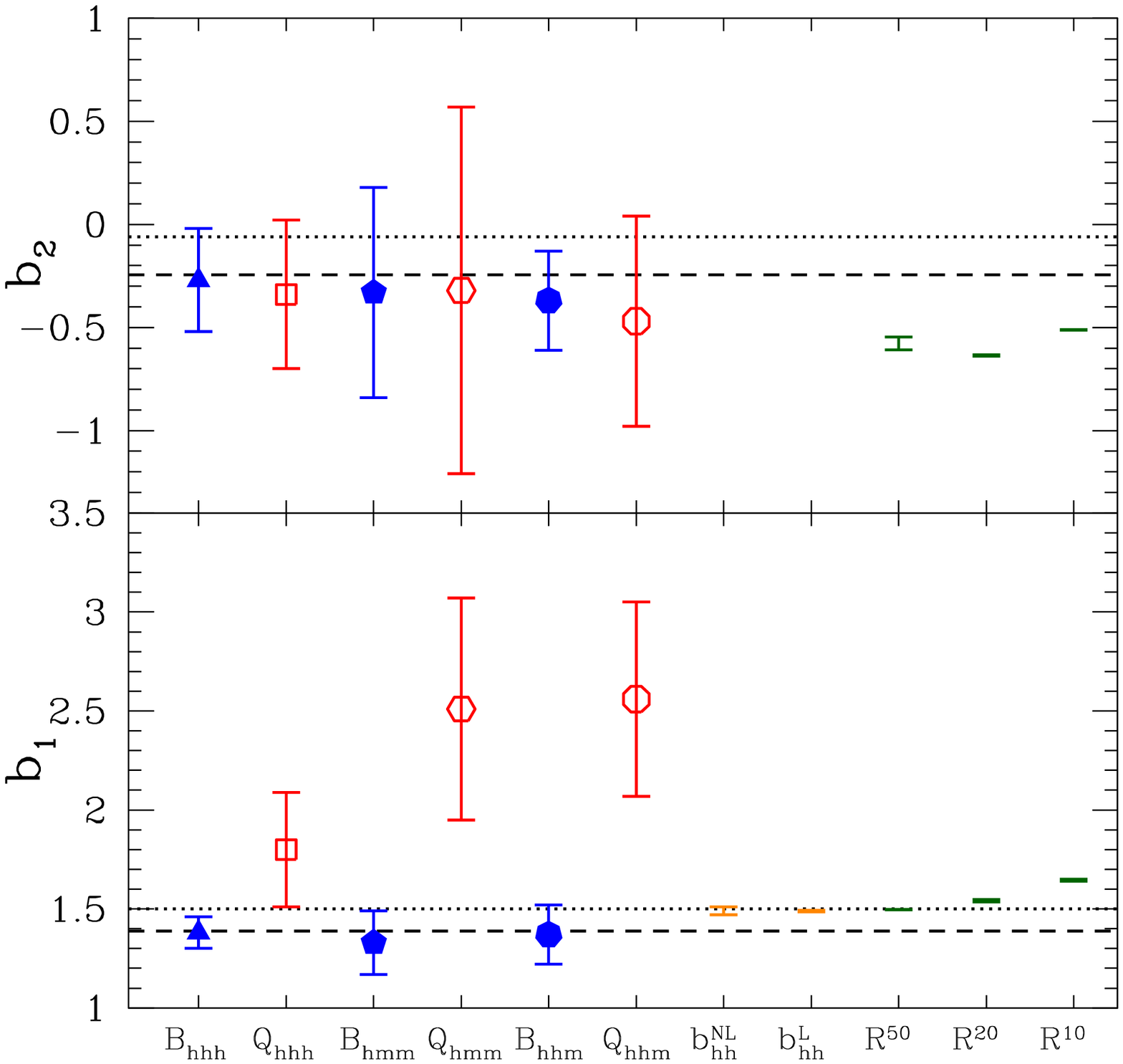}}  
  \caption{\small{Plot summary of bias measurements on $b_1$ and $b_2$
      for the second-order local bias model from different
      estimators:, shot-noise corrected $B$, $Q$, $P$, and finally
      smoothed, $\delta^R$, in comparison with analytical predictions 
      applying the peak-background split ansatz with the \citet{ShethTormen1999} 
      mass function  denoted by the dotted line, as well as the \citet{Warrenetal2006} 
      and \citet{Pillepichetal2010} mass-functions, which are both represented by the single dashed-line.}}
  \label{fig:bests}
\end{figure}

We have evaluated the local halo bias model at second-order using
three different probes: smoothed density fields; power spectra; and
bispectra and reduced bispectra.  A summary of our results for the
best-fitting bias parameters determined from shot-noise corrected
spectra is shown in Figure~\ref{fig:bests}.

%%%%%%%%%%%%%%%%%%%%%%%%%%%%%%%%%%%%%%%%%%%%%%%%%%%%%%%

%%%%%%%%%%%%%%%%%%%%%%%%%%%%%%%%%%%%%%%%%%%%%%%%%%%%%%%
In the figure we also compare our estimates for $b_1$ and $b_2$ with
the analytical predictions for the halo bias parameters obtained from
the peak background split (PBS) ansatz
\citep{Bardeenetal1986,MoWhite1996}.  The average theory bias
parameters are obtained through computing the expression:
\be 
\overline{b}_i =\frac{1}{\overline{n}}
\int_{M_{\rm min}}^{\infty} dM\, n(M)\, b_i(M)\ \ ; \ 
\overline{n} \equiv \int_{M_{\rm min}}^{\infty} dM\, n(M) \ ,\ee

where $n(M)$ is the halo mass function, and $M_{\rm min}$ is set equal to the value
of the minimum halo mass identified in the simulations (see \S\ref{sec:sims}) . We evaluate the above integral using three different fits to N-body simulations by
\citet{ShethTormen1999}, \citet{Warrenetal2006} and \citet{Pillepichetal2010}. The 
corresponding expressions for the bias parameters as a function of halo mass are presented in \citet{Scoccimarroetal2001}
and \citet{Maneraetal2010}.  

Considering, the results for $b_1$ (bottom panel), we see that when
the reduced bispectra, $Q_{\rm hhh}$, $Q_{\rm hhm}$ and $Q_{\rm hmm}$
are used, the recovered parameters are poorly constrained and appear
incompatible with respect to the other estimates and are only weakly
consistent with one another. On the other hand, the estimates from the
bispectra $B_{\rm hhh}$, $B_{\rm hhm}$ and $B_{\rm hmm}$ are in much
better agreement with each other. They are also in close agreement
with the predictions from \citet{Warrenetal2006} and \citet{Pillepichetal2010}, which
both provided an estimate of $b_1=1.39$. However, they slightly undershoot the values
from the effective bias estimates, $b^{\NL}_{\hh}$ and $b^{\rm
  L}_{\hh}$, likewise the smoothed density fields, and the
Sheth-Tormen prediction. The analytical predictions from the
\citet{ShethTormen1999} mass-function yielded $b_1=1.50$ in good
agreement with the power spectrum and density field results smoothed
on a scale $R\sim50 \Mpc$. The recovered values of $b_1$ from the
effective bias in the power spectra and the smoothed density fields
collectively are in broad agreement, but the latter increase with
decreasing smoothing scale.

In the top panel of the figure, we see that the constraints on $b_2$
from the different estimators used in the simulations are reasonably
consistent with one another, albeit with significant error bars.
These estimates also agree well with the PBS prediction from the
\citet{Warrenetal2006} and \citet{Pillepichetal2010} mass functions, which give an average
$b_2=-0.24$.  However, the prediction from the \citet{ShethTormen1999}
mass function gives $b_2=-0.06$, and this appears to be in worse
agreement with the data.

There are a number of possible explanations for the deviations in the
recovered bias parameters. Firstly, the relation between matter and
halo fluctuations may not be local. Indeed, we know it is not
deterministic owing to the scatter in relation $\delta_{\rm h}$
vs. $\delta$. Perhaps, this is a consequence of non-locality.  In this
case we need a more advanced theoretical approach to understand the
halo clustering. One possibility may be that the bias is local in
Lagrangian space \citep{Catelanetal2000, Matsubara2011}.

Secondly, our simple biasing-by-hand test has enabled us to discern
that, for the current set of tests that we have performed, the most
likely explanation at this point is that the tree-level expansions for
$B_{\rm hhh}$ and $Q_{\rm hhh}$ are not sufficiently accurate
enough. Higher-order nonlinear corrections in the modelling must be
included, and if possible all order expansions for $B_{\rm mmm}$ and
$P_{4,\rm m}$ would be invaluable.

Thirdly, as we have argued, the local bias model only makes sense in
the context of smoothing. The bias parameters one recovers from
fitting, depend sensitively on the smoothing scale $R$. For the
biasing-by-hand tests,  the exact smoothing scale was known beforehand.
However, in real data we do not know this {\em a priori}. In all of
the cases, when recovering bias parameters from the bispectra, we
have assumed that we are on sufficiently large scales such that
$W(k_iR)\rightarrow 1$. However, in general $R$ should be a free
parameter and as such, marginalized over in the analysis.

\citet{ManeraGaztanaga2011} performed a similar study of nonlinear
halo bias with the three-point correlation function in configuration
space. In contrast to our analysis, they measured the bias parameters
for different halo mass bins.  They found inconsistencies between the
predictions of the different estimators considered.  They evaluated
the scatter plots of $\delta_h$ vs. $\delta$ as function of smoothing
scale, and found that stability in the local bias parameters
($b_1,b_2)$ occurred for smoothing scales, $R > 30 - 60\,\Mpc$, albeit
with larger errors. They also found that the bias predictions derived
from $\delta_h$ vs. $\delta$ for $R = 60\,\Mpc$ were in good
agreement, to within the errors, with the linear bias measured from
evaluating the two-point correlation function on large and
intermediate scales.  As in the case of our findings, they found the
linear bias measured from evaluating the three-point correlation
function, expressed in terms of the $Q$-amplitudes, was not consistent
with that of the two-point correlation function for the lower mass
bins: $M < 10^{13}\,\Msol$. They were unable to formulate solid
conclusions for larger mass ranges.

\citet{GuoJing2009} explored the differences between estimates of
bias from $Q$ and $P$. They also found that $b_1$ based on analysis of
the mock galaxy catalogues was larger for galaxy reduced bispectra and power
spectra $Q_{\rm g}$ than for $P_{\rm g}$. While \citet{GuoJing2009} also noted that this might be due to
the failure of SPT at tree-level, they also reported that agreement
could be found between estimators if $Q_{\mmm}$ measured directly from
the simulations was used in place of the tree-level expression.
However, when we performed the same test with our data we found no
dramatic reconciliation of the two bias estimates.  The investigation
performed by \citet{GuoJing2009} was carried out using only 4 large 
volume and 3 smaller volume runs. As a result of having too few realizations,
they assumed the Gaussian approximation for the covariance matrix in order
to perform their study at large-scales.  

%%%%%%%%%%%%%%%%%%%%%%%%%%%%%%%%%%%%%%%%%%%%%%%%%%%%%%%

\section{Conclusions}\label{sec:conclusions}

In this paper we have used a sample of 40 large volume $N$-body
simulations, with total volume $V\sim 135 \,{\rm Gpc}^{3}h^{-3}$, to
test the local model of halo biasing, and the extent to which
nonlinearities impact the modelling. We used three different methods
for exploring the bias: smoothed density fields; power spectra; and
bispectra and reduced bispectra. We focused mainly on the results from
the bispectra. All of the reported results were scaled to a single
realization of our simulations, and so are directly relevant for galaxy
surveys with a total volume $V\sim4\Gpccube$.

In \S\ref{sec:SPT} we reviewed the basic results of perturbation
theory and how they connect to density statistics. We then reviewed
the local model of halo biasing, drawing special attention to the
r\^ole that smoothing plays in the theory. The important result being
that even at tree-level, the smoothing explicitly enters the
theory. The expressions for $B_{\rm hhh}$ and $Q_{\rm hhh}$, which are
typically used in all past and current analysis, make the assumption
that smoothing is unimportant. In subsequent sections we argued that
this assumption is not safe.

In \S\ref{sec:sims} we described our suite of $N$-body simulations,
and the halo catalogues used in this study.

In \S\ref{sec:HaloBiasMethods} we made measurements of the relation
between $\delta_{\rm h}(\bx|R)$ and $\delta(\bx|R)$ smoothed on the
set of scales \mbox{$R=\{50,20, 10\}\Mpc$}. To this data we fitted the
local model of halo biasing up to second order, including $b_0$, $b_1$
and $b_2$. We found that the fits were reasonably good, however the
best-fit parameters showed a running with the filter scale $R$. We then
demonstrated, theoretically, why the nonlinear bias parameters from this
approach, could not be made independent of smoothing scale. We then
turned to Fourier space statistics, and used the halo auto- and
cross-power spectra to obtain an effective large-scale bias. We found
that the effective bias estimators $b^{\rm NL}_{\rm hh}$ and $b^{\rm
  NL}_{\rm hm}$ were reasonably scale independent for
$k<0.08\kMpc$. However, on scales smaller than this, $b^{\rm NL}_{\rm
  hh}$ decreased with increasing wavenumber, whereas $b^{\rm NL}_{\rm
  hm}$ remained surprisingly flat.

In \S\ref{sec:BispHaloEst} we estimated the matter auto- and halo
auto-bispectra and reduced bispectra from our simulations.  We
measured these statistics for the triangle configurations
$k_1=\{0.03,\,0.04,\,0.05,\,0.06\}\kMpc$ and with $k_2/k_1=2$ and
$\theta_{12}\in[0,\pi]$. These triangles all lay in the weakly
nonlinear regime $k=0.03-0.18$ $\kMpc$. We modeled these estimates
using tree-level perturbation theory expressions for the matter
bispectrum and nonlinear bias at second order, and assumed smoothing
to be unimportant. Our method for estimation of the bias parameters
followed a standard minimum $\chi^2$ approach. We estimated the covariance matrix
for the full ensemble applying principal component analysis to minimize the intrinsic
noise.  We also performed a jackknife subsampling routine to propagate the error of the estimated
covariance matrix onto the errors of the bias parameters.

We tested how well the measurements of the matter bispectra $B_{\rm
  mmm}$ and reduced bispectra $Q_{\rm mmm}$ could be described by such
modelling. The results obtained for the bias parameters $b_1$ and
$b_2$ showed that the tree-level expressions were a good description
of the data for configurations, $k_1=\{0.03,\,0.04\}\kMpc$, for which
$b_1=1$ and $b_2=0$. However, for smaller scale triangles,
$k_1=\{0.05,\,0.06\}\kMpc$, significant deviations were apparent, and
these were manifest as $b_1\ne1$ and $b_2\ne0$ at high significance.

We then applied the $\chi^2$ test to the halo bispectra $B_{\rm hhh}$
and reduced bispectra $Q_{\rm hhh}$. We found, for the shot-noise
corrected $B_{\rm hhh}$, that the estimated values for $b_1\sim1.40$
and $b_2\sim-0.25$ were reasonably consistent with one
another. However, the fits became progressively poorer as smaller
scales were added, yet the reduced--$\chi^2$ remained $\lesssim2$ for
$k_1=\{0.06\}\kMpc$. For the shot-noise corrected $Q_{\rm hhh}$, we
found that the values of $b_1$ were significantly larger
$b_1\sim1.85$, with large errors, and the values evolved with triangle
configuration scale. However, $b_2\sim-0.3$, appeared to be more
stable, although again with large errors. For triangle configurations with
$k_1\ge 0.04\kMpc$, the fits from $B_{\rm hhh}$ and $Q_{\rm hhh}$ were
inconsistent with each other at the $\sim3\sigma$ level. For both
$B_{\rm hhh}$ and $Q_{\rm hhh}$ shot-noise corrections significantly
influenced the recovered bias parameters.

In \S\ref{sec:CrossSpec} we explored the halo and matter
cross-bispectra, $B_{\rm hhm}$ and $B_{\rm hmm}$, and reduced
bispectra $Q_{\rm hhm}$ and $Q_{\rm hmm}$. We calculated the
tree-level expressions for these quantities symmetrized in all of
their arguments. We then developed estimators for them. We showed that
for $B_{\hmm}$ and $Q_{\hmm}$, provided the matter distribution was densely
sampled, the shot-noise corrections were small.

We applied the $\chi^2$ analysis from \S\ref{sec:BispHaloEst} to these
statistics and recovered the best-fit values for $b_1$ and $b_2$. We
found that for $B_{\rm hhm}$ the shot-noise corrected data were all
reasonably consistent with one another, giving $b_1\sim1.39$ and
$b_2\sim-0.3$. For $B_{\rm hmm}$ we found a similar pattern, except
that for $k_1\ge0.05\kMpc$ where we found $b_2\sim0.0$, but with large
errors. The results for $Q_{\rm hhm}$ and $Q_{\rm hmm}$ appeared to
vary significantly.

Finally in \S\ref{sec:biasbyhand} we explored to what extent the
break-down could be attributed to the absence of terms that were
beyond tree-level in the modelling. In order to do this, we developed
a novel approach, whereby we constructed smoothed biased density
fields from the smoothed matter density field, using the local model
at quadratic order. We showed that if we set $b_1$ and $b_2$ to some
fiducial values, and then measured the smoothed matter and halo
bispectra and their cross-bispectra, then the higher order matter
correlators $P_{4,m}$, $P_{5,m}$ and $P_{6,m}$ could be recovered
exactly. Thus we were able to construct three models: an all order
model; a model that used the exact matter bispectrum and trispectrum;
and a tree-level model.

We applied the $\chi^2$ analysis using these three models and for
bispectra with $k_1=0.04\kMpc$. As expected, the exact model recovered
the correct bias parameters. The model with the exact
$B_{\rm mmm}$ and $P_{4,m}$, was in fact also exact for $B_{\rm
  bmm}$. For $B_{\rm bbm}$ and $B_{\rm bbb}$ the recovered parameters
were close to the true values, but showed evolution with smoothing
scale. Finally, for the tree-level model we showed that there was a
significant evolution in the estimated bias parameters with smoothing
scale and with the type of statistic used. 

We conclude that estimates of nonlinear bias from the bispectrum that
do not attempt to account for higher-order corrections, will most
likely provide biased estimates for the bias parameters $b_1$ and
$b_2$.  Robust modelling of nonlinear bias from bispectra, will, at
the very least, require almost exact models for the matter bispectrum
and trispectrum.

Real space estimates of bias appear to be inconsistent with Fourier
space based ones. We believe that this owes primarily to the mixing of
large- and small-scale wavemodes in real space. We therefore recommend
that perturbative methods should strictly be applied in Fourier
space. We also recommend that measurements focus on the bispectrum and
the associated cross statistics, rather than the reduced bispectra,
since this appears very sensitive to nonlinearities in the modelling
and also shot-noise corrections.

Finally, we emphasize the importance of smoothing in the local
model. Owing to the fact that the smoothing scale associated with the
halo/galaxy distribution in question is not known {\em a priori}, it
must be treated as a nuisance parameter and so marginalized over.

An alternative strategy for recovering information from higher order
statistics, which may be of interest for future consideration, is the
use of `Gaussianizing transformations' or `density clipping'
\citep{Neyrincketal2009,Seoetal2011,Simpsonetal2011}. However, the
theoretical connection between what is measured and what is
interpreted from such approaches still remains to be fully calculated.

%%%%%%%%%%%%%%%%%%%%%%%%%%%%%%%%%%%%%%%%%%%%%%%%%%%%%%%
%%%%%%%%%%%%%%%%%%%%%%%%%%%%%%%%%%%%%%%%%%%%%%%%%%%%%%%

\section*{Acknowledgements}
We thank the anonymous referee for helpful suggestions.  We also thank
Tobias Baldauf, Martin Crocce, Roman Scoccimarro, Emiliano Sefussati,
Ravi Sheth and Masahiro Takada for useful discussions. We thank
V. Springel for making public {\tt GADGET-2} and for providing his
{\tt B-FoF} halo finder, and R.~Scoccimarro for making public his {\tt
  2LPT} code. JEP and CP were supported by funding provided through
the SFB-Transregio 33 “The Dark Universe” by the Deutsche
Forschungsgemeinschaft.  RES acknowledges support from a Marie Curie
Reintegration Grant, the Alexander von Humboldt Foundation and partial
support from the Swiss National Foundation under contract
200021-116696/1.

%%%%%%%%%%%%%%%%%%%%%%%%%%%%%%%%%%%%%%%%%%%%%%%%%%%%%%%
%%%%%%%%%%%%%%%%%%%%%%%%%%%%%%%%%%%%%%%%%%%%%%%%%%%%%%%

% BIBILIOGRAPHY

\bibliographystyle{mn2e}
\bibliography{refs}

%%%%%%%%%%%%%%%%%%%%%%%%%%%%%%%%%%%%%%%%%%%%%%%%%%%%%%%
%%%%%%%%%%%%%%%%%%%%%%%%%%%%%%%%%%%%%%%%%%%%%%%%%%%%%%%

\appendix

\onecolumn

%%%%%%%%%%%%%%%%%%%%%%%%%%%%%%%%%%%%%%%%%%%%%%%%%%%%%%%
%%%%%%%%%%%%%%%%%%%%%%%%%%%%%%%%%%%%%%%%%%%%%%%%%%%%%%%

\section{Bispectrum Estimation: Algorithm}\label{app:bispec}

Briefly, the algorithm that we employ is as follows: firstly the dark
matter density field is computed by assigning the dark matter
particles to a cubical grid using the `Cloud in Cell' (CIC)
technique \citep{HockneyEastwood1988}.  Next the fast Fourier
transform (FFT) of the gridded density field is computed.  Each
Fourier mode is then corrected for convolution with the Fourier
mesh. We do this by dividing out from each mode, the Fourier transform
of the window assignment function of the CIC scheme
\citep{HockneyEastwood1988,Jing2005}:
\be
 \dek(\vk) = \frac{\dek_{\rm g}(\vk)}{W_{\rm CIC}(\vk)} \ ; \ \ 
 W_{\rm CIC}(\vk) \equiv \prod_{i = 1,3}\left[\frac{\sin(\pi k_i/2k_{Ny})}
{\pi k_i/2k_{Ny}}\right]^2 \ ,
\ee
where subscript $\rm g$ denotes gridded quantities, $k_{\rm Ny} = \pi
N_{\rm g}/L$ is the Nyquist frequency of the mesh and $N_{\textrm{g}}$
is the number of Fourier grid cells.

The estimator for the bispectrum can be written
\citep{Scoccimarroetal1998}:
\be
\widehat{\overline{B}}(k_1, k_2, \theta) =  \frac{\Vu^2}{V_B(k_1,k_2,\theta)} 
\int  
\frac{\dq_1}{(2 \pi)^3} 
\frac{\dq_2}{(2 \pi)^3} 
\frac{\dq_3}{(2 \pi)^3} 
(2\pi)^3\dD(\vq_{123}) \delta(\vq_1)\delta(\vq_2)\delta(\vq_3) \ ,
\ee
where $\Vu$ is the sample volume (in our case the simulation volume), the normalization factor, $V_B$, can be written as \citep{Sefusattietal2006,Joachimietal2009}
\be
 V_B(k_1,k_2,\mu)  \equiv  \int 
\frac{\dq_1}{(2 \pi)^3} 
\frac{\dq_2}{(2 \pi)^3} 
\frac{\dq_3}{(2 \pi)^3} 
(2\pi)^3\dD(\vq_{123})
\approx \frac{8\pi^2 k_1 k_2 k_3}{(2\pi)^6} (\Delta k)^3,
\ee

\noindent and we write in shorthand $\dD(\vq_{1\dots n})\equiv\dD(\vq_1+\dots+\vq_n)$. 
A practical implementation of the above estimator may be achieved
through \citep{Smithetal2008b}:
\be
\widehat{\overline{B}}^{\rm d}(k_1,k_2,\theta_{12}) = 
\frac{\Vu^2}{N_{\rm tri}(k_1,k_2,\theta_{12})}
\sum_{(\vn_1, \vn_2)}^{N_{\rm tri}(k_1,k_2,\theta_{12})}
\mathcal{R}e[\delta(\vk_{\vn_1})
  \delta(\vk_{\vn_2})\delta(\vk_{-\vn_1 - \vn_2})],
\label{eq:bispest}
\ee
where superscript ${\rm d}$ denotes discretized quantities; $\vn_i$ denotes
an integer vector from the origin of the $k$-space to each mesh point;
$(\vn_1,\vn_2)$ represents a pair of integer vectors, which lie in
thin shells centred on $k_1$ and $k_2$ and whose angular separation
lies in a narrow angular bin centred on $\theta_{12}$, and for which
$\vk_3= -\vk_1 - \vk_2$. The upper limit of the sum
$N_{\textrm{tri}}(k_1,k_2,\theta_{12})$ represents the total number of
triangles that have such a configuration.

The estimator for the bin-averaged reduced bispectrum,
$\widehat{\overline{Q}}$, is written as:
\be
\widehat{\overline{Q}}^{\rm d} = \widehat{\overline{B}}^{\rm d}/\widehat{\overline{Q}}^{\rm denom,d},
\label{eq:qqest}
\ee
where $\widehat{\overline{Q}}^{\rm denom,d}$ is the estimator for the
bin-averaged cyclical terms of the power spectrum generated from first
computing the bin-averaged power-spectra,
$\widehat{\overline{P}}_i^{\rm d}$. Note that we estimate the power
spectra that enter into this product in a slightly different way than
normal: we use only those modes that go into estimating the particular
$B$ triangle configuration to estimate $\widehat{\overline{Q}}^{\rm
  denom,d}(k_1,k_2,\theta_{12})$. Hence,
\be
\widehat{\overline{P}}^{\rm d}_{i} = 
\frac{\Vu}{N_{\rm tri}(k_1,k_2,\theta_{12})}
\hspace{-0.3cm}\sum_{(\vn_{1},\vn_{2})}^{N_{\rm tri}(k_1,k_2,\theta_{12})}\!\!|\dek(\vk_{\vn_i})|^2  \ ,
\ee
where $i \in \{1,2,3\}$ and where $\widehat{\overline{P}}^{\rm d}_{3}$ is
dependent on the angular bin, since, with $|\vn_1|$ and $|\vn_2|$
fixed, we still have
\mbox{$\cos\theta_{12}={\vn}_1\cdot{\vn}_2/|\vn_1||\vn_2|$} and the
closure criterion implies $\vn_{3}=-\vn_1-\vn_2$ varies as function of
$\theta_{12}$. Therefore,
\be
\widehat{\overline{Q}}^{\rm denom,d} = \widehat{\overline{P}}^{\rm d}_{1}\widehat{\overline{P}}^{\rm d}_{2} + 
\widehat{\overline{P}}^{\rm d}_{2}\widehat{\overline{P}}^{\rm d}_{3} + 
\widehat{\overline{P}}^{\rm d}_{1}\widehat{\overline{P}}^{\rm d}_{3}.
\label{eq:pcycest}
\ee

The estimates of $B^{\rm d}$ and $Q^{\rm d}$ are then corrected for
discreteness, i.e. shot-noise.  For the estimators of interest, the
corrections are \citep{Peebles1980,Smithetal2008b}:
\ba
\widehat{\overline{P}}_{\rm shot} & \equiv & 1/\nbar \ ;\\
\widehat{\overline{B}}_{\rm shot} & \equiv & 
[\widehat{\overline{P}}_{1}^{\rm d} + \widehat{\overline{P}}_{2}^{\rm d} 
+ \widehat{\overline{P}}_{3}^{\rm d}]/{\nbar}  - 2/
{\nbar^2} \ ; \\
\widehat{\overline{Q}}_{\rm shot}^{\rm denom} & \equiv & {2}
[\widehat{\overline{P}}^{\rm d}_{1} + \widehat{\overline{P}}^{\rm d}_{2}
+ \widehat{\overline{P}}_{3}^{\rm d}] /{\overline{n}}
- {3}/{\nbar^2} \ .
\label{eq:shotcorrauto}
\ea
Shot-noise corrected estimates of the statistics are obtained:
\be 
\chi = \chi^{\rm d}-\chi_{\rm shot} \ \label{eq:shotcorr} ,
\ee
where
$\chi\in\{\widehat{\overline{P}},\widehat{\overline{B}},\widehat{\overline{Q}}^{\rm
  denom}\}$ and where $\widehat{\overline{Q}} =
\widehat{\overline{B}}/\widehat{\overline{Q}}^{\rm denom}$. Note that the above
recipe corrects some typos that are present in \citet{Smithetal2008b}.

%%%%%%%%%%%%%%%%%%%%%%%%%%%%%%%%%%%%%%%%%%%%%%%%%%%%%%%
%%%%%%%%%%%%%%%%%%%%%%%%%%%%%%%%%%%%%%%%%%%%%%%%%%%%%%%

\section{Halo cross-bispectra in the local model}\label{app:halobispectra}

As was shown in \Eqn{eq:biasLoc}, at quadratic order, the local model
of nonlinear biasing can be written as:
\be
\delta_{\rm h}(\bk|R) =  b_1(M)\delta(\bk|R) + \frac{b_2(M)}{2}
\int \frac{\dq_1}{(2\pi)^3} \frac{\dq_1}{(2\pi)^3} 
\delta(\bq_1|R)\delta(\bq_2|R) (2\pi)^3 \delta^{D}(\bk_1-\bq_1-\bq_2)\ .
\label{eq:halobias2}\ee
where $\delta(\bq_i|R)\equiv \delta(\bq_i)W(q_iR)$, is the filtered
density.  Using this model we may now proceed to calculate the halo
auto- and halo-mass cross-bispectra.

%%%%%%%%%%%%%%%%%%%%%%%%%%%%%%%%%%%%%%%%%%%%%%%%%%%%%%%

\subsection{Halo-mass-mass bispectrum in the local model}

Let us start with the simplest three-point cross-statistic, the
halo-matter-matter bispectrum, this can be written:
\ba 
\left<\delta_{\rm h}(\bk_1|M,R)\delta(\bk_2|R)\delta(\bk_3|R)\right> 
&  = & 
b_1(M)\left<\delta(\bk_1|R)\delta(\bk_2|R)\delta(\bk_3|R)\right> \nn \\
& & +\frac{b_2(M)}{2}\int \frac{\dq_1}{(2\pi)^3} \frac{\dq_1}{(2\pi)^3}
(2\pi)^3 \delta^{D}(\bk_1-\bq_1-\bq_2)
\left<\delta(\bq_1|R)\delta(\bq_2|R)\delta(\bk_1|R)\delta(\bk_2|R)\right> \ .
\ea
Let us define the smoothed $n$-point spectra as:
\ba \left<\delta(\bk_1|R)\dots\delta(\bk_n|R)\right> & \equiv &
(2\pi)^3\delta(\bk_1+\dots+\bk_{n})\widetilde{\mathcal
  P}_{n}(\bk_1,\dots,\bk_n|R)\ ,\nn \\ 
& = & (2\pi)^3\delta(\bk_1+\dots+\bk_{m})W(k_1R)\dots W(k_nR) 
P_n(\bk_1,\dots,\bk_n) \ea
where $\widetilde{\mathcal P}_{2}\equiv \widetilde{\mathcal
  P}=W^2(kR)P$, $\widetilde{\mathcal P}_{3}\equiv \widetilde{\mathcal
  B} = W(k_1R)W(k_2R)W(k_3R)B$, and where $\widetilde{\mathcal
  P}_{4}\equiv \widetilde{\mathcal T} =
W(k_1R)W(k_2R)W(k_3R)W(k_4R)T$. We may now integrate over $\bq_2$ to
obtain
\be 
\left<\delta_{\rm h}(\bk_1|M,R)\delta(\bk_2|R)\delta(\bk_3|R)\right> 
 =   (2\pi)^3 \delta^{D}(\bk_1+\bk_2+\bk_3) 
\left[\frac{}{} b_1(M)\widetilde{\mathcal B}(\bk_1,\bk_2,\bk_3) 
+\frac{b_2(M)}{2}\int \frac{\dq_1}{(2\pi)^3} 
\widetilde{\mathcal T}(\bq_1,\bk_1-\bq_1,\bk_2,\bk_3) \right] \ .
\ee
On dividing the above expression by $W(k_1R)W(k_2R)W(k_3R)$, then we
find the halo-mass-mass bispectrum can be written as
\be
B_{\rm hmm}(\bk_1,\bk_2,\bk_3) = b_1(M)B_{\rm mmm}(\bk_1,\bk_2,\bk_3) +\frac{b_2(M)}{2}
\int \frac{\dq_1}{(2\pi)^3} \widetilde{W}_{\bq_1,\bk_1-\bq_1} 
T_{\rm mmmm}(\bq_1,\bk_1-\bq_1,\bk_2,\bk_3)\ .
\ee
We may symmetrize the above result by constructing the sum $[B_{\rm
    hmm}+B_{\rm mhm}+B_{\rm mmh}]/3$, and this gives us:
\be B_{\rm hmm}(\bk_1,\bk_2,\bk_3) = b_1(M)B_{\rm
  mmm}(\bk_1,\bk_2,\bk_3) + \frac{b_2(M)}{6}\int
\frac{\dq_1}{(2\pi)^3} \left[ \widetilde{W}_{\bq_1,\bk_1-\bq_1}
  T_{\rm mmmm}(\bq_1,\bk_1-\bq_1,\bk_2,\bk_3) +2\,\cyc \right] \ee
On expanding $B$ and $T$ to fourth order in $\delta$, the above
expression can be approximated as,
\be B^{(0)}_{\rm hmm}(\bk_1,\bk_2,\bk_3) \approx
b_1(M)B^{(0)}_{\rm mmm}(\bk_1,\bk_2,\bk_3) +
\frac{b_2(M)}{3}\left[\widetilde{W}_{\bk_2,\bk_3}
  P^{(0)}(k_2)P_{\rm mm}^{(0)}(k_3)+2\,\cyc \right] \ .\ee
Finally, in the large-scale limit $k_i\rightarrow 0$, or for
arbitrarily small smoothing scales, $k_iR\ll1$, the above expression
becomes,
\be
B^{(0)}_{\rm hmm}(\bk_1,\bk_2,\bk_3) 
\approx  b_1(M)B^{(0)}_{\rm mmm}(\bk_1,\bk_2,\bk_3) 
+ \frac{b_2(M)}{3}\left[ P_{\rm mm}^{(0)}(k_2)P_{\rm mm}^{(0)}(k_3)+2\, \cyc\, \right] 
\ee

The reduced bispectrum $Q_{\rm hmm}$ is given by 
\be 
Q_{\rm hmm}(\bk_1,\bk_2,\bk_3) \equiv \frac{B_{\rm hmm}(\bk_1,\bk_2,\bk_3)}{PP_{\rm hmm}} \ ,
\ee
where 
\be 
PP_{\rm hmm} = 
\frac{2}{3}\left[P_{\rm hm}(k_1)P_{\rm mm}(k_2)+2\,\cyc\,\right]+
\frac{1}{3}\left[P_{\rm hm}(k_1)P_{\rm hm}(k_2)+2\,\cyc\,\right]\ .
\label{eq:PPhmmAp}\ee

In order to calculate the reduced halo-mass cross-bispectrum, then we
need to evaluate the halo-matter power spectrum. In the local model
and up to quadratic order in the bias we have,
\be 
P_{\rm hm}(k)=b_1(M) P_{\rm mm}(\bk_1) +
\frac{b_2(M)}{2} \int \frac{\dq_1}{(2\pi)^3} \widetilde{W}_{\bq_1,\bk_1-\bq_1} 
B_{\rm mmm}(\bq_1,\bk_1-\bq_1,-\bk_1) \ . \label{eq:Phm}
\ee
Using the above expression in \Eqn{eq:PPhmmAp} we find
\ba
PP_{\rm hmm}(\bk_1,\bk_2,\bk_3) & = &  
\left\{\frac{2}{3}\left[b_1(M) P_{\rm mm}(\bk_1) +
\frac{b_2(M)}{2} \int \frac{\dq_1}{(2\pi)^3}
\widetilde{W}_{\bq_1,\bk_1-\bq_1} B_{\rm mmm}(\bq_1,\bk_1-\bq_1,-\bk_1)\right]
P_{\rm mm}(k_2)+2\,\cyc\,\right\} \nn \\
& & +\frac{1}{3}\left\{
\left[b_1(M) P_{\rm mm}(\bk_1) + 
\frac{b_2(M)}{2} \int \frac{\dq_1}{(2\pi)^3}\widetilde{W}_{\bq_1,\bk_1-\bq_1} B_{\rm mmm}(\bq_1,\bk_1-\bq_1,-\bk_1)\right]\right.
\nn \\
& & \times \left.\left[b_1(M) P_{\rm mm}(\bk_2) + 
\frac{b_2}{2} \int \frac{\dq_1}{(2\pi)^3}\widetilde{W}_{\bq_2,\bk_2-\bq_2} B_{\rm mmm}(\bq_2,\bk_2-\bq_2,-\bk_2)\right]
+2\,\cyc\,\right\}
\ea
If we expand $PP_{\rm hmm} $ to fourth order in the density then the
above expression simplifies to:
\be 
PP_{\rm hmm}^{(0)}(\bk_1,\bk_2,\bk_3) \approx
 \frac{b_1(M)}{3}\left[2+b_1(M)\right] 
\left[P^{(0)}_{\rm mm}(k_1)P_{\rm mm}^{(0)}(k_2)+2\,\cyc\,\right] \ .
\ee
Hence we have,
\be 
Q_{\rm hmm}^{(0)}(\bk_1,\bk_2,\bk_3) \approx \frac{3}{2+b_1(M)}Q^{(0)}_{\rm mmm}(\bk_1,\bk_2,\bk_3)
+\frac{b_2(M)}{2b_1(M)+b_1^2(M)} 
\left[\frac{\widetilde{W}_{\bk_1,\bk_2}P^{(0)}_{\rm mm}(k_1)P^{(0)}_{\rm mm}(k_2)+2\,\cyc\,}
{P^{(0)}_{\rm mm}(k_1)P^{(0)}_{\rm mm}(k_2)+2\,\cyc\,}\right]
\ee
Finally, in the limit that $k_iR\ll1$, $\widetilde{W}\rightarrow 1$
and the above result can be approximated by
\be 
Q_{\rm hmm}^{(0)}(\bk_1,\bk_2,\bk_3) \approx \frac{3}{2+b_1(M)}Q^{(0)}_{\rm mmm}(\bk_1,\bk_2,\bk_3)
+\frac{b_2(M)}{2b_1(M)+b_1^2(M)} \ .
\ee

%%%%%%%%%%%%%%%%%%%%%%%%%%%%%%%%%%%%%%%%%%%%%%%%%%%%%%%

\subsection{Halo-halo-mass bispectrum in the local model}

Again, using \Eqn{eq:halobias2}, the halo-halo-mass bispectrum,
symmetrized in the $k_i$ arguments, can be written:
\ba 
B_{\rm hhm}(\bk_1,\bk_2,\bk_3) & = & 
b_1^2(M) B_{\rm mmm}(\bk_1,\bk_2,\bk_3) +\frac{b_1(M)b_2(M)}{3} 
 \int \frac{\dq_1}{(2\pi)^3} 
\left[\widetilde{W}_{\bq_1,\bk_1-\bq_1} T_{\rm mmmm}(\bq_1,\bk_1-\bq_1,\bk_2,\bk_3) +2\,\cyc\,\right] \nn \\
& & + \frac{b_2^2(M)}{12} \int \frac{\dq_1}{(2\pi)^3}  \frac{\dq_2}{(2\pi)^3} 
\left[\widetilde{W}_{\bq_1,\bk_1-\bq_1}  \widetilde{W}_{\bq_2,\bk_2-\bq_2}
P_{\rm 5,m}(\bq_1,\bk_1-\bq_1,\bq_2,\bk_2-\bq_2,\bk_3) 
+2\,\cyc\,\right] \ .
\ea
If we use perturbation theory to expand $P$, $B$, $T$ and $P_{5}$, and
only keep terms that are $4^{\rm th}$ order in the density field, then
the above expression can be approximated by:
\be
B^{(0)}_{\rm hhm}(\bk_1,\bk_2,\bk_3) \approx 
b_1^2(M) B_{\rm mmm}^{(0)}(\bk_1,\bk_2,\bk_3) 
+\frac{1}{3}b_1(M)b_2(M) \left[ \widetilde{W}_{\bk_1,\bk_2} P_{\rm mm}^{(0)}(k_1)P_{\rm mm}^{(0)}(k_2) 
+2\,\cyc\,\right] \ .
\ee
In the large-scale limit $k_iR\rightarrow0$, the above expression can
be approximated as:
\be
B^{(0)}_{\rm hhm}(\bk_1,\bk_2,\bk_3)\approx
b_1^2(M) B_{\rm mmm}^{(0)}(\bk_1,\bk_2,\bk_3) +\frac{1}{3}b_1(M)b_2(M) \left[ P_{\rm mm}^{(0)}(k_1)P_{\rm mm}^{(0)}(k_2) 
+2\,\cyc\,\right] \ .
\ee

The reduced bispectrum is given by,
\be 
Q_{\rm hhm}(\bk_1,\bk_2,\bk_3) \equiv \frac{B_{\rm hhm}(\bk_1,\bk_2,\bk_3)}{PP_{\rm hhm}} \ ,
\ee
where 
\be 
PP_{\rm hhm} = 
\frac{2}{3}\left[P_{\rm hh}(k_1)P_{\rm hm}(k_2)+2\,\cyc\,\right]+
\frac{1}{3}\left[P_{\rm hm}(k_1)P_{\rm hm}(k_2)+2\,\cyc\,\right]\ .
\ee
The halo-mass power spectrum is given by \Eqn{eq:Phm} and the halo
auto-power spectrum is given by:
\ba
P_{\rm hh}(k) & = &
 b_1^2(M) P_{\rm mm}(\bk_1) + b_1(M)b_2(M) 
\int \frac{\dq_1}{(2\pi)^3} 
\widetilde{W}_{\bq_1,\bk_1-\bq_1} B_{\rm mmm}(\bq_1,\bk_1-\bq_1,-\bk_1) \nn  \\
& & +
\frac{b_2^2(M)}{4} \int \frac{\dq_1}{(2\pi)^3}\frac{\dq_2}{(2\pi)^3} 
\widetilde{W}_{\bq_1,\bk_1-\bq_1}\widetilde{W}_{\bq_2,-\bk_1-\bq_2}
T_{\rm mmmm}(\bq_1,\bk_1-\bq_1,\bq_2,-\bk_1-\bq_2)  \ . \label{eq:Phh}
\ea
Expanding the above expression to lowest order in perturbation theory
gives,
\be 
PP_{\rm hhm}^{(0)} \approx 
\frac{2b_1^2(M)}{3}\left[2b_1(M)+1\right]
\left[P_{\rm mm}^{(0)}(k_1)P_{\rm mm}^{(0)}(k_2)+\,2\,\cyc\,\right] \ .
\ee
Using the above expression, we find that the tree-level expression for
the reduced bispectrum can be written:
\be Q_{\rm hhm}^{(0)}(\bk_1,\bk_2,\bk_3) \approx
\frac{3}{2b_1(M)+1}Q^{(0)}_{\rm mmm}(\bk_1,\bk_2,\bk_3)
+\frac{2b_2(M)}{2b_1^2(M)+b_1(M)}
\left[\frac{\widetilde{W}_{\bk_1,\bk_2}P^{(0)}_{\rm mm}(k_1)P^{(0)}_{\rm mm}(k_2)+2\,\cyc\,}
{P^{(0)}_{\rm mm}(k_1)P^{(0)}_{\rm mm}(k_2)+2\,\cyc\,}\right]\ . \ee
In the large-scale limit $k_iR\rightarrow0$, we again have
$\widetilde{W}\rightarrow1$ and 
\be 
Q_{\rm hhm}^{(0)}(\bk_1,\bk_2,\bk_3) \approx 
\frac{3}{2b_1(M)+1}Q^{(0)}_{\rm mmm}(\bk_1,\bk_2,\bk_3)
+\frac{2b_2(M)}{2b_1^2(M)+b_1(M)} \ .
\ee

%%%%%%%%%%%%%%%%%%%%%%%%%%%%%%%%%%%%%%%%%%%%%%%%%%%%%%%

\subsection{Halo-halo-halo bispectrum in the local model}

Again using \Eqn{eq:halobias2}, the halo-halo-halo bispectrum,
symmetrized in the $k_i$ arguments, can be written:
\ba 
B_{\rm hhh}(\bk_1,\bk_2,\bk_3) & = & 
b_1^3(M) B_{\rm mmm}(\bk_1,\bk_2,\bk_3) +\frac{1}{2} b_1^2(M)b_2(M) 
\int \frac{\dq_1}{(2\pi)^3} 
\left[\widetilde{W}_{\bq_1,\bk_1-\bq_1} T_{\rm mmmm}(\bq_1,\bk_1-\bq_1,\bk_2,\bk_3) +2\,\cyc\,\right] \nn \\
& & + \frac{1}{4} b_1b_2^2 \int \frac{\dq_1}{(2\pi)^3}  \frac{\dq_2}{(2\pi)^3} 
\left[\widetilde{W}_{\bq_1,\bk_1-\bq_1} \widetilde{W}_{\bq_2,\bk_2-\bq_2} 
P_{\rm 5,m}(\bq_1,\bk_1-\bq_1,\bq_2,\bk_2-\bq_2,\bk_3) +2\,\cyc\,\right]  \nn \\
& &  + \frac{b_2^3}{8} \int \frac{\dq_1}{(2\pi)^3} \frac{\dq_2}{(2\pi)^3} \frac{\dq_3}{(2\pi)^3} 
\widetilde{W}_{\bq_1,\bk_1-\bq_1} 
\widetilde{W}_{\bq_2,\bk_2-\bq_2}
\widetilde{W}_{\bq_3,\bk_3-\bq_3}
P_{\rm 6,m}(\bq_1,\bk_1-\bq_1,\bq_2,\bk_2-\bq_2,\bq_3,\bk_3-\bq_3) \ . \nn \\
\ea
If we use perturbation theory to expand $P$, $B$, $T$, $P_{5}$, and
$P_6$, and keep only terms that are fourth order in the density field,
then the above expression can be approximated by:
\be
B_{\rm hhh}^{(0)}(\bk_1,\bk_2,\bk_3) \approx
b_1^3(M) B^{(0)}_{\rm mmm}(\bk_1,\bk_2,\bk_3) + b_1^2(M)b_2(M) 
\left[\widetilde{W}_{\bk_1,\bk_2} P_{\rm mm}^{(0)}(k_1)P_{\rm mm}^{(0)}(k_2)  + 2\,\cyc\,\right]\ .
\ee
In the large-scale limit $k_iR\rightarrow0$, the above expression can
be approximated as:
\be
B^{(0)}_{\rm hhh}(\bk_1,\bk_2,\bk_3)\approx
b_1^3(M) B_{\rm mm}^{(0)}(\bk_1,\bk_2,\bk_3) +b^2_1(M)b_2(M) 
\left[ P_{\rm mm}^{(0)}(k_1)P_{\rm mm}^{(0)}(k_2) +2\,\cyc\,\right] \ .
\ee

The reduced halo-halo-halo bispectrum is given by,
\be 
Q_{\rm hhh}(\bk_1,\bk_2,\bk_3) \equiv \frac{B_{\rm hhh}(\bk_1,\bk_2,\bk_3)}{PP_{\rm hhh}} \ ,
\ee
where 
\be 
PP_{\rm hhh} = \left[P_{\rm hh}(k_1)P_{\rm hh}(k_2)+2\,\cyc\,\right]\ ,
\ee
where the halo auto-power spectrum is given by \Eqn{eq:Phh}. On
expanding the above expression to fourth order in the density we find,
\be 
PP_{\rm hhh}^{(0)} \approx b_1^4(M)\left[P_{\rm mm}^{(0)}(k_1)P_{\rm mm}^{(0)}(k_2)+\,2\,\cyc\,\right] \ .
\ee
Using the above expression, we find that the tree-level expression for
the reduced bispectrum can be written:
\be Q_{\rm hhh}^{(0)}(\bk_1,\bk_2,\bk_3) \approx
\frac{1}{b_1(M)}Q^{(0)}_{\rm mmm}(\bk_1,\bk_2,\bk_3)
+\frac{b_2(M)}{b_1^2(M)}
\left[\frac{\widetilde{W}_{\bk_1,\bk_2}P^{(0)}_{\rm mm}(k_1)P^{(0)}_{\rm mm}(k_2)+2\,\cyc\,}
{P^{(0)}_{\rm mm}(k_1)P^{(0)}_{\rm mm}(k_2)+2\,\cyc\,}\right]\ . \ee
In the large-scale limit $k_iR\rightarrow0$, we again have
$\widetilde{W}\rightarrow1$ and 
\be 
Q_{\rm hhh}^{(0)}(\bk_1,\bk_2,\bk_3) \approx 
\frac{1}{b_1(M)}Q^{(0)}_{\rm mmm}(\bk_1,\bk_2,\bk_3)
+\frac{b_2(M)}{b_1^2(M)} \ .
\ee

%%%%%%%%%%%%%%%%%%%%%%%%%%%%%%%%%%%%%%%%%%%%%%%%%%%%%%%

\end{document}